\documentclass[aps,twocolumn,superscriptaddress]{revtex4-1} 
\usepackage[english]{babel}
\usepackage{amsmath, bm, amssymb}
\usepackage{mathtools}
\usepackage{siunitx}
\usepackage{mathrsfs}
\usepackage{braket}
\usepackage[]{hyperref}
\DeclareMathAlphabet{\mathscrlower}{OT1}{pzc}{m}{it} 
\usepackage{prettyref}
\usepackage{graphicx}
\usepackage{booktabs}
\usepackage{threeparttable}
\usepackage{placeins}
\usepackage[stable]{footmisc}
\usepackage{mhchem}
\usepackage{xfrac}
\usepackage[normalem]{ulem}
\usepackage{xcolor}
\usepackage{adjustbox}

\newcommand{\pauli}{\boldsymbol{\sigma}}

\newcommand{\diraccontra}[1]{\boldsymbol{\gamma}^{#1}}

\newcommand{\diraca}{\vec{\boldsymbol{\alpha}}}
\newcommand{\diracg}{\vec{\boldsymbol{\gamma}}}
\newcommand{\diracb}{\boldsymbol{\beta}}

\newcommand{\pos}{\vec{r}}

\newcommand{\momop}{\hat{\vec{p}}}

\newcommand{\efield}{\mathcal{E}}

\newcommand{\Sum}[2]{\sum\limits_{#1}^{#2}}

\newcommand{\parantheses}[1]{\left(#1\right)}

\newcommand{\braces}[1]{\left\{ #1\right\}}
\let\nablatmp\nabla
\renewcommand{\nabla}{\vec{\nablatmp}}
\DeclarePairedDelimiter\abs{\lvert}{\rvert}
\makeatletter
\let\oldabs\abs
\def\abs{\@ifstar{\oldabs}{\oldabs*}}

\newcommand{\op}[1]{\hat{#1}}

\AtBeginDocument{
\heavyrulewidth=.08em
\lightrulewidth=.05em
\cmidrulewidth=.03em
\belowrulesep=.65ex
\belowbottomsep=0pt
\aboverulesep=.4ex
\abovetopsep=0pt
\cmidrulesep=\doublerulesep
\cmidrulekern=.5em
\defaultaddspace=.5em
}
\begin{document}
\title{Global analysis of $\mathcal{CP}$-violation in atoms, molecules and
role of medium-heavy systems}
\thanks{Parts of this work have been published in preliminary form in
Konstantin J. Gaul, PhD thesis, Philipps-Universit\"at Marburg,
\doi{10.17192/z2021.0109}.}
\date{\today}
\author{Konstantin Gaul}
\affiliation{Fachbereich Chemie, Philipps-Universit\"{a}t Marburg, Hans-Meerwein-Stra\ss{}e 4, 35032 Marburg}
\author{Robert Berger}
\affiliation{Fachbereich Chemie, Philipps-Universit\"{a}t Marburg, Hans-Meerwein-Stra\ss{}e 4, 35032 Marburg}
\begin{abstract}
Detection of parity ($\mathcal{P}$) and time-reversal
($\mathcal{T}$) symmetry-odd electric dipole moments (EDMs) within currently achievable resolution
would evidence physics beyond the Standard Model of particle physics.
Via the $\mathcal{CPT}$-theorem, which includes charge conjugation
($\mathcal{C}$), such low-energy searches complement high-energy physics experiments that
probe $\mathcal{CP}$-violation up to the TeV
scale. Heavy-elemental atoms
and molecules are considered to be among the most promising candidates
for a first direct detection of $\mathcal{P,T}$-violation due to
enhancement effects that increase steeply with increasing nuclear
charge number $Z$. However, different $\mathcal{P,T}$-odd sources 
on the subatomic level can contribute to
molecular or atomic EDMs,
which are target of measurements, and this complicates obtaining rigorous
bounds on $\mathcal{P,T}$-violation on a fundamental level.
Consequently, several experiments of
complementary sensitivity to these individual $\mathcal{P,T}$-odd sources
are required for this purpose.
Herein, a simply-applicable qualitative model is developed for
global analysis of the $\mathcal{P,T}$-odd parameter space from
an electronic-structure theory perspective. Rules
of thumb are derived for the choice of atoms and molecules in terms of their
angular momenta and nuclear charge number. Contrary to naive
expectations from $Z$-scaling laws, it is demonstrated that
medium-heavy molecules with $Z\leq54$ can be of great value to tighten
global bounds on $\mathcal{P,T}$-violating parameters, in particular,
if the number of complementary experiments increases. The model is
confirmed by explicit density functional theory calculations of all
relevant $\mathcal{P,T}$-odd electronic structure parameters in
systems that were used in past experiments or are of current interest
for future experiments, respectively: the atoms Xe, Cs, Yb, Hg, Tl, Ra, Fr
and the molecules CaOH, SrOH, YO, CdH, BaF, YbF, YbOH, HfF$^+$, WC,
TlF, PbO, RaF, ThO, ThF$^+$ and PaF$^{3+}$.
\end{abstract}

\maketitle

\section{Introduction}
Permanent electric dipole moments (EDMs) of atoms and molecules are
powerful probes of a simultaneous violation of the fundamental
symmetries parity and time-reversal ($\mathcal{P,T}$-violation)
\cite{demille:2015}. Due to the $\mathcal{CPT}$-theorem
\cite{schwinger:1951,luders:1954,pauli:1955}, which includes also
charge-conjugation $\mathcal{C}$ symmetry, $\mathcal{P,T}$-violation
signals indirectly the violation of combined $\mathcal{CP}$-symmetry
\cite{khriplovich:1997}. This $\mathcal{CP}$-violation is regarded 
to be a crucial ingredient for baryogenesis \cite{sakharov:1967}, but
the current Standard Model of particle physics (SM) predicts
$\mathcal{CP}$-violating effects to be so tiny that
they are considered insufficient to explain
baryogenesis. Physical theories beyond the SM (BSM), in contrast, typically include processes
that can induce larger $\mathcal{CP}$-violating effects \cite{gross:1996}.

In the present work, we focus on ordinary atoms and molecules, which are composed of
electrons and of atomic nuclei made from nucleons, i.e., protons and neutrons that by
themselves consist of up- and down-quarks as well as gluons, but we do not include e.g.\
myonic systems or hyperonic nuclei in our discussions. Possible
$\mathcal{P,T}$-odd effects that appear on the fundamental quark
and electron level together with resulting net $\mathcal{P,T}$-odd
effects acquired on subsequent lower energy-scales can, thus,
contribute to a net $\mathcal{P,T}$-odd EDM of an atom or molecule
(see also Ref.~\cite{fischler:1992}).  EDMs of charged fermionic
elementary particles (see Ref.  \cite{salpeter:1958}),
$\mathcal{P,T}$-odd quark-quark, quark-electron and electron-electron
current interactions can produce $\mathcal{P,T}$-odd moments of
hadrons, nuclei, atoms and molecules. A full Lagrangian of
$\mathcal{P,T}$-odd effects in the electron-quark sector is given,
e.g., in Ref. \cite{fischler:1992}.

Such $\mathcal{P,T}$-odd effects in atoms and molecules usually
increase steeply with increasing nuclear charge  (see e.g. Reviews
\cite{khriplovich:1997,ginges:2004}) and, thus, on-going and new
experiments are usually developed with heavy elements or heavy-elemental
compounds, starting with Xe ($Z=54$), the last element in the 5th row of the
periodic table. 

\begin{figure*}
\centering
\includegraphics[width=0.75\textwidth]{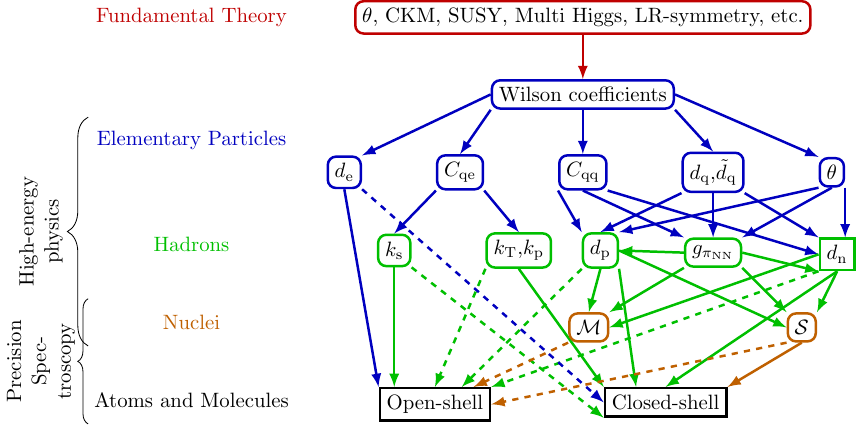}
\caption{Overview of the sources of $\mathcal{P,T}$-violation on the
fundamental particle level predicted in the SM and various
BSM theories. Arrows on solid lines indicate main contributions
and arrows on dashed lines less important contributions. Similar
overviews considering parts of this scheme can be found in Refs.
\cite{ginges:2004,poseplov:2005,chupp:2019,cairncross:2019}. See section
\ref{sec:sources} for the definition of the various symbols.}
\label{fig: sources_ptviolation}
\end{figure*}

\begin{figure}
\centering
\includegraphics[width=0.5\textwidth]{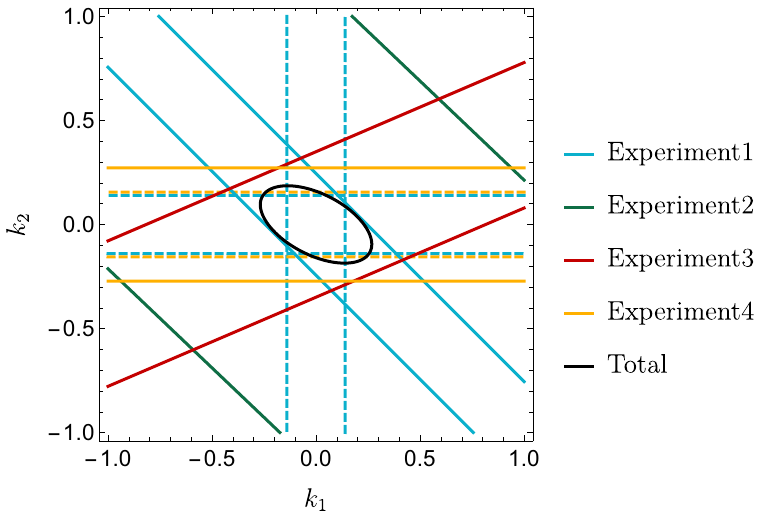}
\caption{Exemplary demonstration of bounds in a multivariate
measurement model following Ref.~\cite{jcgm102:2011} featuring two fundamental sources of
$\mathcal{P,T}$-violation with interaction strength $k_1$ and $k_2$.
Assuming Gaussian probability distributions of experimental
uncertainties, from an individual experiment $i$ we obtain a coverage 
region in the $\mathcal{P,T}$-odd parameter space which is not excluded by
the measurement in form of a parabola $\sigma_i^{-2}(W_{i1} k_1 + W_{i2}k_2)^2=P^2$, with 
$W_{ij}$ denoting the specific enhancement factor for source $j$
in experiment $i$. The slope in the $k_1,k_2$ plane is
determined by the ratio of enhancement factors via $-W_{i2}/W_{i1}$ for each
experiment $i$, whereas the width of the infinite coverage region is determined by the focal
width of the parabola, which is $2P\sigma_i/\sqrt{W_{i1}^2+W_{i2}^2}$. In total, 
the resulting coverage region of more than one experiment has an elliptical shape and is,
for equal standard uncertainties $\sigma_i=\sigma$, 
of the form
$\sigma^{-2}\sum_i (W_{i1} k_1 + W_{i2} k_2)^2=P^2$, where $i$ runs over the
number of experiments. The global coverage region is then equal to the area
of this ellipse, which is $P^2\sigma^2\pi/\sqrt{(\sum_i W_{i1}W_{i2}
)^2-\sum_{ij} W_{i1}^2W_{j2}^2}$. Results are shown for 
$W_{11}=W_{12}=10$ for experiment 1, $W_{21}=-2,W_{22}=-2.1$ for experiment 2,
$W_{31}=-3, W_{32}=7$ for experiment 3, $W_{41}=0.001, W_{42}=9$ for 
experiment 4 and with 95\,\% confidence level (CL) ($P^2\approx5.99$).
For comparison the single source limits are shown for experiments 1
and 4 as dotted lines ($P^2\approx3.84$).} 
\label{fig: multivar_measure}
\end{figure}

As many different sources can in principle contribute to the $\mathcal{P,T}$-odd atomic or
molecular EDM that is target of measurements, the interpretation of
corresponding EDM experiments in terms of fundamental parameters is all but trivial. Present limits on
parameters on the level of elementary particles are commonly derived in a so-called
single source model, i.e., only \emph{one} fundamental $\mathcal{P,T}$-odd
parameter is assumed to be existent and bound by a given experiment. Within this crude
approximation it is concluded, for example, that the ThO experiment sets a limit of
$d_\mathrm{e}<\SI{1.1e-29}{\elementarycharge\centi\metre}$ on the EDM
of the electron (eEDM)~\cite{andreev:2018}. This limit would imply an
exclusion of several BSM models or a restriction of their parameter
spaces.
There appears to be no good reason to believe, however, that such a
single-source scenario is realistic. Multivariate measurement models
should be applied, instead, to extract limits on fundamental
parameters on the elementary particle level (see discussion in
Refs.~\cite{barr:1992,jung:2013,engel:2013,chupp:2015,fleig:2018,chupp:2019,gaul:2019,mohanmurthy:2021,ramseymusolf:2021}
and Figure \ref{fig: multivar_measure}).

Once a non-zero atomic or molecular EDM is successfully measured for the 
first time, its interpretation and implications for BSM physics require 
complementary experiments and likely also further advances in the 
theoretical description of $\mathcal{P,T}$-odd effects in atoms and molecules.

In this paper we address, purely from the electronic structure
perspective, the question, which molecular or atomic systems are best
to be chosen for future experiments in order to provide complementary
information on the \emph{full} $\mathcal{P,T}$-odd parameter space.
This means we do not exclude \emph{a priori} any possible fundamental
source of $\mathcal{P,T}$-violation except $\mathcal{P,T}$-odd
electron-electron current interactions, but abstract from specific
further enhancements that may emerge on the nuclear structure level.

The purpose of the present paper is not to obtain rigorous bounds on
the fundamental particle level so far, but rather to provide a framework to
identify systems that have a suitable \emph{electronic structure} to
improve the present status of measurements. 
For our purpose, we apply a simple phenomenological model for the
description of the electronic structure based on effective atomic
one-electron wave-functions. Within this model we minimize the
coverage region in the global $\mathcal{P,T}$-odd parameter space
employing a multivariate measurement model \cite{jcgm102:2011}.
Variables in these minimizations involve the electronic
angular momentum, the nuclear angular momentum and the nuclear charge
number of the systems that would be used in experiments. We support our
conclusions from this global minimization by explicit density
functional theory calculations of systems that were or are relevant
for past and future experiments that search for
$\mathcal{P,T}$-violation. The model derived and conclusions drawn
serve as a qualitative guide for design of future experiments and can
aid to identify well-suited complementary molecular systems that can
provide new information on the $\mathcal{P,T}$-odd parameter space.
Our model and provided \emph{ab initio} data can be combined in future
studies with a more sophisticated treatment of
nuclear structure to obtain rigorous bounds on the fundamental
particle level in a global analysis.

\section{Theory}
\subsection{Sources of $\mathcal{P,T}$-violation and effective electronic Hamiltonian\label{sec:sources}}
The hypothetical fundamental sources of $\mathcal{P,T}$-violation
comprise the permanent EDMs of the electron $d_\mathrm{e}$ and quarks
$d_\mathrm{q}$, color EDMs of quarks $\tilde{d}_\mathrm{q}$, the
strong $\mathcal{CP}$-violating coupling angle $\theta$ as well as
$\mathcal{P,T}$-odd electron-electron, electron-quark and quark-quark
currents characterized by the coupling constants $C_\mathrm{ee}$,
$C_\mathrm{eq}$ and $C_\mathrm{qq}$, respectively
\cite{bernreuther:1991,fischler:1992,ginges:2004,poseplov:2005,engel:2013,chupp:2019,cairncross:2019}.
We neglect in the following possible $\mathcal{P,T}$-odd
electron-electron current interactions. The contribution of all other
sources to the net $\mathcal{P,T}$-odd EDM of atoms and molecules via
various mechanisms are shown in Fig.~\ref{fig: sources_ptviolation}.
On the level of electronic structure theory, the direct sources of
$\mathcal{P,T}$-violation in an atom or molecule are a permanent
electric dipole moment of the electron (eEDM) $d_\mathrm{e}$, a
permanent electric dipole moment of the proton (pEDM) $d_\mathrm{p}$,
a permanent electric dipole moment of the neutron (nEDM)
$d_\mathrm{n}$, interactions with net $\mathcal{P,T}$-odd moments of
the nucleus, namely the collective Schiff moment $\mathcal{S}$, the
nuclear magnetic quadrupole moment (NMQM) $\mathcal{M}$ and higher
moments as well as $\mathcal{P,T}$-odd scalar-pseudoscalar
nucleon-electron current (SPNEC) interactions, tensor-pseudotensor
nucleon-electron current (TPNEC) interactions, and pseudoscalar-scalar
nucleon-electron current (PSNEC) interactions with coupling constants
$k_\mathrm{s}$, $k_\mathrm{T}$, and $k_\mathrm{p}$, respectively. We
use here the following definitions of nucleon-electron current
interaction parameters: $k_\mathrm{s}=\frac{A}{Z}C_\mathrm{s}$,
$k_\mathrm{T}=\Braket{\pauli_z}C_\mathrm{T}$ and
$k_\mathrm{p}=\Braket{\pauli_z}C_\mathrm{p}$, where
$\Braket{\pauli_z}$ is the nucleon spin expectation value (see Ref.
\cite{chupp:2019} for the definitions of the constants $C$). 

For the present purpose we want to develop a model that is independent
of the choice of specific elements of the periodic table. Thus, we
should express $\mathcal{P,T}$-violation parameters on the level of
hadrons, which are the lowest feasible energy scale for this purpose.
Furthermore we will concentrate on \emph{electronic} structure only
and want to keep our model to be independent of \emph{nuclear}
structure theory as the latter can not provide presently equally
reliable predictions of $\mathcal{P,T}$-violation in many systems that
are considered for EDM experiments. To achieve a qualitative
description of the $\mathcal{P,T}$-odd nuclear moments $\mathcal{S}$,
$\mathcal{M}$ in terms of the short-range nucleon EDMs (NEDMs)
$d_\mathrm{p}$, $d_\mathrm{n}$ (see e.g. Refs.
\cite{ginges:2004,chupp:2019} for details) and an effective
$\mathcal{P,T}$-odd nucleon-nucleon and nucleon-pion current (NNC)
interaction constant $g_{\pi\mathrm{NN}}\sim g(\bar{g}_0 + a_1
\bar{g}_1 - 2 a_2\bar{g}_2)$ where $a_1$ and $a_2$ depend on nuclear
structure in principle but are assumed herein to be constant, we use the
following crude approximation (see Reviews
\cite{khriplovich:1997,ginges:2004,engel:2013,chupp:2019} for a
discussion):
\begin{align}
\mathcal{S}_\mathrm{EDM} &\approx R_\mathrm{vol}
(a_\mathrm{p}d_\mathrm{p}+a_\mathrm{n}d_\mathrm{n}) \label{eq: nucstruc_start}\\
\mathcal{S}_\pi &\approx \tilde{\mathcal{S}} g_{\pi\mathrm{NN}}\\
\mathcal{M}_\mathrm{EDM} &\approx \tilde{\mathcal{M}}_\mathrm{EDM}
(a_\mathrm{p}d_\mathrm{p}+a_\mathrm{n}d_\mathrm{n})\\
\mathcal{M}_\pi &\approx \tilde{\mathcal{M}}_\pi g_{\pi\mathrm{NN}}\,.
\end{align}  
Here it is assumed that there is a single external valence nucleon
and we set $a_\mathrm{p}$ to zero for a valence neutron or one for a
valence proton and vice versa for $a_\mathrm{n}$.
We follow closely Refs. \cite{khriplovich:1997,ginges:2004} to
approximate the individual nuclear structure factors as
\begin{align}
R_\mathrm{vol} &\sim \frac{3}{50} A^{2/3}
(\SI{1.2}{\femto\meter})^2\frac{3+2I}{3+3I}\\
\tilde{\mathcal{S}} &\sim
\left(0.02\frac{18 e}{20\pi\sqrt{35}} \frac{0.14 Z}{A^{1/3}}
(\SI{1.2}{\femto\meter}A^{1/3})^3 
\nonumber\right.\\&\left.
+ \frac{3e q_\mathrm{ext}}{50}\SI{0.14}{\femto\meter}(\SI{1.2}{\femto\meter}
A^{1/3})^2\right)\frac{3+2I}{3+3I}\\
\tilde{\mathcal{M}}_\mathrm{EDM} &\sim
\frac{-3}{2m_\mathrm{p}}\frac{1}{I+1}\\
\tilde{\mathcal{M}}_\pi &\sim
\frac{3e(\mu/\mu_\mathrm{N}-q_\mathrm{ext})}{2m_\mathrm{p}}\SI{0.14}{\femto\meter} \frac{1}{I+1}
\label{eq: nucstruc_end}
\end{align}
For details see in particular  eqs.  (177), (182), (183), (185), (189)
and (190) of Ref.  \cite{ginges:2004}.  Here $I$ is the quantum number
of the  nuclear angular momentum, $A$ the nuclear mass number, $Z$
is the nuclear charge number, $\mu$ the nuclear magnetic moment,
$\mu_\mathrm{N}$ is the nuclear magneton, $q_\mathrm{ext}$ the charge number of the
external nucleon, $e$ the elementary electric charge and
$m_\mathrm{p}$ 
the proton mass. In addition to the valence nucleon contribution we
introduced the first term of $\tilde{\mathcal{S}}$, which stems from
the collective Schiff moment in a deformed nucleus, where
$\beta_2\beta_3^2\frac{\si{\electronvolt}}{\abs{E_+-E_-}}\approx2\times10^{-8}$
is the typical size of the quadrupole and octupole deformations
$\beta_{2,3}$, weighted by the splitting of states of opposite parity
$\abs{E_+-E_-}$ as suggested in eqs. (203) and (204) of Ref.
\cite{ginges:2004} and we assumed the effective nucleon-nucleon
current interaction to be of strength $\sim\SI{7e6}{}g_\mathrm{\pi
NN}$. We included a fixed additional factor of 0.02 to guarantee that
the collective Schiff moment contribution becomes only dominant for
nuclei with $Z\gtrsim50$, as for lighter nuclei octupole deformations
are seldom. The second term in $\tilde{\mathcal{S}}$ stems from the
external nucleon and is according to Ref. \cite{ginges:2004} (like the
first term) proportional to
$\SI{2e-21}{\centi\meter}\times\SI{7e6}{}g_\mathrm{\pi
NN}\approx\SI{0.14}{\femto\meter}g_\mathrm{\pi NN}$. In all equations
above we take the case of a nucleus with total angular momentum
quantum number $I=l-1/2$, where $l$ is the
orbital angular momentum quantum number of the nucleus. The NMQM is
defined such that the NMQM tensor is
$\bm{\mathcal{M}}=\mathcal{M}\bm{\mathsf{T}}$ with $\bm{\mathsf{T}}$
being a second-rank tensor with components
$T_{ij}=I_{i}I_{j}+I_{j}I_{i}-\frac{2}{3}\delta_{ij}I(I+1)$, where
$I_{j}$ are the components of the nuclear angular momentum $\vec{I}$.  Here and in
the following we use the relation $A= 0.004467 Z^2 + 2.163 Z - 1.168 $
\cite{andrae:2000} to express the nuclear and electronic structure parameters as
function of $Z$ only. 

All sources described above contribute to the total atomic or
molecular EDM Hamiltonian $H^\mathcal{P,T}$ that gives rise to the
atomic or molecular permanent EDM:
\begin{equation}
d_\mathrm{atom,mol} = \frac{H^\mathcal{P,T}}{\vec{\efield}\cdot\vec{F}},
\end{equation}
where $\vec{\efield}$ is the external electric field of strength
$\efield$ that polarizes the atom or molecule and $\vec{F}$ is the
total angular momentum. Assuming the external electric field to be
oriented along the $z$-axis in the laboratory frame, i.e. $\vec{\efield}=\begin{pmatrix}0,
0, \efield\end{pmatrix}^\mathsf{T}$, we can write
for Hund's coupling cases (a), (b) and (c)
$\vec{\efield}\cdot\vec{F}=\efield F$ with
$F=\lambda_z (\Omega+\mathcal{I})$, where
$\Omega=\vec{S}'\cdot\vec{\lambda}$,
$\mathcal{I}=\vec{\lambda}\cdot\vec{I}$~\cite{kozlov:1995}. Here
$\vec{\lambda}$ is a unit vector pointing from the heavy to the light
atom, $\vec{I}$ is the nuclear angular momentum of the heaviest
nucleus in the molecule and $\vec{S}'$ is the  effective electronic
spin. For coupling case (a) and (c) we have
$\vec{S}'=\vec{J}_\mathrm{e}$ and for coupling case (b)
$\vec{S}'=\vec{J}_\mathrm{e}=\vec{L}+\vec{S}$, where
$\vec{J}_\mathrm{e}$ is the total electronic angular momentum and
$\vec{L}$, $\vec{S}$ are the orbital and spin electronic angular
momenta, respectively. 

In our considerations, we include atoms and linear molecules with open
and closed electronic shells and account also for different
possibilities of nuclear spin. Linear polyatomic molecules are treated
as rigid entities so that additional vibrational angular momentum
terms are neglected in the present study. In open-shell atoms and molecules that
have a nucleus with nuclear angular momentum quantum number $I> {}^1/_2$ all sources of
$\mathcal{P,T}$-violation discussed above can contribute. We define
the effective Hamiltonian $H^\mathcal{P,T}$ for such a system  for
Hund's coupling case (b) with the projection of the
\emph{electronic} orbital angular momentum on the molecular axis
being $\Lambda=0$ and for cases (a) and (c) following Refs.
\cite{kozlov:1995,hinds:1980} as
\begin{widetext}
\begin{equation}
\begin{aligned}[t]
{H^\mathcal{P,T}} &=
\Omega
\parantheses{W_{\mathrm{s}}k_\mathrm{s}+W_{\mathrm{d}}d_\mathrm{e}}
+\Theta
\parantheses{
W_{\mathcal{M}}\tilde{\mathcal{M}}_{\pi}g_{\pi\mathrm{NN}}
+W_{\mathcal{M}}\tilde{\mathcal{M}}_{\mathrm{EDM}}(a_\mathrm{n}d_\mathrm{n}+a_\mathrm{p}d_\mathrm{p})
}
+ \text{higher nuclear moments\dots}\\                                     
&+\mathcal{I}
\parantheses{
W_{\mathrm{T}} k_\mathrm{T}
+W_{\mathrm{p}} k_\mathrm{p}
+W^\mathrm{m}_{\mathrm{s}}\gamma k_\mathrm{s}
+W_{\mathcal{S}}\tilde{\mathcal{S}}_{A} g_{\pi\mathrm{NN}}
+W^\mathrm{m}_{\mathrm{d}}\gamma d_\mathrm{e}
+W_{\mathrm{m}}\parantheses{\eta_{\mathrm{p}} a_\mathrm{p}d_\mathrm{p}
+\eta_{\mathrm{n}} a_\mathrm{n} d_\mathrm{n}}
+W_{\mathcal{S}}R_\mathrm{vol}(a_\mathrm{n}d_\mathrm{n}+a_\mathrm{p}d_\mathrm{p}) 
}\,,
\label{eq: ptodd_spinrot}
\end{aligned}
\end{equation}
\end{widetext}
where the magnetogyric ratio is $\gamma=\mu/I$ and
$\eta_{\mathrm{p}}=\frac{\mu_\mathrm{N}}{A}+\frac{\mu}{Z}$,
$\eta_{\mathrm{n}}=\frac{\mu_\mathrm{N}}{A}+\frac{\mu}{A-Z}$.  For an
atom in an external electric field, we set $\Omega=J_\mathrm{e}$ and
$\mathcal{I}=I$. Here,
$\Theta=\vec{\lambda}^\mathsf{T}\cdot\bm{\mathsf{T}}\cdot\vec{S}'$.
For closed-shell atoms and molecules, all terms that depend on the
effective electron spin, i.e.\ those proportional to $\Omega$ and
$\Theta$, vanish in this approximation. The various constants $W_i$ are electronic structure coupling
constants enhancing $\mathcal{P,T}$-violation in atoms and molecules,
that need to be determined by electronic structure calculations.
Explicit forms of the $\mathcal{P,T}$-odd electronic structure
coupling constants $W_i$ are given in the appendix and can be found
e.g.\ in Refs. \cite{khriplovich:1997,kozlov:1995,hinds:1980} and
references therein. In the equation above, we can set
$\eta_{\mathrm{p}}\approx \eta_{\mathrm{n}}$ (see Appendix
\ref{analytic_enhancement} near \prettyref{eq: Hmp}). This allows us to define an effective
valence nucleon EDM $d_\mathrm{N} = (a_\mathrm{n} d_\mathrm{n}+
a_\mathrm{p} d_\mathrm{p})$ as
an effective model parameter being independent from details of the nuclear
structure. With this choice we do not distinguish between an external
valence proton and neutron in the following. We leave the 
nuclear structure unresolved and will only report sensitivities to the NEDM. Of
course, the sensitivities of odd-even nuclei to the pEDM and
even-odd nuclei to the nEDM are more pronounced than vice versa, but this
effect can be considered separately, if needed. It is, however, not expected 
to be important for our model.

The total atomic or molecular $\mathcal{P,T}$-odd EDM is

\begin{multline}
d_\mathrm{atom,mol} = \alpha_\mathrm{d}d_\mathrm{e} +
(\alpha_\mathrm{m} + \alpha_{\mathcal{S},\mathrm{EDM}} +
\alpha_{\mathcal{M},\mathrm{EDM}}) d_\mathrm{N} \\
+ \alpha_\mathrm{s} k_\mathrm{s} + \alpha_\mathrm{T} k_\mathrm{T} +
\alpha_\mathrm{p} k_\mathrm{p} 
+ (\alpha_{\mathcal{S},\pi} +
\alpha_{\mathcal{M},\pi}) g_{\pi\mathrm{NN}}\,
\label{eq: totalEDM}
\end{multline}
with 
\begin{align}
\alpha_\mathrm{d}&=\tilde{\Omega}W_{\mathrm{d}}+\tilde{\mathcal{I}}\gamma W^\mathrm{m}_{\mathrm{d}}\\
\alpha_\mathrm{m}&=\tilde{\mathcal{I}}\eta W_\mathrm{m}\\
\alpha_{\mathcal{S},\mathrm{EDM}}&=\tilde{\mathcal{I}}W_\mathcal{S}R_\mathrm{vol}\\
\alpha_{\mathcal{M},\mathrm{EDM}}&=\tilde{\Theta}W_\mathcal{M}\mathcal{M}_\mathrm{EDM}\\
\alpha_\mathrm{s}&=\tilde{\Omega}W_{\mathrm{s}}+\tilde{\mathcal{I}}\gamma W^\mathrm{m}_{\mathrm{s}}\\
\alpha_\mathrm{T}&=\tilde{\mathcal{I}}W_\mathrm{T}\\
\alpha_\mathrm{p}&=\tilde{\mathcal{I}}W_\mathrm{p}\\
\alpha_{\mathcal{S},\pi}&=\tilde{\mathcal{I}}W_\mathcal{S}\tilde{S}\\
\alpha_{\mathcal{M},\pi}&=\tilde{\Theta}W_\mathcal{M}\mathcal{M}_\pi\,
\end{align}
where
$\tilde{\Omega}=\abs{\Omega}/(\efield\lambda_{z}\abs{\Omega+\mathcal{I}})$
and for $\tilde{\mathcal{I}}$ and $\tilde{\Theta}$ analogously. Here,
we have to note that the definition of $\tilde{\Theta}$ is only an
approximation because the polarization of the quadrupole moment contribution is
not necessarily the same as for the dipole moment contributions.
Exploiting this fact, one can gain further complementarity of
experiments by measuring various rotational transitions and by this
disentangle e.g.\ eEDM and NMQM interactions as was suggested in
Ref.~\cite{kurchavov:2022}. For the following discussion we do not
consider the explicit polarization of dipoles and quadrupoles
separately but rather use the approximate polarization factor
$\lambda_{z}$.

\subsection{Measurement model}
Without further approximation, at least six \emph{different}
experiments are required in order to set bounds on all parameters that
are considered herein to be fundamental without information on nuclear
and hadronic structure. This \emph{minimal complete measurement model} can be
described by a system of linear equations:
\begin{align}
\hbar
\vec{\omega}
=
\bm{W} 
\vec{x}_\mathcal{P,T}\\
\vec{d}
=
\bm{\alpha} 
\vec{x}_\mathcal{P,T}
\,,
\end{align}
where
$\vec{\omega}^\mathsf{T}=(\Delta\omega_1,\dots,\Delta\omega_6)$
represents the vector of measured $\mathcal{P,T}$-odd frequency shifts for six
different experiments which is related to the vector of measured
$\mathcal{P,T}$-odd EDMs by $\hbar \omega_i=d_i \efield_i F_i$. The vector
$\vec{x}_\mathcal{P,T}^\mathsf{T} = ( d_\mathrm{e}, d_\mathrm{N} ,
k_\mathrm{s}, k_\mathrm{T},  k_\mathrm{p}, g_{\pi\mathrm{NN}})$ corresponds
to the six $\mathcal{P,T}$-odd parameters and the matrix of
sensitivity coefficients
$\bm{W}=(\vec{W}_1,\vec{W}_2,\vec{W}_3,\vec{W}_4,\vec{W}_5,\vec{W}_6)$
has for each experiment $i$ a row $\vec{W}_{i}^\mathsf{T}$ with
elements
\begin{equation}
\begin{aligned}
W_{i1} &= \Omega_i W_{\mathrm{d},i}+\mathcal{I}_i\gamma_iW^\mathrm{m}_{\mathrm{d},i}\\
W_{i2} &= \mathcal{I}_i(R_{\mathrm{vol},i} W_{\mathcal{S},i} + \eta_i W_{\mathrm{m},i}) + \Theta_iW_{\mathcal{M},i} \tilde{\mathcal{M}}_{\mathrm{EDM},i} \\
W_{i3} &= \Omega_i W_{\mathrm{s},i}+\mathcal{I}_i\gamma_i W^\mathrm{m}_{\mathrm{s},i}\\
W_{i4} &= \mathcal{I}_iW_{\mathrm{T},i} \\
W_{i5} &= \mathcal{I}_iW_{\mathrm{p},i} \\
W_{i6} &= \mathcal{I}_i W_{\mathcal{S},i}\tilde{\mathcal{S}}_{i} + \Theta_iW_{\mathcal{M},i} \tilde{\mathcal{M}}_{\pi,i}\,
\end{aligned}
\label{eq: w_parameters}
\end{equation}
and the elements of $\bm{\alpha}$ are related to those of $\bm{W}$ by
$W_{ij}=\alpha_{ij}F_i \efield_i$ for all $i$ and $j$.

In order to construct a simple, easily applicable model, we focus on
three important properties: i) the electronic angular momentum
$\Omega$, ii) the nuclear angular momentum $\mathcal{I}$ of the
heaviest nucleus and iii) the nuclear charge number $Z$ of the
heaviest nucleus in the fully polarized system. Here we assume always
the case of maximally coupled angular momenta
$F=\abs{\Omega}+\abs{\mathcal{I}}$. Note, that this choice
disadvantages systems with higher angular momentum in all
considerations of sensitivity in the following.

According to differences in the effective Hamiltonian, we can discuss
four classes of atoms and linear molecules with different sensitivity
to the $\mathcal{P,T}$-odd parameter space:
\begin{enumerate}
\item[I] Open-shell systems ($\Omega>0$) with a closed-shell nucleus
with $\mathcal{I}=0$,
\item[II] Closed-shell systems ($\Omega=0$) with an open-shell nucleus with
$\mathcal{I}=\sfrac{1}{2}$,
\item[III] Open-shell systems with an open-shell nucleus with
$\mathcal{I}=\sfrac{1}{2}$,
\item[IV] Open-shell systems with an open-shell nucleus with
$\mathcal{I}\geq1$.
\end{enumerate}
Class I systems are sensitive to eEDM and SPNEC interactions only.
Class II, III and IV systems, in contrast, are receptive to all
parameters $\vec{x}_\mathcal{P,T}$ but with varying sensitivities to
individual parameters: Class II systems feature suppressed sensitivity
to eEDM and SPNEC interactions in comparison to class I systems,
whereas class III and IV systems are similarly receptive to eEDM and
SPNEC interactions as class I systems. Due to the possibility of
contributions from an NMQM, however, class IV systems differ from
class III systems in their sensitivity to NEDM and NNC interactions.

In total there are $\binom{4+6-1}{6}-\sum_{k=1}^4\binom{k+2-1}{2}=64$
possibilities to choose six systems out of the four classes defined
above, if we take into account that a maximum of two systems can be chosen
from class I in order to prevent the system of equations to be
undetermined.

Assuming Gaussian probability distributions for the uncertainties of
the measurements, we receive ellipsoidal coverage regions of shape
\begin{equation}
(\vec{x}_\mathcal{P,T}-\vec{x}_{\mathcal{P,T},0})^\mathsf{T}\bm{U}_\mathcal{P,T}^{-1}(\vec{x}_\mathcal{P,T}-\vec{x}_{\mathcal{P,T},0})=P^2\,,
\end{equation}
where $\vec{x}_{\mathcal{P,T},0}$ contains the resulting values for
each $\mathcal{P,T}$-odd parameter in $\vec{x}_\mathcal{P,T}$,
$P=2.45$ in a two-dimensional parameter space and $P=3.55$ in a
six-dimensional parameter space for an ellipsoidal region representing
95~\% confidence level (CL) \cite{jcgm102:2011}. The covariance matrix
$\bm{U}_{\mathcal{P,T}}$ can be obtained from the covariance matrix of
measured frequencies $\bm{U}_\omega$ via the matrix product
$\bm{W}^{-1}\bm{U}_\omega\parantheses{\bm{W}^{-1}}^\mathsf{T}$.  If
the measurements are assumed to be uncorrelated, $\bm{U}_\omega$ is a
diagonal matrix containing the squared standard uncertainties of the
experiments $\sigma^2_{\omega,i}$.

The quality of a set of $M$ measurements for restriction of the total
parameter space can be found by calculation of the volume of the
ellipsoidal coverage region in $N$ dimensions (for details on the
volume of an $N$-dimensional ellipsoid see the appendix):
\begin{align}
V&=P \frac{2
\pi^{N/2}}{N\Gamma\parantheses{N/2}}\mathrm{det}\parantheses{\bm{U}_\mathcal{P,T}^{-1}}^{-1/2}\\
&=
P\frac{2\pi^{N/2}}{N\Gamma\parantheses{N/2}}\mathrm{det}\parantheses{\bm{W}^\mathsf{T}\bm{U}_\omega^{-1}\bm{W}}^{-1/2}\,,
\label{eq: ndim_volume}
\end{align}
where $\Gamma(x)$ is the gamma function. In case of $M=N=6$ independent
measurements, the determinant of the inverse of the $\mathcal{P,T}$-odd
covariance matrix reads 
\begin{equation}
\mathrm{det}\parantheses{\bm{U}^{-1}_\mathcal{P,T}}=\mathrm{det}\parantheses{\bm{W}^\mathsf{T}\bm{U}_\omega^{-1}\bm{W}}
=\mathrm{det}\parantheses{\bm{W}}^2 \prod\limits_{i=1}^6
\frac{1}{\sigma_{\omega,i}^2}\,,
\end{equation}
and, thus, the volume is inversely proportional to the absolute value of
the determinant of the sensitivity coefficient matrix:
\begin{equation}
V=P \frac{\pi^{3}}{6}\frac{\prod\limits_{i=1}^6
\abs{\sigma_{\omega,i}}}{\abs{\mathrm{det}\parantheses{\bm{W}}}}.
\end{equation}
For the case that there are more than six measurements, for equal standard
uncertainties $\sigma_{\omega,i}=\sigma_{\omega,0}$ the volume is 
$P \frac{\pi^{3}}{6}\frac{\abs{\sigma_{\omega,0}}^M}{\sqrt{\mathrm{det}\parantheses{
\bm{W}^\mathsf{T}\bm{W}}}}$, and for unequal $\sigma_{\omega,i}$ and
$M>N$ we have to
explicitly employ eq. (\ref{eq: ndim_volume}) with $N=6$.

\begin{figure*}
\includegraphics[width=\textwidth]{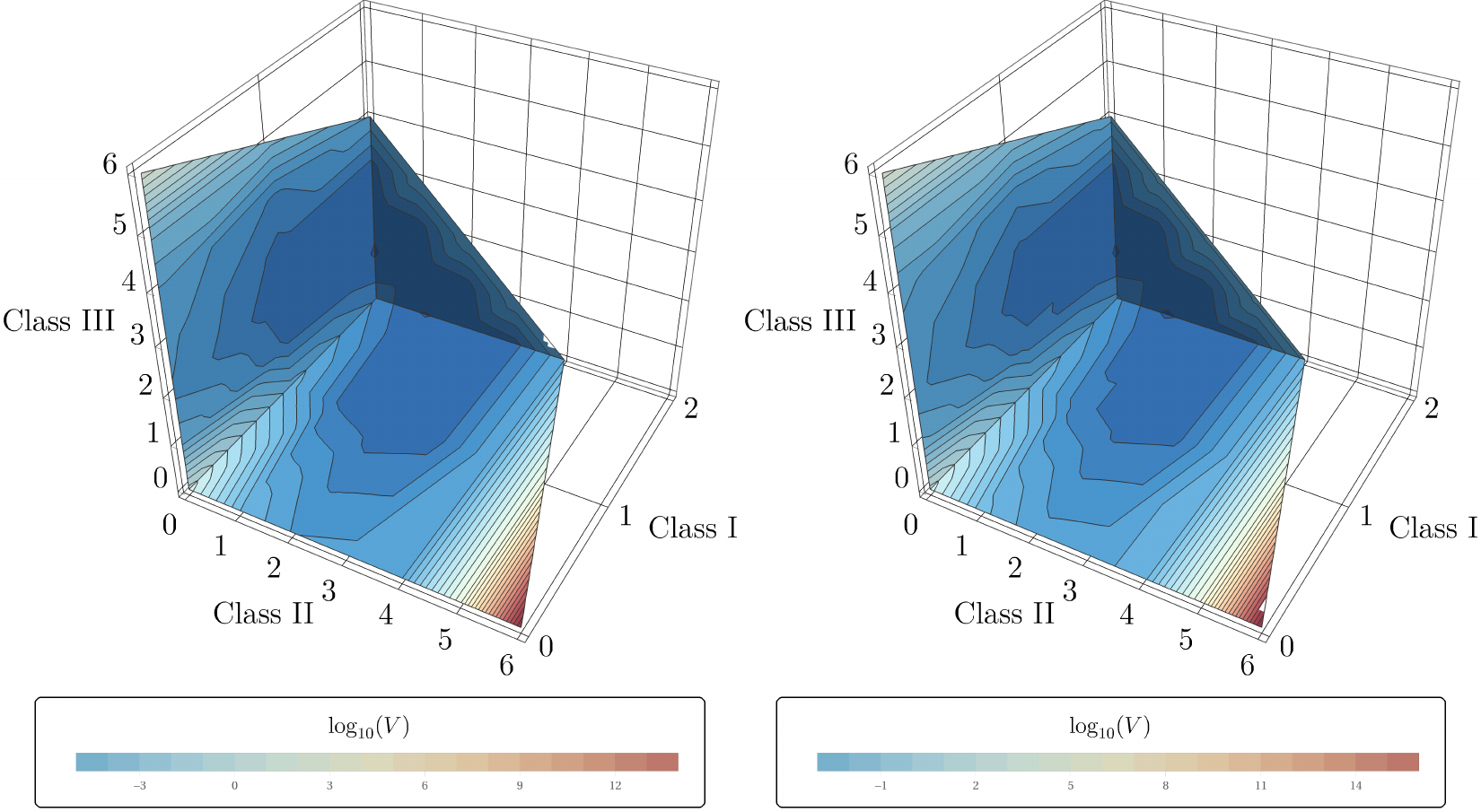}%
\caption{Results for a global minimization of the coverage volume in
$\mathcal{P,T}$-odd parameter space with respect to nuclear charges of
the heavy atoms in the molecules. The results for the minimum value of
the coverage volume for the optimal nuclear charge distributions are
presented as a three-dimensional projection on the planes for
different numbers of molecules from classes I to III defined above.
The number of molecules of class IV is determined by the difference
between the number of molecules from other classes and the number of
total experiments (six).  Results shown on the left were received for
a minimization with condition $20\le Z\le 100$ and those on the right
were received with restriction $20\le Z \le 90$.}
\label{fig: ptodd_global_minimization}
\end{figure*}

\begin{figure*}
\includegraphics[width=\textwidth]{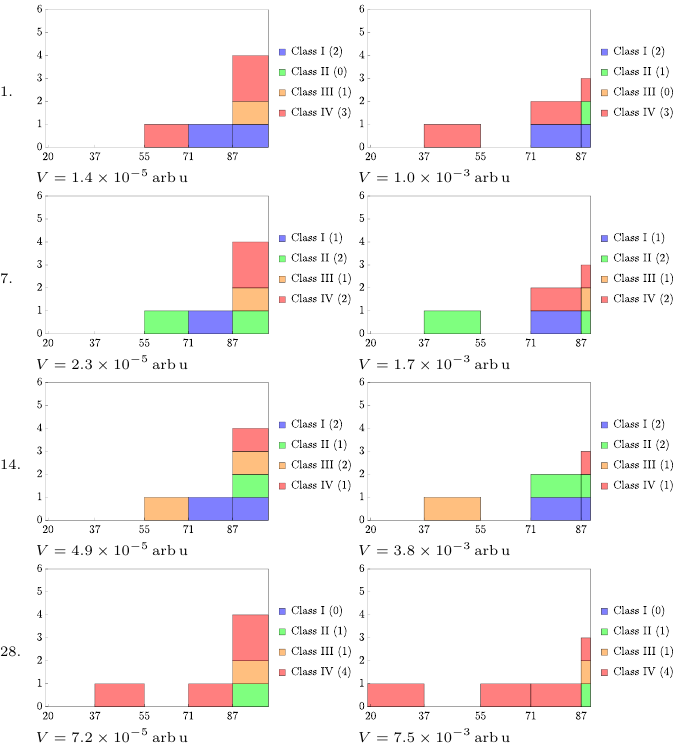}
\caption{Examples of optimal nuclear charge distributions for the
global minimum using different combinations of classes for
optimization in the regions of $20\leq Z\leq100$ (left) and $20\leq
Z\leq90$ (right). The binning is chosen such that bins represent the
rows of the periodic table. The first bin corresponds to the 4th row
of the periodic table ($20\leq Z\leq36$) except Na, the second bin corresponds
to the 5th row of the periodic table ($37\leq54$), the third and
fourth bin correspond to the first ($55\leq Z\leq70$) and second half
($71\leq Z\leq86$) of the 6th row of the periodic table and the last
bin includes the elements of the 7th row of the periodic table up to
$Z=100$ or $Z=90$.}
\label{fig: ptodd_global_charge_distribution}
\end{figure*}

\section{Results}
\subsection{Results of the global minimization}
With the model defined in the previous section, we can now attempt to
minimise the coverage region that would be obtained from six independent
measurements. 

In order to achieve this in a simple qualitative approach, we focus on
the relative size of the coverage region $V$ as a function of
$\Omega_i$, $\mathcal{I}_i$ and $Z_i$ of six systems $i$ and aim to
achieve a measure for the impact of the electronic structure. In order
to not be biased by experimental details, we assume equal uncertainties
for all experiments ($\sigma_{\omega,i}=1$ for all $i$). By doing so the
coverage volume is completely determined by
$V\sim\abs{\mathrm{det}\parantheses{\bm{W}}}^{-1}$. 

We approximate all electronic structure enhancement factors $W$ as
described in Appendix \ref{analytic_enhancement} as simple analytic
functions of the nuclear charge number.  

To find the optimal combination of systems, we need to
minimize the coverage volume globally:
\begin{equation}
\underset{\small\braces{\Omega_i,\mathcal{I}_i}\in\braces{\frac{n}{2}|n\in\mathbb{N}_0},\,\braces{\braces{Z_i}\in\mathbb{N}_0:
20\le Z_i 
\le
100}}{\mathrm{min}}V\parantheses{\braces{\Omega_i,\mathcal{I}_i,Z_i}}\,.
\end{equation}
For this constrained global minimization, we represent each class of
systems with minimal spin, i.e., we set $\Omega=\sfrac{1}{2}$ for
class I and $\mathcal{I}=0$, $\Omega=0$ and $\mathcal{I}=\sfrac{1}{2}$
for class II, $\Omega=\sfrac{1}{2}$ and $\mathcal{I}=\sfrac{1}{2}$ for
class III, $\Omega=\sfrac{1}{2}$ and $\mathcal{I}=1$ for class IV. At
this point it has to be noted that other choices of nuclear or
electronic spin may have an influence on the results of the
optimization. However, tests in which all nonzero spins were set to
one were in qualitative agreement with the present results. We assume
all nuclear magnetic moments $\mu$ to be $1\,\mu_\mathrm{N}$. A detailed discussion
of the global minimization procedure and approximations used can be
found in the Appendix.

The detailed numerical results of the minimization are shown in 
\prettyref{tab: results_global_min_90} and
\prettyref{tab: results_global_min_100} in
the Appendix. A qualitative visualization of the results is
provided in \prettyref{fig: ptodd_global_minimization} in arbitrary
units. Therein the size of the volume in the $\mathcal{P,T}$-odd
parameter space is shown as a function of the number of systems from
classes I to III. 

We ranked the results according to the size of the coverage volume.
Qualitatively the two optimizations up to $Z=100$ and $Z=90$ are in
agreement and in the following we will restrict the discussion to the
data set optimized for $20<Z<90$ as this coincides with a momentarily more
realistic range of nuclear charge numbers. The best six results (see
\prettyref{fig: ptodd_global_minimization}, and \prettyref{tab:
results_global_min_90} and \prettyref{tab: results_global_min_100} in
the appendix) differ only slightly in the coverage volume and the
first 32 results are all within almost one order of magnitude.  After
place 32 the coverage volume increases steeper, with a big jump from
place 58 to 59. These worst results include only combinations with at
least five systems of the same class. From the 32 best results one
may deduce following rules for a good choice of complementary systems
in dependence of electronic and nuclear angular momenta for the
minimal number of experiments:
\begin{enumerate}
\item The trivial requirements are: 
$N_\mathrm{I}\leq2$,
$N_\mathrm{I}+N_\mathrm{II}+N_\mathrm{III}+N_\mathrm{IV}\geq6$
\item $N_\mathrm{II,III,IV}\leq3$
\item If $N_A=3$, $N_B\leq2$ $\forall
A,B\in\braces{\mathrm{II,III,IV}}$
\end{enumerate}
From this we can conclude that we need, as naively expected, systems
that have fundamentally different sensitivities to the different
sources of $\mathcal{P,T}$-violation. One should note that the rules are only valid for exactly six experiments. 
If more experiments are included, it appears beneficial to have fewer
experiments of class I than of the other classes and to distribute the
number of experiments somewhat equally over classes II, III and IV. 

Beside the classes or different angular momenta of the molecules, we
can also deduce the optimal distribution of nuclear charge numbers.
Exemplary, the distribution of nuclear charge numbers for the
different classes of molecules is shown for specific choices of
molecules from the four classes in \prettyref{fig:
ptodd_global_charge_distribution} for both regions of $Z$ studied. The
distributions for all other cases can be found in the Supplementary
Material. From this we find the simple rule, that \emph{if the number
of systems of the same class increases, the new system should be
chosen with a different $Z$}. In particular, \emph{if there are three
or more systems of the same class, medium-heavy or even lighter
systems with $Z\leq54$ from the 5th row of the periodic table or
earlier are beneficial for a restriction of the coverage region}.
Although this rule contradicts naive expectations, as for systems with
lower $Z$ individual enhancement factors are small, it highlights the
importance of complementarity of chosen systems.

Currently more than six experiments were already performed and many
new experiments with different systems are planned for the near
future. Thus, in the actual experimental situation the system of
equations is over-determined. This means that systems with lower $Z$
will probably become even more important in future, as according to
the above rules of thumb, complementary new experiments should be
chosen with considerably different nuclear charge number if there are
already experiments of the same class. This we will demonstrate in
more detail in the following section.

\begin{table}
\begin{threeparttable}
\caption{Experimental standard uncertainty $\sigma_{d}$ on atomic and molecular
$\mathcal{P,T}$-odd EDMs and external electric field
strength $\efield$ applied in molecular experiments multiplied by the
degree of polarization of the molecular axis $\lambda_z$.}
\label{tab: exp_data}
\begin{tabular}{
l
l
l
S[table-format= 2]
S[table-format= 1.1]
S[table-format= 1.1]
S[table-format= 4.0,scientific-notation=fixed,fixed-exponent = 0]
S[round-precision=1,round-mode=figures,table-format= 1e-2,scientific-notation=true]
}
\toprule
\multicolumn{2}{c}{System} & 
Class &
{$Z$} &
{$\Omega$\tnote{a}} &
{$\mathcal{I}$} &
{$\efield\lambda_z/\frac{\si{\volt}}{\si{\centi\meter}}$} &
{$\sigma_{d}/\si{\elementarycharge\centi\meter}$\tnote{b}}
\\
\midrule
\ce{^{129}Xe},$^1S_0$           &{\cite{allmendinger:2019,sachdeva:2019}}&II &54&0  &0.5& {-}    & 6.9e-28\\
\ce{^{133}Cs},$^2S_{1/2}$       &{\cite{murthy:1989}    }&IV &55&0.5&3.5& {-}    & 6.9e-24\\
\ce{^{171}Yb},$^1S_0$           &{\cite{zheng:2022}     }&II &70&0  &2.5& {-}    & 6.9e-24\\
\ce{^{174}YbF},$^2\Sigma_{1/2}$ &{\cite{hudson:2002,hudson:2011}}&I  &70&0.5&0  & 5.58e3 & 2.7e-21 \\
\ce{^{180}HfF+},$^3\Delta_{1}$  &{\cite{roussy:2023}    }&I  &72&1  &0  & 58     & 8.5e-22\\
\ce{^{199}Hg},$^1S_0$           &{\cite{garner:2016}    }&II &80&0  &0.5& {-}    & 3.1e-30\\
\ce{^{205}Tl},$^2P_{1/2}$       &{\cite{regan:2002}     }&III&81&0.5&0.5& {-}    & 4.3e-25\\
\ce{^{205}TlF},$^1\Sigma_{0}$   &{\cite{cho:1991}       }&II &81&0  &0.5& 16e3   & 2.9e-23\\
\ce{^{207}PbO},$^3\Sigma_{1}$   &{\cite{eckel:2013}     }&III&82&1  &0.5& 100    & 2.4e-18\\
\ce{^{225}Ra},$^1S_0$           &{\cite{parker:2015,bischof:2016}    }&II &88&0  &0.5& {-}    & 6e-24\\
\ce{^{232}ThO},$^3\Delta_{1}$   &{\cite{andreev:2018}   }&I  &90&1  &0  & 80     & 3.9e-21\\
\bottomrule
\end{tabular} 
\begin{tablenotes}
\item[a] For atoms, this entry corresponds to $J_\mathrm{e}$.
\item[b] For open-shell molecules computed 
as $\sigma_\mathrm{d}=\sigma_{d_\mathrm{e}}\mathcal{E}_\mathrm{eff}/(\efield\lambda_z)$
from experimental standard uncertainty on the eEDM
$\sigma_{d_\mathrm{e}}=\SI{5.9e-28}{\elementarycharge\centi\meter}$
for YbF,
$\sigma_{d_\mathrm{e}}=\SI{2.1e-30}{\elementarycharge\centi\meter}$
for \ce{HfF+},
$\sigma_{d_\mathrm{e}}=\SI{9.7e-27}{\elementarycharge\centi\meter}$
for \ce{PbO},
$\sigma_{d_\mathrm{e}}=\SI{4.0e-30}{\elementarycharge\centi\meter}$
for \ce{ThO}, the value applied for the internal effective electric
field that $d_\mathrm{e}$ experiences
$\mathcal{E}_\mathrm{eff}=\SI{26}{\giga\volt\per\centi\meter}$ for
YbF, $\mathcal{E}_\mathrm{eff}=\SI{23}{\giga\volt\per\centi\meter}$
for \ce{HfF+},
$\mathcal{E}_\mathrm{eff}=\SI{25}{\giga\volt\per\centi\meter}$ for
\ce{PbO},
$\mathcal{E}_\mathrm{eff}=\SI{78}{\giga\volt\per\centi\meter}$ for ThO
and the applied external electrical field multiplied with the degree
of polarisation $\efield\lambda_z$.
For Tl computed from $\alpha_\mathrm{d}=-585$ and
$\sigma_{d_\mathrm{e}}=\SI{7.4e-28}{\elementarycharge\centi\meter}$ as
$\sigma_{d}=\alpha_\mathrm{d}\sigma_{d_\mathrm{e}}$. The experimental
combined standard uncertainties were computed from statistical (stat.)
and systematic (sys.) uncertainties as
$\sigma_{d}=\sqrt{\sigma_\mathrm{stat.}^2+\sigma_\mathrm{sys.}^2}$.
\end{tablenotes}
\end{threeparttable}
\end{table}

\begin{table}
\begin{threeparttable}
\caption{Assumed achievable experimental uncertainties on atomic and molecular
$\mathcal{P,T}$-odd EDMs $\sigma_{d}$ chosen differently for atomic and
molecular experiments and assumed external electric field
strength $\efield\lambda_z$ needed to polarize molecular systems in future
experiments multiplied by the maximum degree of polarization as discussed in
the text.}
\label{tab: proposed_data}
\begin{tabular}{
l
l
S[table-format= 2]
S[table-format= 1.1]
S[table-format= 1.1]
S[table-format= 5.0]
S[round-precision=1,round-mode=figures,table-format= 1e-2,scientific-notation=true]
}
\toprule
System & 
Class &
{$Z$} &
{$\Omega$\tnote{a}} &
{$\mathcal{I}$} &
{$\efield\lambda_z/\frac{\si{\volt}}{\si{\centi\meter}}$} &
{$\sigma_{d}/\si{\elementarycharge\centi\meter}$ }
\\
\midrule
\ce{^{137}BaF},$^2\Sigma_{1/2}$ &IV   &56&0.5&1.5& 10000 & 1e-23\\
\ce{^{173}YbOH},$^2\Sigma_{1/2}$&IV   &70&0.5&2.5&   100 & 1e-23\\
\ce{^{183}WC},$^3\Delta_1$      &III  &74&1  &0.5&   100 & 1e-23\\
\ce{^{210}Fr},$^2S_{1/2}$       &IV   &87&0.5&6.5&   {-} & 1e-28\\
\ce{^{223}RaF},$^2\Sigma_{1/2}$ &IV   &88&0.5&1.5& 10000 & 1e-23\\
\ce{^{225}RaF},$^2\Sigma_{1/2}$ &III  &88&0.5&0.5& 10000 & 1e-23\\
\ce{^{226}RaF},$^2\Sigma_{1/2}$ &I    &88&0.5&0  & 10000 & 1e-23\\
\ce{^{229}ThF+},$^3\Delta_1$    &IV   &90&1  &2.5&   100 & 1e-23\\
\ce{^{229}PaF^3+},$^2\Phi_{5/2}$&II/IV&91&2.5&2.5&  1000 & 1e-23\\
\ce{^{43}CaOH},$^2\Sigma_{1/2}$ &IV   &20&0.5&3.5&   100 & 1e-23\\
\ce{^{87}SrOH},$^2\Sigma_{1/2}$ &IV   &38&0.5&4.5&   100 & 1e-23\\
\ce{^{89}YO},$^2\Sigma_{1/2}$   &III  &39&0.5&0.5& 10000 & 1e-23\\
\ce{^{111}CdH},$^2\Sigma_{1/2}$ &IV   &48&0.5&0.5& 10000 & 1e-23\\
\bottomrule
\end{tabular} 
\begin{tablenotes}
\item[a] For atoms, this entry corresponds to $J_\mathrm{e}$.
\end{tablenotes}
\end{threeparttable}
\end{table}

\begin{table*}
\begin{threeparttable}
\caption{Electronic structure sensitivity coefficients of atoms and
molecules in former, planned and in this work proposed experimental
searches for $\mathcal{P,T}$-odd EDMs $\alpha_i=W_i/(F \efield)$ as
defined in eq.  (\ref{eq: w_parameters}), computed at the level of
ZORA-cGKS-BHandH using eqs. (\ref{eq: nucstruc_start})-(\ref{eq:
nucstruc_end}) to estimate nuclear structure parameters. For molecules
assumptions for the applied external electric fields and angular
momenta are chosen $F=|\Omega|+|\mathcal{I}|$, with $\Omega$ and
$\mathcal{I}$ provided in are given in Table \ref{tab: proposed_data} and
details on different contributions to the individual $\alpha_i$ values
as well as individual electronic and nuclear structure parameters are
provided in the Appendix.}
\label{tab: all_data}
\begin{tabular}{l
S[round-precision=2,round-mode=figures,table-format=-1.1e-2,scientific-notation=true]
S[round-precision=2,round-mode=figures,table-format=-1.1e-2,scientific-notation=true]
S[round-precision=2,round-mode=figures,table-format=-1.1e-2,scientific-notation=true]
S[round-precision=2,round-mode=figures,table-format=-1.1e-2,scientific-notation=true]
S[round-precision=2,round-mode=figures,table-format=-1.1e-2,scientific-notation=true]
S[round-precision=2,round-mode=figures,table-format=-1.1e-2,scientific-notation=true]
}
\toprule
System & 
{$\alpha_1$ }&
{$\alpha_2$ }&
{$\alpha_3/\si{\elementarycharge\centi\meter}$ }&
{$\alpha_4/\si{\elementarycharge\centi\meter}$ }&
{$\alpha_5/\si{\elementarycharge\centi\meter}$ }&
{$\alpha_6/\si{\elementarycharge\centi\meter}$ }
\\
\midrule
\multicolumn{3}{l}{Former experiments}\\
\midrule
\ce{^{129}Xe},$^1S_0$           & -1.6300000000e-03 &  7.3187648514e-05 & -7.3320000000e-23 &  5.4419600000e-21 &  1.6432560000e-23 &  4.4494239740e-18 \\
\ce{^{133}Cs},$^2S_{1/2}$       &  2.9395355000e+01 & -4.1251342660e-04 &  7.7257312500e-20 & -1.4233585000e-20 & -4.3063635000e-23 & -8.7606734935e-18 \\
\ce{^{171}Yb},$^1S_0$           & -3.7920000000e-04 & -3.4897127142e-04 &  7.8908000000e-23 & -3.1020768000e-20 & -1.0789578000e-22 & -2.8182468358e-17 \\
\ce{^{174}YbF},$^2\Sigma_{1/2}$ & -8.3159813620e+06 &         {-}         & -3.0051315412e-14 &        {-}         &        {-}         &        {-}         \\
\ce{^{180}HfF+},$^3\Delta_{1}$  &  4.0406312069e+08 &         {-}         &  1.4776858621e-12 &        {-}         &        {-}         &         {-}         \\
\ce{^{199}Hg},$^1S_0$           & -2.5580000000e-03 & -5.7372311743e-04 & -5.4094000000e-22 & -4.7120580000e-20 & -1.7574524000e-22 & -5.5463648919e-17 \\
\ce{^{205}Tl},$^2P_{1/2}$       & -6.4050693000e+02 & -2.7898462000e-05 & -3.0738931400e-18 &  1.0774390000e-20 &  3.7640630000e-23 & -5.3283321000e-18 \\
\ce{^{205}TlF},$^1\Sigma_{0}$   &  1.3611250000e+03 &  1.0634924875e+01 &  4.2670000000e-17 &  1.0212500000e-15 &  3.7977737500e-18 &  1.1003925558e-12 \\
\ce{^{207}PbO},$^3\Sigma_{1}$   & -2.1959987333e+07 &  6.2441977303e+01 & -1.0516343333e-13 &  4.9046533333e-15 &  1.8510173333e-17 &  6.1713385893e-12 \\
\ce{^{225}Ra},$^1S_0$           & -5.3796000000e-02 & -1.9312886661e-03 &  2.2302600000e-21 & -1.4539576000e-19 & -5.6878190000e-22 & -1.9623881803e-16 \\
\ce{^{232}ThO},$^3\Delta_{1}$   &  1.0282208500e+09 &         {-}         &  6.0038223750e-12 &        {-}         &        {-}         &         {-}         \\
\midrule
\multicolumn{3}{l}{Planned experiments} \\                                                                                 
\midrule
\ce{^{137}BaF},$^2\Sigma_{1/2}$ & -3.1158445000e+05 &  3.7850710283e-01 & -8.2600100000e-16 & -5.8777500000e-17 & -1.7985950000e-19 & -4.5929046538e-14 \\
\ce{^{173}YbOH},$^2\Sigma_{1/2}$& -7.8015680000e+07 &  2.1968438678e+02 & -2.8200206667e-13 & -1.6165700000e-14 & -5.6704633333e-17 & -1.7301409565e-11 \\
\ce{^{183}WC},$^3\Delta_1$      &  3.0537626667e+08 & -1.8165317237e+02 &  1.2333851333e-12 & -1.2774266667e-14 & -4.5936000000e-17 & -1.5843120390e-11 \\
\ce{^{210}Fr},$^2S_{1/2}$       &  1.2972177077e+02 & -2.7870909801e-03 &  7.0911573846e-19 & -1.3322700000e-19 & -5.1802436308e-22 & -1.3540943211e-16 \\
\ce{^{223}RaF},$^2\Sigma_{1/2}$ & -2.6295312000e+06 & -3.3557093023e+00 & -1.4649835500e-14 & -5.5142700000e-16 & -2.1754605000e-18 & -7.9037823875e-13 \\
\ce{^{225}RaF},$^2\Sigma_{1/2}$ & -5.2590624000e+06 & -5.7830061661e+00 & -2.9299671000e-14 & -3.6761800000e-16 & -1.4503070000e-18 & -5.8794121318e-13 \\
\ce{^{226}RaF},$^2\Sigma_{1/2}$ & -1.0518124800e+07 &         {-}         & -5.8599342000e-14 &        {-}         &        {-}         &        {-}         \\
\ce{^{229}ThF+},$^3\Delta_1$    &  9.9131668571e+07 & -1.6949489583e+03 &  5.6304122857e-13 & -1.0874171429e-13 & -4.3093662857e-16 & -1.3459138912e-10 \\
\ce{^{229}PaF^3+},$^2\Phi_{5/2}$&  1.0673694000e+06 & -1.3515558629e+02 &  6.7040720000e-15 & -1.2130130000e-14 & -4.8679482000e-17 & -1.5118749567e-11 \\
\midrule
\multicolumn{3}{l}{Proposed experiments (this work)}\\
\midrule
\ce{^{43}CaOH},$^2\Sigma_{1/2}$ & -7.1064000000e+05 &  2.2687557060e+01 & -7.6120000000e-16 & -2.6482500000e-16 & -4.6822500000e-19 & -2.2997802000e-15 \\
\ce{^{87}SrOH},$^2\Sigma_{1/2}$ & -4.1668020000e+06 &  9.1433713959e+01 & -7.8846200000e-15 & -1.9846200000e-15 & -4.8884800000e-18 & -6.2119469940e-13 \\ 
\ce{^{89}YO},$^2\Sigma_{1/2}$   & -2.8952050000e+05 & -1.8177275542e-01 & -5.5559100000e-16 & -1.8454000000e-17 & -4.6092000000e-20 & -9.0489207320e-15 \\
\ce{^{111}CdH},$^2\Sigma_{1/2}$ & -1.2651235000e+06 & -2.2151383653e-01 & -3.0340130000e-15 & -1.7635000000e-17 & -5.0128000000e-20 & -1.2223141542e-14 \\
\bottomrule
\end{tabular} 
\end{threeparttable}
\end{table*}

\begin{figure*}
\includegraphics[width=\textwidth]{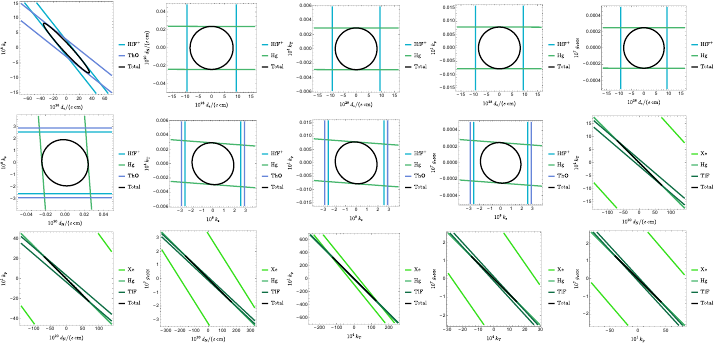}
\caption{Cuts of all two-dimensional subspaces of the full
six-dimensional $\mathcal{P,T}$-odd parameter space. Experimental
uncertainties are chosen as given in Table \ref{tab: exp_data} and
sensitivity coefficients computed at the level of ZORA-cGKS-BHandH
shown in Table \ref{tab: all_data} are used, where nuclear structure
parameters were estimated following eqs. (\ref{eq:
nucstruc_start})-(\ref{eq: nucstruc_end}). We furthermore assumed a
an overall theoretical uncertainty of 20\%\ and scaled all $W$
by a factor of $0.8$ to account for this in a worst case scenario. We
chose a six dimensional ellipsoid of 95\%\ CL ($P=3.55$).}
\label{fig: current_status}
\end{figure*}

\begin{table}[h!]
\begin{threeparttable}
\caption{Comparison of global bounds on $\mathcal{P,T}$-violation with
single-source model bounds employing experimental limits on atomic and
molecular EDMs from Table \ref{tab: exp_data} and molecular and atomic
enhancement factors from Table \ref{tab: all_data} that contribute to
the total atomic or molecular EDM (\ref{eq: totalEDM}) using the crude
estimations for nuclear structure parameters in eqs. (\ref{eq:
nucstruc_start})-(\ref{eq: nucstruc_end}). Global bounds are obtained
from six dimensional ellipsoidal coverage regions of 95\%\ CL
($P=3.55$). Single-source bounds are obtained from the single
experiment that is most sensitive to the parameter given in the last
column for 95\%\ CL ($P=1.96$). We furthermore assumed a 
conservative theoretical uncertainty of 20\%\ and scaled all $W$ by a factor of $0.8$ to
account for this in a worst case scenario. All values are rounded up.}
\label{tab: limits}
\begin{tabular}{l
S[round-precision=1,round-mode=places,table-format=1.0e-2,scientific-notation=true]
S[round-precision=1,round-mode=places,table-format=1.0e-2,scientific-notation=true]
l
}
\toprule
Source
&\multicolumn{1}{c}{Global bound}
&\multicolumn{1}{c}{Single-source}
&\multicolumn{1}{c}{Best system}
\\
\midrule
$d_\mathrm{e}/(e\,\mathrm{cm})$& 4e-29& 6e-30 & \ce{HfF+}\\
$d_\mathrm{N}/(e\,\mathrm{cm})$& 7e-22& 2e-26 & \ce{Hg}\\
$k_\mathrm{s}$                 & 9e-9 & 2e-9  & \ce{HfF+}\\
$k_\mathrm{T}$                 & 3e-5 & 2e-10 & \ce{Hg}\\
$k_\mathrm{p}$                 & 8e-2 & 5e-8  & \ce{Hg}\\
$g_\mathrm{\pi NN}$            & 2e-8 & 2e-13 & \ce{Hg}\\
\bottomrule           
\end{tabular}         
\end{threeparttable}  
\end{table}

\subsection{Implications on planned experiments}
To illustrate the rules we found in our simple analytic global
minimization of the $\mathcal{P,T}$-odd parameter space with respect
to the electronic properties of atoms and molecules used in EDM
experiments, we discuss \emph{ab initio} electronic structure
calculations of the various enhancement factors in systems from 
previous and planned EDM experiments. Furthermore, based on
the above findings we propose systems with lower $Z$ than usually
considered for new experiments.

\begin{table}[h!]
\begin{threeparttable}
\caption{Reduction factor of the current total coverage volume in the
$\mathcal{P,T}$-odd parameter space by the different contributing
experiments. For already performed experiments a modified volume
$\tilde{V}_i$ is obtained as the current total coverage volume excluding
the system and the reduction factor by this system is $f_i=V/\tilde{V}_i$,
whereas for planned or proposed experiments $\tilde{V}_i$ is obtained as
the current total coverage volume including the system in addition and
the reduction factor by this system is $f_i=\tilde{V}_i/V$. All volumes for
planned and proposed experiments are computed with the assumed
experimental uncertainties $\sigma_d$ and polarization field strength
$\tilde{\efield\lambda_z}$ given in Table \ref{tab: proposed_data}. We
furthermore assumed a conservative theoretical uncertainty of
20\%\ and scaled all $W$ by a factor of $0.8$ to account for this in a
worst case scenario. \emph{Ab initio} results are compared to those
derived from the scaled phenomenological model (see text for
details).}
\label{tab: compare_newexps}
\begin{tabular}{l
S[round-precision=1,round-mode=places,table-format=1.1e-2,scientific-notation=true]
S[round-precision=1,round-mode=places,table-format=1.1e-2,scientific-notation=true]
}
\toprule
& 
{\emph{Ab initio} } &
{Phenomenology}
\\
\midrule
$V/\si{\elementarycharge\squared\centi\meter\squared}$&
1.16946e-82&7.1195e-82\\
\midrule
System&\multicolumn{2}{c}{$f$}\\
\midrule
\multicolumn{3}{l}{Former experiments}\\
\midrule

\ce{^{129}Xe},$^1S_0$           & 0.011312335254956238    &   0.0012409687468745891             \\ 
\ce{^{133}Cs},$^2S_{1/2}$       & 1.0 & 1.0 \\
\ce{^{171}Yb},$^1S_0$           & 0.04620651278271144     &   0.015928301104313872              \\  
\ce{^{174}YbF},$^2\Sigma_{1/2}$ & 1.0 & 1.0 \\
\ce{^{180}HfF+},$^3\Delta_{1}$  & 0.006774431752818054    &   0.007312392988064499              \\  
\ce{^{199}Hg},$^1S_0$           & 0.000011221356339693272 &   0.0000274308055628344             \\ 
\ce{^{205}Tl},$^2P_{1/2}$       & 1.0 & 1.0 \\
\ce{^{205}TlF},$^1\Sigma_{0}$   & 0.0012816007102166435   &   0.0053852994158133235             \\ 
\ce{^{207}PbO},$^3\Sigma_{1}$   & 1.0 & 1.0 \\
\ce{^{225}Ra},$^1S_0$           & 1.0 & 1.0 \\
\ce{^{232}ThO},$^3\Delta_{1}$   & 0.0029607394325669536   &   0.0030568641260849145             \\ 

\midrule
\multicolumn{3}{l}{Planned experiments} \\                                                                                 
\midrule
\ce{^{137}BaF},$^2\Sigma_{1/2}$  &  0.06531227933812786     &   0.03885536121815887      \\   
\ce{^{173}YbOH},$^2\Sigma_{1/2}$ &  0.00017164992360264603  &   0.00004658989041675972   \\ 
\ce{^{183}WC},$^3\Delta_1$       &  0.007094029954088857    &   0.0035680146241801472     \\  
\ce{^{210}Fr},$^2S_{1/2}$        &  0.0004052106642747813   &   0.000058062655269839265   \\ 
\ce{^{223}RaF},$^2\Sigma_{1/2}$  &  0.015841964894210244    &   0.15537879559185003    \\     
\ce{^{225}RaF},$^2\Sigma_{1/2}$  &  0.10216697407354293     &   0.027581718525803677     \\   
\ce{^{226}RaF},$^2\Sigma_{1/2}$  &  0.27396826172889865     &   0.2691371221628265     \\     
\ce{^{229}ThF+},$^3\Delta_1$     &  0.0001351622744749119   &   0.00004373016248910274   \\ 
\ce{^{229}PaF^3+},$^2\Phi_{5/2}$ &  0.00592350363895661     &   0.010041886687407344     \\   
\midrule
\multicolumn{3}{l}{Proposed experiments (this work)}\\
\midrule               
\ce{^{43}CaOH},$^2\Sigma_{1/2}$  &  0.002794104212920359   &   0.018431618393853145  \\   
\ce{^{87}SrOH},$^2\Sigma_{1/2}$  &  0.00064809626379276    &   0.0007418355482683739   \\ 
\ce{^{89}YO},$^2\Sigma_{1/2}$    &  0.6951267741707459     &   0.540752386339604    \\    
\ce{^{111}CdH},$^2\Sigma_{1/2}$  &  0.8555677619777894     &   0.2564192221181516   \\       
\bottomrule           
\end{tabular}         
\end{threeparttable}  
\end{table}

Modern EDM experiments are listed in Table \ref{tab:
exp_data}, where we have neglected those experiments that are older than the Cs
experiment. Although the Cs experiment is dated as well (performed in
1989) we keep it, as it is so far the only class IV experiment. In total,
three class I experiments, five class II experiments, two class III
and one class IV experiments were performed so far. Various isotopes are considered for
upcoming experiments: Among others, measurements are planned with
$^{210}$Fr (row 7, IV) \cite{wundt:2012,shitara:2021}, $^{183}$WC (row
6, III) \cite{lee:2013}, $^{173/174}$YbOH (row 6, IV/I)
\cite{kozyryev:2017a}, $^{137/138}$BaF (row 6, IV/I)
\cite{aggarwal:2018,vutha:2018}, $^{226/225/223}$RaF (row 7, I/III/IV)
\cite{isaev:2010,Isaev:13,Isaev:2013,kudashov:2014,garciaruiz:2020,gaul:2020,udrescu:2021}
and \ce{^{229}ThF+} (row 7, IV)
\cite{barker:2012,denis:2015,skripnikov:2015,gresh:2016,ng:2022}.  Moreover,
recently pronounced sensitivity to $\mathcal{P,T}$-violation in
highly charged actinide-containing molecules such as \ce{PaF^3+} was
demonstrated \cite{zulch:2022,zulch:2023} (row 7, IV).

Of these we choose only for RaF all three isotopes mentioned explicitly to illustrate the
influence of different molecular classes on the size of the coverage
region in $\mathcal{P,T}$-odd parameter space. All selected systems
that are considered within this work are shown in Table \ref{tab:
proposed_data}. 

From our results of the model minimization of the global
$\mathcal{P,T}$-odd parameter space, we suggest to do EDM experiments
with lighter molecules from the fifth row with isotopes with
$\mathcal{I}\geq\sfrac{1}{2}$. A promising system is $^{89}$YO
(class III), which was sub-Doppler cooled \cite{ding:2020} and for
which we computed enhancement of $k_\mathrm{s}$ and $d_\mathrm{e}$
earlier \cite{gaul:2019}. 
To include also an even lighter system, we take into account the 
laser-coolable polyatomic molecule CaOH
\cite{isaev:2016,isaev:2018}, for which sub-Doppler cooling and
control of individual quantum states was achieved recently
\cite{vilas:2022,anderegg:2023}, for our analysis with the class IV isotopologue
$^{43}$CaOH. Furthermore, the related
molecule SrOH \cite{kozyryev:2017} could be promising, if
the isotopologue $^{87}$SrOH is chosen (see also
Ref.~\cite{gaul:2020a}).  If experimentally controllable, $^{111}$CdH
would be a promising relatively light species, due to large chemical
enhancement of $\mathcal{P,T}$-odd properties as suggested in
Refs.~\cite{gaul:2019,talukdar:2020}. 

To illustrate the impact of experiments which are performed
or planned in our global analysis, we can use existing experimental
data and results from \emph{ab initio} calculations of the electronic
structure parameters to compute the current ellipsoidal coverage
region in the $\mathcal{P,T}$-odd parameter space.  Not for all
electronic structure parameters do \emph{ab initio} calculations exist in
the literature.  Thus, we computed all relevant electronic structure
parameters at the level of ZORA-cGKS-BHandH (for details see
Appendix). This method agrees typically within about $15~\%$ with
available \emph{ab initio} data in the literature (see Appendix for
details). The uncertainty of the electronic structure theory plays for
the current situation only a minor role as (1) experimental
uncertainties limit the experiments and so far all measurements agree
with zero, (2) uncertainties of nuclear structure calculations are
much larger and (3) uncertainties are usually below 20~\% even for
more complicated systems and as of that do not change bounds within a
confidence level of $90~\%$ considerably. Nonetheless, it maybe be
worth to note that systems with an easier electronic structure such as
RaF or BaF could have advantages in comparison to systems with a more
complicated electronic structure such as ThO or \ce{ThF^+}, once a
non-zero measurement would be achieved.

We shall still use the rather crude approximations for nuclear
structure factors $\mathcal{M}$ and $\mathcal{S}$ shown in eqs.
(\ref{eq: nucstruc_start})-(\ref{eq: nucstruc_end}), as reliable
nuclear structure \emph{ab initio} calculations are lacking for many
relevant nuclei. In contrast to the simple model we employ the mass
numbers and nuclear magnetic moments of the considered
isotopes with explicitly including the signs. Due to the approximation of nuclear structure, the limits
presented in the following can not be directly taken as strict limits
on $\mathcal{P,T}$-violation, but should instead be used for
comparison among different experiments and approaches. Further, at the
present stage we also do not distinguish between $d_\mathrm{p}$ and
$d_\mathrm{n}$ but use $d_\mathrm{N}$ as in the analytic model. 

All resulting $\alpha_i$ we obtained are listed in Table \ref{tab:
all_data}. Individual electronic structure parameters needed to
compute $\alpha_i$ are listed in Table \ref{tab: elecstruc} and
estimates of nuclear structure parameters are given in Table \ref{tab:
nucstruc}, both in the Appendix.

\subsubsection{Current experimental status}
We computed the current total coverage volume using $\alpha_i$ from
Table \ref{tab: all_data} and experimental uncertainties and external
electrical fields (for molecular experiments) shown in Table \ref{tab:
exp_data} to be
$V=\SI{1.2e-82}{\elementarycharge\squared\centi\meter\squared}$ and
visualize the current restriction of the $\mathcal{P,T}$-odd parameter
space by showing the edges of the six-dimensional ellipsoid in all
two-dimensional sub-spaces (i.e.\ at the planes for which all other
sources are zero) in Figure \ref{fig: current_status}. From the
two-dimensional sub-spaces one sees that currently the most
determining experiments are with ThO, HfF$^{+}$, Xe, Hg and TlF. 
The importance of the ThO and Hg measurements for
constraining the coverage region in the $d_\mathrm{e}$-$k_\mathrm{s}$
sub-space was highlighted in Ref. \cite{fleig:2018} as well, although
the role of ThO became severely reduced by virtue of the latest
HfF$^{+}$ results \cite{roussy:2023}. The reduction of the total
volume by inclusion of individual experiments is shown in Table
\ref{tab: compare_newexps}.  We find that currently the Hg, Xe, TlF,
HfF$^+$ and ThO give the most important restrictions
of the parameter space with reduction of the volume by two orders of
magnitude or more. But also Yb contributes a more than one order of
magnitude reduction of the total coverage
volume. The reported experiments with Tl, Cs, PbO, Ra and YbF have in
the present analysis no significant influence on
the total coverage volume. In case of Cs, Tl,
PbO and Ra the experimental uncertainty achieved, was in comparison
too, whereas the YbF experiment is inferior to the \ce{HfF+}
experiment that has an almost equal inclination in the
$d_\mathrm{e}-k_\mathrm{s}$ parameter space. The latter underlines the
relevance of complementarity of new experiments and the role of a global
analysis. 

Furthermore, the importance of a global analysis is highlighted with Table
\ref{tab: limits}, wherein global bounds are opposed to supposedly
best single-source bounds. The latter underestimate in some
cases global bounds by several orders of magnitude. 

\subsubsection{Influence of planned experiments}
We assume for all upcoming molecular experiments to reach experimental
uncertainties with
$\sigma_d=\SI{1e-23}{\elementarycharge\centi\meter}$ and for Fr with
$\sigma_d=\SI{1e-28}{\elementarycharge\centi\meter}$, which are
improved compared to the \ce{HfF+} and Tl experiments, respectively.
We choose for upcoming molecular experiments typical external electric
fields for polarization of strength
$\efield=\SI{10}{\kilo\volt\per\centi\meter}$ for
$^2\Sigma_{1/2}$-states (see also
Refs.~\cite{aggarwal:2018,petrov:2020} for BaF and RaF) and of
strength $\efield=\SI{100}{\volt\per\centi\meter}$ for $\Delta$ states
and polyatomic molecules (see also \cite{lee:2013,kozyryev:2017a})
were we assume full polarization $\efield=\efield\lambda_z$. For the
$\Phi$ state of \ce{PaF^3+} we assume that a smaller electric field is
needed than for $\Sigma$ states such as in RaF and take the value
$\efield=\SI{1000}{\volt\per\centi\meter}$. We summarize these
assumptions in Table \ref{tab: proposed_data}. Both uncertainties and
external fields needed for full polarization may differ in actual
future experiments and full polarization may actually not be
reachable.  Nonetheless, for the chosen values, only the order of
magnitude has an influence on the current results and this should not
spoil our analysis (see for discussions on actual experimental
uncertainties of some of the planned experiments also
Refs.~\cite{lee:2013,kozyryev:2017a,aggarwal:2018}). Recently it was
suggested to perform EDM experiments by employing superposition states
without the need for external electric field for polarization
\cite{anderegg:2023}. This would lift the apparent advantages of
$^3\Delta_{1}$ states.

We again assess the potential impact of individual planned and proposed
experiments by their reduction factor of the total coverage volume as shown
in Table \ref{tab: compare_newexps}. The reduction factors
have no absolute meaning, as the crude assumptions on experimental
uncertainties, used external field strengths and nuclear structure will severely influence the
absolute bounds of single experiments. In particular, octupole deformed nuclei
such as $^{223/225}$Ra, $^{229}$Th or nuclei that are discussed to
have huge octupole deformations such as $^{229}$Pa may reduce the
total coverage volume by many orders of magnitude below the values
discussed in this work. E.g.\ employing an estimate of
$\tilde{\mathcal{S}}(\ce{^{229}Pa})\approx\SI{-104}{\elementarycharge\femto\meter\cubed}$
from Ref.~\cite{flambaum:2020b} would lead to a larger $\alpha_6$ and
a volume reduction of $f\approx\SI{6.4e-6}{}$, which is about three
orders of magnitude stronger than without considering this large
nuclear structure enhancement and superior to the result of
\ce{^{229}ThF+}, for which $\tilde{\mathcal{S}}$ would be of a similar
size as the used value according to Ref.~\cite{flambaum:2020b}.
Furthermore, the discussed $\mathcal{P,T}$-odd parameter space
contains only the eEDM as elementary $\mathcal{P,T}$-odd parameter.
All other parameters contain more than one parameter on the elementary
particle level. In particular $d_\mathrm{N}$ and $g_\mathrm{\pi NN}$
are effective parameters that are even on the nucleon level composed
of several $\mathcal{P,T}$-odd parameters. As of that in combination
with nuclear structure theory the $\mathcal{P,T}$-odd parameter space
should be of higher dimension and more than six experiments are
required to define a finite coverage volume.

Our results provide a relative measure for possible restrictions of
the coverage volume in $\mathcal{P,T}$-odd parameter space due to
electronic structure effects of a proposed system. All planned
experiments have a considerable influence on the coverage volume. Most
notable is that a zero result from a class I experiment with RaF would
have a smaller influence as this reduces only the two-dimensional
$d_\mathrm{e}-k_\mathrm{s}$ parameter space, which is, in comparison,
relatively well restricted so far as discussed above. Table \ref{tab:
compare_newexps} demonstrates the benefit of experiments with
medium-heavy systems such as CaOH, SrOH or YO. Only CdH seems to bring
less advantages, maybe because its nuclear charge number ($Z=48$) is
already too close to e.g.\ that of Xe ($Z=54$).  Considering that the
experimental uncertainty is probably even lower for these systems, one
could expect still higher benefits for experiments with YO and SrOH or
CaOH. This paves the way for directly accessible high-precision
studies that have the potential to advance current limits on
$\mathcal{P,T}$-violation and tighten rigorous bounds.

\subsection{Quality of phenomenological predictions}
Finally, with our \emph{ab initio} data we can determine the quality of
the simple model used for the global optimization.  In the present,
simple model we completely neglected many-body effects, for
instance from the chemical environment in a molecule on the electronic
structure enhancement factor of an atom. Although chemical
substituents can have a pronounced effect on the absolute enhancement
factors, predictions of the present model can be expected
to be qualitatively good, as we have shown in our previous work (see
Ref.  \cite{gaul:2019}) for the two dimensional
$d_\mathrm{e}$-$k_\mathrm{s}$-parameter space of class I molecules
when compared to results computed \emph{ab initio} for each molecular
system.  In order to compare our phenomenological global model to our
\emph{ab inito} calculations, we eliminate the unknown effective quantum
numbers $\nu$ in the definition of the $\alpha_i$ by computing the matrix of ratios
relative to $\alpha_{1i}$:
\begin{equation}
\tilde{\alpha}_{ji} = \alpha_{ji}/\alpha_{1i}
\end{equation} 
The matrix of ratios $\tilde{\bm{\alpha}}$ is then multiplied with a
diagonal matrix containing $\bm{\alpha}_{1,\mathrm{ab
initio}}=\mathrm{diag}(\alpha_{11},\dots,\alpha_{1N})$ from \emph{ab
initio} calculations. From this we find a numerical phenomenological total
coverage volume from the simple model used in the global minimization,
which we can compare to the \emph{ab initio} total coverage volume in
Table \ref{tab: compare_newexps}. We find for many systems an
excellent agreement, however, in particular the effect of some
molecular systems on the size of the coverage volume is overestimated
in the simple phenomenological approach, showing the need for \emph{ab
initio} data for deducing robust limits from experiments. Nonetheless,
the qualitative trend is for most systems well represented by the
phenomenological model, justifying our present conclusions. 

\section{Conclusion}
We presented a global analysis of the complete $\mathcal{P,T}$-odd
parameter space from the perspective of electronic structure theory.
For this purpose, we developed a simple analytical electronic
structure model to determine the optimal choice of complementary
systems for a minimization of the coverage volume in the
$\mathcal{P,T}$-odd parameter space as function of nuclear charge
number, electronic and nuclear angular momenta. To support and
illustrate this model and its results, we performed density functional
theory calculations for all atomic and molecular systems that have
been used in modern experiments to restrict $\mathcal{P,T}$-violation
and for many of those that are momentarily planned to be used for this
purpose in future. Our results show that it is most important to have
complementary experiments, i.e.\ experiments with systems for which
different terms appear in the effective Hamiltonian. In contradiction
to what is commonly assumed and naively expected from $Z$-scaling laws of
individual enhancement factors, we find that systems with relatively low 
nuclear charge number ($Z\leq54$) can improve rigorous
bounds in the $\mathcal{P,T}$-odd parameter space by several orders of 
magnitude due to different constraining behaviour compared to their heavier
counterparts. This
motivates experiments with well-understood lighter molecules, in which
the heavier atom has a nonzero nuclear spin for detecting
$\mathcal{P,T}$-odd fundamental parameters, such as $^{89}$YO,
$^{87}$SrOH or even $^{43}$CaOH. 

Momentarily, our model treats nuclei only on a rather crude level and
does not even distinguish explicitly between odd-even and even-odd nuclei.  Nonetheless, our present
model can easily be extended to include nuclear structure effects in a more
sophisticated way for identifying complementarity on the elementary level
of strong $\mathcal{CP}$-violation and 
in systems with long isotope chains such as \ce{RaF}. The
findings of the current work clearly open-up an important direction in the
search for $\mathcal{P,T}$-violation with high-precision spectroscopy:
well understood, well-controllable molecules such as CaOH, YO and SrOH can
greatly enrich the set of previous and upcoming experiments with heavy-elemental
systems and have the potential to rigorously tighten in a global analysis bounds on
the fundamental sources of $\mathcal{P,T}$-violation.
\bigskip

\begin{acknowledgments}
Computer time provided by the center for scientific computing (CSC)
Frankfurt is gratefully acknowledged. We thank participants of the
766th WE-Heraeus-Seminar on ``High‐Precision Measurements and Searches
for New Physics'' in May 2022, of the DPG-spring meeting in February
2023 and of the XVIIIth and XIXth Symmetry in Science Symposia in
August 2019 and 2023 for discussion of our results. We are grateful to
Skyler Degenkolb for feedback, comments and discussions on our manuscript.
We also thank ECT* Trento for support at the Workshop "EDMs:
complementary experiments and theory connections" during which this work
has been discussed.
\end{acknowledgments}

\bibliography{AK.bib}

\appendix
\section{Volume of an $N$-dimensional
ellipsoid\label{suppl_globalptodd_vol}}
An $N$-dimensional ellipsoid can be described by $N$ semi-axes $a_i$.
In Euclidean space its volume is 
\begin{equation}
V=\frac{2\pi^{N/2}}{N\Gamma(N/2)}\prod\limits_{i=1}^N a_i,
\end{equation} 
which reduces to the volume of an $N$-ball if all semi-axes are of 
equal length.
In general an ellipsoid centered at $\vec{x}_0$ can be represented as
quadric:
\begin{equation}
\parantheses{\vec{x}-\vec{x}_0}^\mathsf{T}
\bm{A}
\parantheses{\vec{x}-\vec{x}_0} = b,
\end{equation}
where the eigenvectors of $\bm{A}$ are the principal axes of the
ellipsoid and the eigenvalues $\tilde{a}_{ii}$ of $b^{-1}\bm{A}$ are the inverse, squared
semi-axes of the ellipsoid $\tilde{a}_{ii}=a_i^{-2}$. Thus, the equivalence 
\begin{equation}
\frac{2\pi^{N/2}}{N\Gamma(N/2)}\prod\limits_{i=1}^N a_i
~\Leftrightarrow~
\frac{2\pi^{N/2}}{N\Gamma(N/2)}\parantheses{\mathrm{det}(\tilde{\bm{a}})}^{-1/2}\,,
\end{equation}
where 
\begin{equation}
b^{-1}\bm{A}\bm{U}=\bm{U}\tilde{\bm{a}};\qquad\bm{U}^\dagger\bm{U}=\bm{1},
\end{equation}
holds and $\tilde{\bm{a}}$ is a diagonal matrix containing the
eigenvalues of $b^{-1}\bm{A}$. Here, $\bm{U}$ is the matrix of
eigenvectors of $\bm{A}$. As a unitary transformation does not change
the determinant of a matrix, the volume can be written as
\begin{equation}
V=b^{1/2}\frac{2\pi^{N/2}}{N\Gamma(N/2)}\parantheses{\mathrm{det}(\bm{A})}^{-1/2}\,.
\end{equation}
\section{Effective electronic $\mathcal{P,T}$-odd operators and
enhancement factors}
Employed formulas for the various electronic structure enhancement factors
$W$ are listed in Table~\ref{tab: ws}.
\begin{table*}
\begin{threeparttable}
\caption{Form of electronic enhancement factors $W$ as employed in this
work. Here
$\Braket{}$ denotes the expectation value for a given many-electron
wave function and $\Ket{0}$, $\Ket{a}$ denote wave functions of a
reference and excited electronic states with energies $E_0$ and $E_a$
respectively. $c$ is the speed of light, $\hbar$ is the reduced Planck
constant, $\mu_0$ is the magnetic constant, $\epsilon_0$ is the
electric constant, $G_\mathrm{F}$ is the Fermi constant for which we
employ the value \SI{2.22249e-14}{\hartree\bohr\cubed}. $\pos_a$,
$\momop_a=-\imath\hbar\nabla_a$
and $\hat{\vec{\ell}}_{ab}=-\imath\hbar\pos_{ab}\times\nabla_a$ are the position operator, momentum operator
and angular momentum operator of a particle $a$ relative to particle $b$, respectively. The
relative position of two particles is $\pos_{ab}=\pos_a-\pos_b$ and
the distance operator is $r_{ab}=\left|\pos_{ab}\right|$.
$\rho_A$ is the normalized nuclear charge density distribution of
nucleus $A$. $\vec{I}/I$ is the direction in which the nuclear
angular momentum is pointing. The Dirac $\gamma$-matrices are defined as
$\diracg=\begin{pmatrix}\bm{0}&\vec{\pauli}\\-\vec{\pauli}&\bm{0}\end{pmatrix}$,
$\diraccontra{0}=\begin{pmatrix}\bm{1}&\bm{0}\\\bm{0}&-\bm{1}\end{pmatrix}$,
$\diracb=\diraccontra{0}$, $\diraca=\diraccontra{0}\diracg$ and
$\diraccontra{5}=\imath\diraccontra{0}\diraccontra{1}\diraccontra{2}\diraccontra{3}$.}
\label{tab: ws}
\begin{tabular}{lll}
\toprule
$W_\mathrm{d}$           =& $ \frac{\Braket{\frac{2c}{e\hbar}\Sum{i=1}{N_\mathrm{elec}}\imath\diraccontra{0}_i\diraccontra{5}_i\momop^2_i}}{\Omega}  $ & \cite{martensson-pendrill:1987}\\ 
$W^\mathrm{m}_\mathrm{d}$=& $
\frac{\vec{I}}{I}\cdot\Braket{\frac{2c\mu_0}{4\pi\hbar}\Sum{i=1}{N_\mathrm{elec}}\frac{\imath\diraccontra{0}_i\diraccontra{5}_i}{r_{iA}^3}\hat{\vec{\ell}}_{iA}}+\frac{\vec{I}}{I}\cdot2\mathrm{Re}\braces{\Sum{a}{}\frac{\Braket{0|\frac{2c}{e\hbar}\Sum{i=1}{N_\mathrm{elec}}\imath\diraccontra{0}_i\diraccontra{5}_i\momop^2_i|a}\Braket{a|\frac{\mu_0}{4\pi}\Sum{i=1}{N_\mathrm{elec}}\frac{\pos_{iA}\times\diraca_i}{r_{iA}^3}|0}}{E_0-E_a}}  $ & \cite{martensson-pendrill:1987}\\
$W_\mathcal{S}$          =& $ \frac{\vec{I}}{I}\cdot\Braket{-\frac{e}{\epsilon_0}\Sum{i=1}{N_\mathrm{elec}}(\nabla_i\rho_A(\pos_i))}  $ & \cite{hinds:1980,flambaum:2002}\\
$W_\mathrm{m}$           =& $ \frac{\vec{I}}{I}\cdot\Braket{4\frac{c\mu_0}{4\pi\hbar}\Sum{i=1}{N_\mathrm{elec}}\frac{1}{r_{iA}^{3}}\hat{\vec{\ell}}_{iA}\times\diraca_i}  $ & \cite{hinds:1980}\\
$W_\mathcal{M}$          =& $ \frac{\vec{I}^\mathsf{T}}{I}\cdot\frac{\Braket{\frac{ce\mu_0}{4\pi}\frac{3}{2}\Sum{i=1}{N_\mathrm{elec}}\frac{\pos_{iA}\cdot\parantheses{\diraca\times\pos_{iA}}^\mathsf{T}}{r_{iA}^5}}}{\Omega}\cdot\frac{\vec{I}}{I}  $ & \cite{flambaum:1985}\\
$W_\mathrm{s}$           =& $ \frac{\Braket{\frac{-Z_AG_\mathrm{F}}{\sqrt{2}}\Sum{i=1}{N_\mathrm{elec}}\imath\diraccontra{0}_i\diraccontra{5}_i\rho_A(\pos_i)}}{\Omega}  $ & \cite{sushkov:1984}\\
$W^\mathrm{m}_\mathrm{s}$=& $
\frac{\vec{I}}{I}\cdot2\mathrm{Re}\braces{\Sum{a}{}\frac{\Braket{0|\frac{-Z_AG_\mathrm{F}}{\sqrt{2}}\Sum{i=1}{N_\mathrm{elec}}\imath\diraccontra{0}_i\diraccontra{5}_i\rho_A(\pos_i)|a}\Braket{a|\frac{\mu_0}{4\pi}\Sum{i=1}{N_\mathrm{elec}}\frac{\pos_{iA}\times\diraca_i}{r_{iA}^3}|0}}{E_0-E_a}}  $ & \cite{flambaum:1985}\\
$W_\mathrm{T}$           =& $ \frac{\vec{I}}{I}\cdot\Braket{\sqrt{2}G_\mathrm{F}\Sum{i=1}{N_\mathrm{elec}}\imath\diracg\rho_A(\pos_i)}  $ & \cite{hinds:1980}\\
$W_\mathrm{p}$           =& $ \frac{\vec{I}}{I}\cdot\Braket{-\frac{G_\mathrm{F}\mu_\mathrm{N}}{\sqrt{2}ec}\Sum{i=1}{N_\mathrm{elec}}\diracb_i(\nabla_i\rho_A(\pos_i))}  $ & \cite{flambaum:1985}\\
\bottomrule
\end{tabular}
\end{threeparttable}
\end{table*}

\section{Analytic expressions for electronic $\mathcal{P,T}$-odd
enhancement factors\label{analytic_enhancement}} 
In the following, we summarize atomic matrix elements of
$\mathcal{P,T}$-odd operators as functions of the nuclear charge
number $Z$ in terms of relativistic enhancement factors. We discuss
leading order matrix elements only. In most cases this corresponds to
mixing of $\mathrm{s}_{1/2}$ and $\mathrm{p}_{1/2}$ orbitals and in
some cases involve also $\mathrm{p}_{3/2}$ orbitals. These orbitals
are always characterized by effective principal quantum numbers
$\nu_{l_j}$ for the valence electron, that are then assumed to contain
all information on atomic or molecular orbitals. 

In the minimization
of the volume in the global $\mathcal{P,T}$-odd parameter space
calculations presented in the paper, all $\nu_{l_j}$ are set to one.
In contrast to principal quantum numbers, the effective quantum
numbers do not vary
much for elements from different rows of the periodic table and for
different angular momenta, and are, thus, only weakly dependent on the
nuclear charge number $Z$. Hence, the approximation
$\nu_{l_j}\approx1$ does not influence the relative volume of
combinations of systems with different nuclear charge numbers
significantly but will only have a similar influence on the absolute
value of the coverage volume for all combinations of $Z$, which is in
the phenomenological model anyway arbitrary.

A detailed discussion of all matrix elements except that
of the magnetic NEDM interaction $W_\mathrm{m}$ can be found in
Ch. 8 of Ref.~\cite{khriplovich:1997} and specific references for each
property will be given below.

The radial part of the valence single-electron atomic wave functions
in the screened potential of point-like nucleus
employed to arrive at the enhancement factors is (see
Ref.~\cite{khriplovich:1997})
\begin{widetext}
\begin{align}
f_\kappa(r)=&
\frac{\kappa}{\abs{\kappa}r\sqrt{Za_0\nu_\kappa^3}}\parantheses{(\kappa +
\gamma_\kappa)
J_{2\gamma_\kappa}\parantheses{\sqrt{\frac{8Zr}{a_0}}}
 - \sqrt{\frac{2Zr}{a_0}}
   J_{2\gamma_\kappa-1}\parantheses{\sqrt{\frac{8Zr}{a_0}}}}\,,
\label{eq: wave_function1}
\\
g_\kappa(r)=&
\frac{\kappa Z \alpha}{\abs{\kappa}r\sqrt{Za_0\nu_\kappa^3}}
J_{2\gamma_\kappa}\parantheses{\sqrt{\frac{8Zr}{a_0}}}\,,
\label{eq: wave_function2}
\end{align}
\end{widetext}
\sloppy where $f$ and $g$ are the radial functions of the large and
small components, respectively, of the four-component wavefunction. Here, $\kappa$ is the
relativistic angular momentum quantum number with values $\kappa =
(-1)^{j+1/2-l}(j+\sfrac{1}{2})$ with the total electronic angular
momentum quantum number $j=\abs{\kappa}-\sfrac{1}{2}=s+l$ and the
electronic orbital and spin angular momentum quantum numbers
$s=-\frac{\kappa}{2\abs{\kappa}}$ and $l=j-s$ and the inverse Lorentz factor
is $\gamma_\kappa=\sqrt{\kappa^2-\alpha^2Z^2}$. $J$ is the Bessel
function of first kind.  The full four-component atomic wave function
is
\begin{equation}
\psi_{\kappa,m}(r,\vartheta,\varphi) = \begin{pmatrix} 
f_\kappa(r)\Omega_{\kappa,m}(\vartheta,\varphi) \\
\imath g_\kappa(r)\Omega_{-\kappa,m}(\vartheta,\varphi) \\
\end{pmatrix}\,,
\end{equation}
with the angular wave function
\begin{equation}
\Omega_{\kappa,m}(\vartheta,\varphi) =
\frac{1}{\sqrt{2l+1}}
\begin{pmatrix}
2s
\sqrt{l+\frac{1}{2}+2sm}Y_{l,m-\frac{1}{2}}(\vartheta,\varphi)\\
\sqrt{l+\frac{1}{2}-2sm}Y_{l,m+\frac{1}{2}}(\vartheta,\varphi)
\end{pmatrix}\,,
\end{equation}
with the spherical harmonics $Y_{l,m}$. This atomic wave function is
only a good approximation for nuclear charge numbers $20<Z<100$
\cite{khriplovich:1997}. In particular there is a pole at
$Z=\alpha\approx137$.

\sloppy Following Ref.~\cite{khriplovich:1997} closely, the electronic
enhancement factors are presented in terms of units with $\hbar=1$,
$c=1$, $4\pi\epsilon_0=1$ and length is given in units of \si{cm}: the
Bohr radius is $a_0=\SI{5.29e-9}{\centi\meter}$, the fine structure
constant is $\alpha=e^2\approx\frac{1}{137.036}$, the proton mass is
$m_\mathrm{p}=\SI{4.75511e13}{\per\centi\meter}$, the nuclear charge
radius is $r_\mathrm{nuc}=\SI{1.2e-13}{\centi\meter}\,A^{1/3}$, the Fermi constant is
$G_\mathrm{F}=\SI{1.027e-5}{}/m_\mathrm{p}^2$, the elementary charge
is $e=\sqrt{\alpha}$, the electron mass is
$m_\mathrm{e}=\frac{1}{\alpha a_0}$ and energy is given in units of
Rydberg $\mathrm{Ry}= \frac{m_\mathrm{e}\alpha^2}{2}$.

The leading order electronic enhancement factor of the eEDM was
derived in its full form in Ref.~\cite{flambaum:1976}:
\begin{equation}
W_\mathrm{d}(Z)=\frac{4}{3}\frac{\alpha}{a_0^2} \frac{\alpha^2Z^3}{(\nu_{\mathrm{p}_{1/2}}\nu_{\mathrm{s}_{1/2}})^{3/2}}
    \frac{3}{\gamma_{1/2}(4 \gamma_{1/2}^2-1)}\,.
\end{equation}
This factor diverges for $Z =\frac{\sqrt{3}}{2\alpha}\approx118.7$
and, thus, leads to large deviations for $Z>100$ (see discussion
in Refs.~\cite{dinh:2009,gaul:2019}).

The magnetic contribution to it emerges from the second order
contribution due to the nuclear hyperfine interaction and due to the
magnetic field of the nuclei directly interacting with the electrons.
It was calculated in leading order as matrix element of
$\mathrm{s}_{1/2}$ and $\mathrm{p}_{1/2}$ orbitals in
Ref.~\cite{flambaum:1985} and the full analytic expression can be
found in Ref.~\cite{khriplovich:1997}:
\begin{equation}
W^\mathrm{m}_\mathrm{d}(Z)= -\frac{14}{3} \frac{e
\mu/\mu_\mathrm{N}}{m_\mathrm{p}a_0^3}\frac{\alpha Z^2}{(\nu_{\mathrm{p}_{1/2}}\nu_{\mathrm{s}_{1/2}})^{3/2}} (R(Z,A) - 1)\,,
\end{equation}
where $R(Z,A)
=\frac{4}{\Gamma^2(2\gamma_{1/2}+1)}(2Zr_\mathrm{nuc}/a_0)^{2\gamma_{1/2}-2}$
is a relativistic enhancement factor.

The matrix elements of the SPNEC interaction were calculated in
Ref.~\cite{sushkov:1978}:
\begin{equation}
W_\mathrm{s}(Z)=
\frac{G_\mathrm{F}m_\mathrm{e}^2\alpha\,\mathrm{Ry}}{\sqrt{2}\pi}
\frac{\alpha Z^3}{(\nu_{\mathrm{p}_{1/2}}
\nu_{\mathrm{s}_{1/2}})^{3/2}} 
    R(Z,A) \gamma_{1/2}\,.
\end{equation}
A refined relativistic enhancement factor was calculated in
Ref.~\cite{dzuba:2011}:
\begin{multline}
W_\mathrm{s}(Z)=
\frac{G_\mathrm{F}m_\mathrm{e}^2\alpha\,\mathrm{Ry}}{\sqrt{2}\pi}
\frac{\alpha Z^3}{(\nu_{\mathrm{p}_{1/2}}
\nu_{\mathrm{s}_{1/2}})^{3/2}} \\
\times    R(Z,A) \frac{\gamma_{1/2}+1}{2}f_0(Z)\,,
\end{multline}
with $f_0(Z) = (1 - 0.56 \alpha^2 Z^2)/(1 - 0.283 \alpha^2 Z^2)^2$.
This refined form of $W_\mathrm{s}$ was employed in all calculations
in the present work.

The second order contribution that is induced due to hyperfine
coupling to magnetic nuclei was
derived in Ref.~\cite{khriplovich:1997} to be in leading order:
\begin{multline}
W^\mathrm{m}_\mathrm{s}(Z)=-\frac{8}{3}
\frac{G_\mathrm{F}m_\mathrm{e}^2\alpha\,\mathrm{Ry}}{\sqrt{2}\pi}
\frac{\alpha \mu/\mu_\mathrm{N}}{m_\mathrm{p} r_\mathrm{nuc}}\\
\times\frac{\alpha Z^3}{(\nu_{\mathrm{p}_{1/2}}
\nu_{\mathrm{s}_{1/2}})^{3/2}}
 R(Z,A)\,.
\end{multline}

Matrix elements of the TPNEC interaction were discussed in
Ref.~\cite{flambaum:1985} and the full analytic calculation was
presented in Ref.~\cite{khriplovich:1997}:
\begin{equation}
W_\mathrm{T}(Z) = -\frac{4G_\mathrm{F} m_\mathrm{e}^2
\alpha\,\mathrm{Ry}}{\sqrt{2}\pi}
\frac{\alpha Z^2}{(\nu_{\mathrm{p}_{1/2}}\nu_{\mathrm{s}_{1/2}})^{3/2}}
 R(Z,A) \frac{(2 +\gamma_{1/2})}{3}\,.
\end{equation}

Matrix elements of the PSNEC interaction were discussed in
Ref.~\cite{flambaum:1985} and a more detailed analytic calculation is
presented in Ref.~\cite{khriplovich:1997}:
\begin{equation}
W_\mathrm{p}(Z) = -\frac{2G_\mathrm{F} m_\mathrm{e}^2
\alpha\,\mathrm{Ry}}{3\sqrt{2}\pi m_\mathrm{p}r_\mathrm{nuc}}
\frac{\alpha^2 Z^3}{(\nu_{\mathrm{p}_{1/2}}\nu_{\mathrm{s}_{1/2}})^{3/2}}
 R(Z,A)\,.
\end{equation}

The interactions of atomic
electrons with the $\mathcal{P,T}$-odd nuclear moments, namely the Schiff
moment and the NMQM, were derived in Ref.~\cite{sushkov:1984} and
read in leading order
\begin{equation}
W_\mathcal{S}(Z) =
\frac{1}{a_0^4}
\frac{Z^2}{(\nu_{\mathrm{p}_{1/2}}\nu_{\mathrm{s}_{1/2}})^{3/2}}
R(Z, A)\frac{3\gamma_{1/2}}{2\gamma_{1/2}+1}\,,
\end{equation}
and
\begin{widetext}
\begin{equation}
W_\mathcal{M}(Z)=
-\frac{4}{3}\frac{m_\mathrm{e}\alpha\,\mathrm{Ry}}{e^2 a_0} 
\frac{\alpha
Z^2}{(\nu_{\mathrm{p}_{3/2}}\nu_{\mathrm{s}_{1/2}})^{3/2}}
\frac{720 \Gamma\parantheses{
\gamma_{1/2}+\gamma_{3/2}-2
}}
{\Gamma\parantheses{
3+\gamma_{1/2}-\gamma_{3/2}
}\Gamma\parantheses{
3-\gamma_{1/2}+\gamma_{3/2}
}\Gamma\parantheses{
3+\gamma_{1/2}+\gamma_{3/2}
}}\,.
\end{equation}
\end{widetext}
It shall be noted that the
$\mathrm{s}_{1/2}$-$\mathrm{p}_{1/2}$ matrix element
vanishes for $W_\mathcal{M}$.

To our knowledge there is no analytic expression for the enhancement of
the pEDM and nEDM due to magnetic electrons. For the pEDM the
operator appears as \cite{hinds:1980,quiney:1998a}
\begin{equation}
\label{eq: Hmp}
\op{H}_\mathrm{m,p} =\frac{2e}{m_\mathrm{p}}\parantheses{\frac{1}{A} + \frac{\mu/\mu_\mathrm{N}}{Z}} \frac{1}{r^3} \diraca\times\op{\vec{\ell}}\,,
\end{equation}
where $\diraca\times\op{\vec{\ell}}$ is a pure angular operator. For
$\op{H}_\mathrm{m,n}$ the prefactor has to be changed to
$\frac{2e}{m_\mathrm{p}}\parantheses{\frac{1}{A} +
\frac{\mu/\mu_\mathrm{N}}{N}}$.  In the present approach we use $A= 0.004467 Z^2 +
2.163 Z - 1.168$, giving $N$ not too different from $Z$ for lighter
nuclei with $Z<50$. For heavier nuclei $W_\mathrm{m}$ is anyway much
smaller than $W_\mathrm{S}$. Thus, we assume  
$\op{H}_\mathrm{m,n}\approx\op{H}_\mathrm{m,p}$.

Using the wave functions (\ref{eq: wave_function1}) and (\ref{eq:
wave_function2}) we computed the matrix elements of operator (\ref{eq:
Hmp}). The
$\mathrm{s}_{1/2}$-$\mathrm{p}_{1/2}$ matrix elements vanish.
The  $\mathrm{s}_{1/2}$-$\mathrm{p}_{3/2}$ matrix element of
$W_\mathrm{m}$ is found to be:
\begin{multline}
\vec{W}_{\mathrm{m}}(Z)\approx\frac{2e}{m_\mathrm{p}}
\parantheses{\frac{1}{A} + \frac{\mu/\mu_\mathrm{N}}{Z}}\\\times\iiint\mathrm{d}r
\mathrm{d}\vartheta \mathrm{d}\varphi\,r^2\sin\vartheta 
\psi_{0,1/2}^\dagger \frac{1}{r^3} \diraca\times\op{\vec{\ell}}
\psi_{-2,3/2}\\
=\frac{2e}{m_\mathrm{p}}\parantheses{\frac{1}{A} + \frac{\mu/\mu_\mathrm{N}}{Z}}\int\mathrm{d}r\,
r^{-1}f_{-2}(r)g_{-1}(r)\begin{pmatrix}\sqrt{6}\\0\\0\end{pmatrix}\,.
\end{multline}
For the chosen wave function the only non-vanishing component is the
$x$ component for which radial integration
results in the enhancement factor 
\begin{multline}
W_{\mathrm{m}}(Z) = \frac{2e}{m_\mathrm{p}}\parantheses{\frac{1}{A} +
\frac{\mu/\mu_\mathrm{N}}{Z}}
  \frac{2\sqrt{6}}{a_0^3(\nu_{\mathrm{s}_{1/2}}\nu_{\mathrm{p}_{3/2}})^{3/2}} \\ \times\frac{15}{16\gamma_{1/2}^2-1} \frac{\sin\parantheses{\pi (
\gamma_{3/2} - \gamma_{1/2})}}{\alpha\pi}\,,
\end{multline}
where the sine approaches zero for $Z\rightarrow0$. The sine was expanded in a series in $\alpha^2
Z^2=1-\gamma_{1/2}^2$ around zero and gives
$-\frac{\pi}{4}\alpha^2Z^2$ in leading order. Dividing by
$-\frac{\pi}{4}\alpha^2Z^2$ yields a relativistic
enhancement factor that approaches one for $Z\rightarrow0$. The series
converges in second order to $\sin\parantheses{\pi (
\gamma_{3/2} - \gamma_{1/2})}$ for $Z<100$ and reads 
\begin{multline}
\sin\parantheses{\pi (\gamma_{3/2} - \gamma_{1/2})} \\\approx
-\frac{\pi}{4}\alpha^2 Z^2 \parantheses{1+\frac{7}{16} (1-\gamma_{1/2}^2)
+ \frac{93-4\pi^2}{384} (1-\gamma_{1/2}^2)^2}\,.
\end{multline}
Finally, we arrive at the electronic enhancement factor:
\begin{widetext}
\begin{equation}
W_\mathrm{m}(Z) =
-\frac{2e}{m_\mathrm{p}}\parantheses{\frac{1}{A} + \frac{\mu/\mu_\mathrm{N}}{Z}}
  \frac{\alpha  Z^2 \sqrt{3}}{\sqrt{2}a_0^3(\nu_{\mathrm{s}_{1/2}}\nu_{\mathrm{p}_{3/2}})^{3/2}} 
\frac{15\parantheses{1+\frac{7}{16} (1-\gamma_{1/2}^2)
+ \frac{93-4\pi^2}{384} (1-\gamma_{1/2}^2)^2}}{16\gamma_{1/2}^2-1}\,.
\end{equation}
\end{widetext}
Due to the prefactor $\frac{1}{A} +
\frac{\mu/\mu_\mathrm{N}}{Z}=Z^{-1}\parantheses{\frac{1}{N/Z+1} +
\mu/\mu_\mathrm{N}}$ this electronic enhancement factor scales as $\alpha Z$.

The relativistic enhancement factors and the scaling with nuclear charge number
$Z$ for all electronic structure parameters discussed above are summarized in
\prettyref{tab: ws2}.
\begin{table*}
\begin{threeparttable}
\caption{Summary of scaling of electronic enhancement factors $W$ with nuclear
charge number $Z$ and relativistic enhancement factors.} 
\label{tab: ws2}
\begin{tabular}{llll}
\toprule
$W_i$                    &  $\alpha Z$-scaling& Relativistic enhancement                      & Reference \\
\midrule
$W_\mathrm{d}$           &  $\alpha^2Z^3$     & $\frac{3}{\gamma_{1/2}(4\gamma_{1/2}^2-1)}$   &  \cite{sandars:1966,flambaum:1976}\\
$W^\mathrm{m}_\mathrm{d}$&  $\alpha  Z^2$     & $R(Z,A) - 1$                                  &  \cite{flambaum:1985}\\ 
$W_\mathcal{S}$          &  $        Z^2$     & $R(Z,A) \frac{3\gamma_{1/2}}{2\gamma_{1/2}+1}$&  \cite{sushkov:1984} \\
$W_\mathrm{m}$           &  $\alpha  Z  $     & $\frac{15\parantheses{1+\frac{7}{16} (1-\gamma_{1/2}^2)+ \frac{93-4\pi^2}{384} (1-\gamma_{1/2}^2)^2}}{16\gamma_{1/2}^2-1}$ & This work.\\
$W_\mathcal{M}$          &  $\alpha  Z^2$     & $\frac{720 \Gamma\parantheses{\gamma_{1/2}+\gamma_{3/2}-2 }} {\Gamma\parantheses{3+\gamma_{1/2}-\gamma_{3/2}}\Gamma\parantheses{3-\gamma_{1/2}+\gamma_{3/2}}\Gamma\parantheses{3+\gamma_{1/2}+\gamma_{3/2} }}$ & \cite{sushkov:1984}\\
$W_\mathrm{s}$           &  $\alpha  Z^3$     & $R(Z,A) \frac{\gamma_{1/2}+1}{2}f_0(Z)$       &\cite{sushkov:1978,dzuba:2011}\\
$W^\mathrm{m}_\mathrm{s}$&  $\alpha^2Z^3$     & $R(Z,A)$                                      &\cite{khriplovich:1997}\\
$W_\mathrm{T}$           &  $\alpha  Z^2$     & $R(Z,A) \frac{(2 +\gamma_{1/2})}{3}$          &\cite{flambaum:1985}\\
$W_\mathrm{p}$           &  $\alpha^2Z^3$     & $R(Z,A)$                                      &\cite{flambaum:1985}\\
\bottomrule
\end{tabular}
\end{threeparttable}
\end{table*}
\section{Methodology}

\subsection{Global minimization procedure}
A minimization of $V$ with respect to $Z_i$ was performed for all 64
combinations of $\Omega_i$ and $\mathcal{I}_i$ described above (four
different classes).  Here, first, we used the restriction $20\le
Z\le100$, as the employed atomic models give reasonable estimates for
this region only (see Ref.~\cite{gaul:2019}), and, second, we used the
restriction $20\le Z\le90$ as so far conducted experiments are limited
to $Z\le90$ (Th as heaviest element). 

\begin{table}
\begin{threeparttable}
\caption{Mathematica 11 algorithms and options used for a global minimization of the volume in the
$\mathcal{P,T}$-odd parameter space with respect to nuclear charges $Z_i$. For
remaining options not explicitly listed, default values were employed.}
\label{tab: global_algorithms}
\footnotesize
\begin{tabular}{
S
l
S[table-format=1e-1,round-mode=places, scientific-notation=true, round-precision =1]
S
}
\toprule
{Index\tnote{c}} &
{Method} &
{Tolerance\tnote{a}} &
{Search points\tnote{a}} \\
\midrule
1 & {Simulated Annealing}  & 0.00001 & 75 \\
2 & {Simulated Annealing}  & 0.00000001 & 100\\ 
3 & {Simulated Annealing}  & 0.0000000001 & 75\\ 
4 & {Differential Evolution      }  & 0.00001 & 75\\ 
5 & {Differential Evolution      }  & 0.00000001 & 100\\ 
6 & {Differential Evolution      }  & 0.0000000001 & 75\\ 
7 & {Nelder-Mead        }  & 0.0000000001 & {-}\\
8 & {Automatic\tnote{b} }  & {-} & {-}\\
\bottomrule
\end{tabular}
\begin{tablenotes}
\item[a] \emph{Tolerance} is the threshold at which a point is accepted or
withdrawn and \emph{search points} refers to the number of initially generated
points. Several other values for these two options have been tested
for selected sets of molecular classes and the herein covered range
was found to be most stable.
\item[b] The standard Mathematica algorithm with default options.
\item[c] Algorithms were evaluated in the given order. The result
was accepted as minimum, if the resulting volume in the
$\mathcal{P,T}$-odd parameter space was smaller than that found by the
previous algorithm.
\end{tablenotes}
\end{threeparttable}
\end{table}

The global minimization was performed with Mathematica
11~\cite{mathematica11}, employing different build-in optimization
algorithms for constrained global minimizations, which are listed in
\prettyref{tab: global_algorithms}. In order to speed up the minimization
considerably, the determinant
$\abs{\mathrm{det}\parantheses{\bm{W}}}^{-1}$ was described via
exponential expansions: 
\begin{equation}
\bm{W}=\mathrm{diag}\parantheses{W'_{\mathrm{d},1},\dots,W'_{\mathrm{d},7}}\cdot\tilde{\bm{W}}
\end{equation}
with $W'_{\text{d},i}=\Omega
W_{\mathrm{d},i}+\mathcal{I}W^\mathrm{m}_{\mathrm{d},i}$ and 
\begin{equation} 
\tilde{\bm{W}}_{ki} = \frac{W_{k,i}}{W'_{\mathrm{d},i}}
\end{equation}
is the matrix of $\mathcal{P,T}$-odd ratios with respect to the
electronic structure enhancement of the eEDM. Such $\mathcal{P,T}$-odd
ratios are expected to have an exponential dependence on the nuclear
charge~\cite{gaul:2019} and thus can be written as exponential
series expansion in very good approximation:
\begin{equation}
\frac{W_{k,i}}{W'_{\mathrm{d},i}} \approx
\exp\braces{\Sum{m=0}{3}a_{ki,m}Z^m}
\end{equation}

The minimizations were performed within the domain of natural numbers
and within the domain of real numbers. In the latter case the obtained
real numbers $Z_i$ were rounded to the nearest integer for the
calculation of the volume and its derivatives. In cases where
different algorithms arrived at different results, the result with the
smallest volume was employed.

The results were checked by calculation of the gradient and the
Hessian in the domain of natural numbers and the result was accepted
as minimum for $\abs{\nabla_Z V}/V(\vec{Z}_\mathrm{min}) < 0.5$ and if
the smallest eigenvalue of the Hessian $h_\mathrm{min}$ satisfies
$h_\mathrm{min}/V(\vec{Z}_\mathrm{min})>-10^{-2}$ and
$h_\mathrm{min}/h_\mathrm{max}>-10^{-1}$ and at least one other
eigenvalue of the Hessian has $h_i/h_\mathrm{max}>10^{-1}$. As a
completely vanishing gradient could not be achieved, small negative
eigenvalues of the Hessian were accepted. This limits the numerical
accuracy of the resulting volume to one significant figure, which is
acceptable as we aim only for a qualitative description.

\subsubsection{Note on the influence of neglected nuclear structure
effects} 

The nuclear magnetic moment $\mu$, that appears in $\mathcal{M}_\pi$,
$W_\mathrm{d}^\mathrm{m}$, $W_\mathrm{s}^\mathrm{m}$ and
$W_\mathrm{m}$, was set in all calculations to $\mu=1\,\mu_\mathrm{N}$. It shall be
noted that $\mu$ can have different signs for different nuclei, which
is not considered in the present study. This could have an influence
on the qualitative results, if the sign of the overall effect is
influenced by this. E.g., if two class II molecules would have heavy
nuclei of comparable nuclear charge $Z_1\approx Z_2$ but nuclear
magnetic moments of opposite sign $\mu_1\sim-\mu_2$, the coverage
volume should reduce, but in the present model a rather larger volume
would be obtained. This subtlety is to be considered for the choice of
complementary systems and can lead to a considerable reduction of the
coverage volume. In this sense our present study represents a worst
case scenario and we can conclude that such individual effects do not
influence the rules of thumb we presented.

\subsection{\emph{Ab initio} calculation of electronic structure parameters}
Quasi-relativistic electronic densities were computed at the level of
two-component zeroth order regular approximation (ZORA) within the
complex generalized Hartre--Fock (cGHF) and Kohn-Sham (cGKS)
approximations using a modified version
\cite{berger:2005,nahrwold:09,isaev:2012,gaul:2020,bruck:2023,colombojofre:2022,zulch:2022}
of a two-component program \cite{wullen:2010} based on Turbomole \cite{ahlrichs:1989}. The gauge
dependence of the ZORA Hamiltonian was alleviated by a model potential
as suggested by van W\"ullen \cite{wullen:1998} which was applied with
additional damping \cite{liu:2002}.  All elements were described with
the core-valence correlated Dyall basis set of quadruple zeta quality
(dyall.cv4z) \cite{dyall:2006} that was augmented for all elements
with an additional set of 12 s and 6 p functions for the description
of the wave function within the nucleus. These additional sets of
steep functions were composed as an even-tempered series starting at
$10^9\,\si{\bohr\squared}$ and progressing by division by $1.5$ for s
functions and $2.5$ for p functions. This very large basis was chosen
to guarantee a balanced description of electronic ground states and
excited states as well as virtual spinors that can be relevant in a
linear response calculation.

Density functional theory (DFT) calculations were performed within the
local density approximation (LDA) using the X$\alpha$ exchange
functional \cite{dirac:1930,slater:1951} and the VWN-5 correlation
functional \cite{vosko:1980}. Furthermore, as usually results for
$\mathcal{P,T}$-odd properties from more sophisticated methods such as
Coupled Cluster approaches lie between HF and LDA
\cite{gaul:2017,gaul:2020,gaul:2020a}, we employed the
hybrid LDA functional with 50\,\% Fock exchange by Becke (BHandH)
\cite{becke:1993}. We note that at the level of ZORA-GKS-LDA the
electronic density of some compounds could not be converged, which is
possibly due to some quasi-degeneracies of electronic states. In these
cases no LDA results are given.

In all calculations, the nucleus was described as normalized spherical Gaussian nuclear charge
density distribution $\rho_A \left( \vec{r} \right) =
\frac{\zeta_A^{3/2}}{\pi ^{3/2}} \text{e}^{-\zeta_A \left| \vec{r} -
\vec{r}_A \right| ^2}$ with $\zeta_A = \frac{3}{2 r^2
_{\text{nuc},A}}$. The root-mean-square radius $r_{\text{nuc},A}$
$r_{\text{nuc},A}$ was chosen as
suggested by Visscher and Dyall \cite{visscher:1997}, where we used the
isotopes  $^{1}$H, $^{12}$C, $^{15}$N, $^{16}$O and $^{19}$F in all
calculations. Specific isotopes used for other elements are given in
the results section.  
 
Excited state orbitals were obtained by
self-consistent field (SCF) calculations choosing occupation numbers regarding to maximum
overlap with the determinant of the initial guess (initial guess
maximum overlap method, IMOM) \cite{gilbert:2008,barca:2018}. As
initial guess we used the cGKS or cGHF determinant, that was found
with occupation of energetically lowest spinors. If the change in the
differential density with respect to the previous cycle was below
$10^{-3}/a_0^{-3}$ the standard MOM was used, where occupation numbers
are chosen with respect to maximum overlap with the determinant of the
previous cycle \cite{gilbert:2008}.

Molecular structure parameters were optimized up to an energy change
of less than $10^{-6}~E_{\text{h}}$ as convergence criterion.
Electronic densities were optimized until the change in energy and
relative spin-orbit coupling contribution was below
$10^{-10}~E_{\mathrm{h}}$ between two consecutive iterations. The
optimized bond length are given in Table \ref{tab: bond_length}. All
molecules are linear, i.e.\ M--O--H bond angles for M $=$ Ca, Sr and Yb
are \SI{180}{\degree}.

\begin{table}
\begin{threeparttable}
\caption{Bond length optimized at the level of ZORA-GHF or ZORA-GKS
with the LDA or BHandH functionals.}
\label{tab: bond_length}
\begin{tabular}{
l
l
S[table-format = 1.3, round-mode=places, round-precision=2]
S[table-format = 1.3, round-mode=places, round-precision=2]
S[table-format = 1.3, round-mode=places, round-precision=2]
}
\toprule
&&\multicolumn{3}{c}{$r_\mathrm{e}/$\AA}\\
\cline{3-5}
    Molecule & Bond & {HF} & {BHandH} & {LDA} \\
\midrule
       \ce{YbF}&        &      2.058442&       2.013164&\\
      \ce{HfF+}&        &      1.824771&       1.804032&\\
       \ce{TlF}&        &      2.078732&       2.071354&       2.070578\\
       \ce{PbO}&        &      2.138740&       2.138759&\\
       \ce{ThO}&        &      1.845215&       1.840054&\\
       \ce{BaF}&        &      2.196955&       2.154360&       2.116272\\
      \ce{YbOH}& Yb--O  &      2.078784&       2.030209\\
               & O--H   &      0.932705&       0.947039\\
        \ce{WC}&        &      1.969962&       1.692552&       1.686519\\
       \ce{RaF}&        &      2.268314&       2.229397&       2.194240\\
      \ce{ThF+}&        &      2.002692&       1.980143&\\
    \ce{PaF^3+}&        &      1.858753&       1.848980&\\
     \ce{CaOH} & Ca--O  &      2.002971&       1.965805&       1.925486\\
               & O--H   &      0.932498&       0.947061&       0.965590\\
      \ce{SrOH}& Sr--O  &      2.130712&       2.095193&       2.059779\\      
               & O--H   &      0.933716&       0.947738&       0.965590\\      
        \ce{YO}&        &      1.770802&       1.767913&       1.774594\\
       \ce{CdH}&        &      1.779603&       1.761862&       1.753357\\
\bottomrule
\end{tabular}
\end{threeparttable}  
\end{table}          

All molecular properties were computed with the toolbox approach
presented in Ref.~\cite{gaul:2020} using the expressions for the
various $W$ given in Table~\ref{tab: ws} in the Appendix. Linear
response equations were solved as described in
Refs.~\cite{bruck:2023,colombojofre:2022}. In atomic mean-field
calculations a very weak external electric field of strength
$\efield_z=0.00024\,E_\mathrm{h}/(e a_0)$ was applied along the z axis
to polarize the electronic shells. The third order properties
$\alpha^\mathrm{m}_\mathrm{d}$ and $\alpha^\mathrm{m}_\mathrm{s}$ were
computed in second order response theory using the polarized wave
function and dividing the response function subsequently by
$\efield_z$. All second order properties, i.e. all other $\alpha$,
were computed as expectation value with the polarized wave function
dividing subsequently by $\efield_z$. The linearity of the polarized
wave function in $\efield_z$ was checked by comparing to results from
second order response theory calculations of all second order
$\alpha$s. These differently achieved results are in agreement within
the numerical precision of the our approach.

\begin{table*}
\begin{threeparttable}
\caption{Nuclear angular momenta and magnetic moments of employed isotopes as reported in
Ref.~\cite{stone:2005}. $\eta_{\mathrm{p}}=\frac{\mu_\mathrm{N}}{A}+\frac{\mu}{Z}$,
$\eta_{\mathrm{n}}=\frac{\mu_\mathrm{N}}{A}+\frac{\mu}{A-Z}$, NMQMs and Schiff moments as estimated following
eqs. (\ref{eq: nucstruc_start})-(\ref{eq: nucstruc_end}).}
\label{tab: nucstruc}
\begin{tabular}{
l
S[table-format = 1.1]
S[table-format = -1.6]
S[table-format = -1.5, round-mode=places, round-precision=5]
S[table-format = 1.5, round-mode=places, round-precision=5]
S[table-format = 1.5, round-mode=places, round-precision=5]
S[table-format = -1.5, round-mode=places, round-precision=5]
S[table-format = -1.5, round-mode=places, round-precision=5]
}
\toprule
    Isotope &    {$I_A$} & {$\mu_A/\mu_\mathrm{N}$} & {$\eta_A/\mu_\mathrm{N}$} & {$R_{\mathrm{vol},A}/(\mathrm{fm}^2)$} &  {$\tilde{\mathcal{S}}_A/(e\,\mathrm{fm}^3)$} &  {$\tilde{\mathcal{M}}_{\mathrm{EDM},A}/(c\,\mathrm{fm})$} &  {$\tilde{\mathcal{M}}_{\pi,A}/(c\,e\,\mathrm{fm}^2)$} \\
\midrule
 \ce{43Ca} &        3.5 &   -1.31764 &       -0.034033  &        0.785523 &         0.181989 &        -0.070100 &        -0.001105 \\
 \ce{87Sr} &        4.5 &    -1.0936 &       -0.010824  &        1.233741 &         0.543079 &        -0.057355 &        -0.000750 \\
  \ce{89Y} &        0.5 &    -0.1374 &        0.007713  &        1.530927 &         0.709942 &              {-} &              {-} \\
\ce{111Cd} &        0.5 &  -0.594886 &       -0.000434  &        1.773821 &         0.986294 &              {-} &              {-} \\
\ce{129Xe} &        0.5 &   -0.77797 &       -0.002621  &        1.960745 &         1.226507 &              {-} &              {-} \\
\ce{133Cs} &        3.5 &  +2.582025 &        0.054465  &        1.667558 &         1.082373 &        -0.070100 &         0.001326 \\
\ce{137Ba} &        1.5 &   +0.93737 &        0.018872  &        1.836894 &         1.191591 &        -0.126180 &         0.001415 \\
\ce{173Yb} &        2.5 &     -0.648 &       -0.000511  &        2.043832 &         1.657290 &        -0.090129 &        -0.000698 \\
 \ce{183W} &        0.5 &      +0.52 &        0.010235  &        2.475494 &         2.122017 &              {-} &              {-} \\
\ce{199Hg} &        0.5 &  +0.505885 &        0.009276  &        2.617761 &         2.425913 &              {-} &              {-} \\
\ce{205Tl} &        0.5 &  +1.638214 &        0.025103  &        2.670118 &         2.537301 &              {-} &              {-} \\
\ce{207Pb} &        0.5 &   +0.59258 &        0.009572  &        2.687457 &         2.552763 &              {-} &              {-} \\
\ce{210Fr} &          6 &      +4.40 &        0.055337  &        2.180379 &         2.223466 &        -0.045064 &         0.001832 \\
\ce{223Ra} &        1.5 &     +0.271 &        0.006492  &        2.541794 &         2.591065 &        -0.126180 &         0.000409 \\
\ce{225Ra} &        0.5 &    -0.7338 &       -0.000912  &        2.841076 &         2.896148 &              {-} &              {-} \\
\ce{229Th} &        2.5 &       0.46 &        0.007676  &        2.463985 &         2.568833 &        -0.090129 &         0.000496 \\
\ce{229Pa} &        2.5 &       1.96 &        0.025905  &        2.463985 &         2.626847 &        -0.090129 &         0.001035 \\
\bottomrule
\end{tabular}
\end{threeparttable}  
\end{table*}          

\begin{table*}
\begin{adjustbox}{max width=.9\textwidth}
\begin{threeparttable}
\caption{Electronic structure parameters needed to compute the EDM
enhancement parameters $\alpha_i$ as obtained at different levels of
theory within the ZORA-GHF or ZORA-GKS approach for all discussed
atoms and molecules. Note that magnetic moments and other
nucleus-dependent constants are not included in the electronic
structure parameters presented below. For closed-shell atoms and
molecules $\alpha_\mathrm{d}^\mathrm{m}$ and
$\alpha_\mathrm{s}^\mathrm{m}$ or rather $W_\mathrm{d}^\mathrm{m}$ and
$W_\mathrm{s}^\mathrm{m}$ are shown, whereas for open-shell atoms and
molecules $\alpha_\mathrm{d}$ and $\alpha_\mathrm{s}$ or rather
$W_\mathrm{d}$ and $W_\mathrm{s}$ are shown.}
\label{tab: elecstruc}
\begin{tabular}{
l
l
S[table-format = -3.4, round-mode=figures, round-precision=3] 
S[table-format = -2.4, round-mode=figures, round-precision=3]
S[table-format = -1.4, round-mode=figures, round-precision=3]
S[table-format = -3.4, round-mode=figures, round-precision=3]
S[table-format = -1.4, round-mode=figures, round-precision=3]
S[table-format = -1.4, round-mode=figures, round-precision=3]
S[table-format = -2.4, round-mode=figures, round-precision=3]
}
\toprule
\multicolumn{2}{l}{Atoms}\\
\toprule
         System &          Method & 
{$\alpha_{\mathrm{d}}$ or} &%
{$\alpha_{\mathrm{s}}\times10^{20}/(e\,\mathrm{cm})$ or} &
{$\alpha_\mathrm{T}\times10^{20}/(e\,\mathrm{cm})$} &
{$\alpha_\mathrm{p}\times10^{23}/(e\,\mathrm{cm})$} &
{$\alpha_\mathrm{m}\times10^{6}\,\mu_\mathrm{N}$}    & 
{$\alpha_\mathcal{S}\times10^{17}/(\mathrm{cm}/\mathrm{fm}^3)$} & 
{$\alpha_\mathcal{M}\times10^{-9}/(c\,e\,\mathrm{cm}^2)$}                 \\
&&
{$\alpha_{\mathrm{d}}^{\mathrm{m}}\,\mu_{\mathrm{N}}$} & 
{$\alpha_{\mathrm{s}}^{\mathrm{m}}\times10^{20}\,\mu_{\mathrm{N}}/(e\,\mathrm{cm})$} &
\\
\midrule
\multicolumn{3}{l}{Former experiments}\\
\midrule
\ce{^{129}Xe},$^1S_0$     &              HF &        0.001104 &        0.004915 &        0.568415 &        1.714637 &     -808.472000 &        0.376277\\
                          &          BHandH &        0.001047 &        0.004712 &        0.544197 &        1.643256 &     -784.934380 &        0.362772\\
                          &             LDA &        0.000940 &        0.004375 &        0.508217 &        1.535840 &     -749.376070 &        0.343963\\
\\
\ce{^{171}Yb},$^1S_0$     &              HF &        0.168661 &        0.027485 &       -3.408783 &      -11.861518 &     3073.828200 &       -1.884429 &       \\
                          &          BHandH &        0.001462 &       -0.030443 &       -3.102077 &      -10.789578 &     2768.625800 &       -1.700515 &       \\
                          &             LDA &       -0.003106 &       -0.026691 &       -2.560286 &       -8.902710 &     2257.169600 &       -1.396565 &       \\
\\
\ce{^{133}Cs},$^2S_{1/2}$ &              HF &      294.240120 &       77.326014 &       -2.271936 &       -6.877596 &     3023.245700 &       -1.405594 &       37.990704\\
                          &          BHandH &      235.162840 &       61.805850 &       -1.626695 &       -4.921558 &     2154.609800 &       -0.998814 &       30.116696\\
                          &             LDA &      190.588768 &       50.112892 &       -1.170474 &       -3.538644 &     1538.914700 &       -0.711310 &       24.128638\\
\\
\ce{^{199}Hg},$^1S_0$     &              HF &        0.021420 &       -0.056018 &       -5.999778 &      -22.380208 &     3461.903900 &       -2.903644\\
                          &          BHandH &       -0.002527 &       -0.053465 &       -4.712058 &      -17.574523 &     2670.858800 &       -2.286300\\
                          &             LDA &       -0.006156 &       -0.043227 &       -3.599285 &      -13.425104 &     1998.382600 &       -1.747121\\
\\
\ce{^{205}Tl},$^2P_{1/2}$ &              HF &    -1636.509800 &     -790.341740 &        1.837818 &        6.334505 &     2493.725100 &       -0.594584 \\
                          &          BHandH &    -1281.007220 &     -614.775440 &        2.154878 &        7.528126 &     2244.681000 &       -0.419999 \\
                          &             LDA &    -1023.739120 &     -488.089080 &        2.522158 &        8.913764 &     1937.055900 &       -0.210515 \\
\\
\ce{^{225}Ra},$^1S_0$     &              HF &       -0.872692 &       -0.691544 &      -17.021675 &      -66.611637 &     8077.723800 &       -8.004431\\
                          &          BHandH &        0.036656 &       -0.151966 &      -14.539576 &      -56.878190 &     6818.109600 &       -6.775857\\
                          &             LDA &       -0.019364 &       -0.157423 &      -12.300873 &      -48.102418 &     5696.314400 &       -5.677246\\
\midrule
\multicolumn{3}{l}{Planned experiments} \\                                                                                 
\midrule
\ce{^{210}Fr},$^2S_{1/2}$ &              HF &     2113.033800 &     1155.132300 &      -20.444977 &      -79.570318 &    10718.414000 &      -10.047622 &      149.780156\\
                          &          BHandH &     1686.383020 &      921.850460 &      -14.432925 &      -56.119306 &     7343.994600 &       -6.946928 &      115.656526\\
                          &             LDA &     1361.543460 &      744.324360 &      -10.318455 &      -40.072529 &     5037.999700 &       -4.835116 &       90.265538\\
\toprule
\multicolumn{2}{l}{Molecules}\\
\toprule
           System &          Method & 
{$W_\mathrm{d}/(\mathrm{GV}/\mathrm{cm})$ or} &
{$W_\mathrm{s}/(\mathrm{peV})$ or} &
{$W_\mathrm{T}/(\mathrm{peV})$} & 
{$W_\mathrm{p}/(\mathrm{feV})$} & 
{$W_\mathrm{m}/(\mathrm{kV}/(\mathrm{cm}\,\mu_\mathrm{N}))$} & 
{$W_\mathcal{S}/(\mathrm{nV}/\mathrm{fm}^3)$} & 
{$W_\mathcal{M}/(\mathrm{EV}/(c\,\mathrm{cm}^2))$}               \\
&&
{$W^\mathrm{m}_\mathrm{d}/(\mathrm{GV}/(\mathrm{cm}\,\mu_\mathrm{N}))$} & 
{$W^\mathrm{m}_\mathrm{s}/(\mathrm{peV}/(\mu_\mathrm{N}))$} & 
\\
\midrule
\multicolumn{3}{l}{Former experiments}\\
\midrule
\ce{^{174}YbF},$^2\Sigma_{1/2}$ &              HF &      -46.671160 &     -168.861934 &\\
                                &          BHandH &      -46.403176 &     -167.686330 &\\
\\                                                                                                                                                                               
\ce{^{180}HfF+},$^3\Delta_1$    &              HF &       26.320507 &       96.606401 &\\
                                &          BHandH &       23.435892 &       85.706620 &\\
\\                                                                                                                                                                               
\ce{^{205}TlF},$^1\Sigma_{0+}$  &              HF &        0.008375 &        0.246848 &       19.411119 &       72.245704 &     -776.542110 &        8.420163 &\\
                                &          BHandH &        0.006647 &        0.208375 &       16.339997 &       60.764386 &     -602.326730 &        6.938983 &\\
                                &             LDA &        0.005083 &        0.162894 &       12.797609 &       47.555343 &     -430.552110 &        5.328858 &\\
\\                                                                                                                                                                               
\ce{^{207}PbO},$^3\Sigma_{1}$   &              HF &      -30.640182 &     -147.172590 &       18.294474 &       69.018421 &     -988.651540 &        8.959367 &\\
                                &          BHandH &      -32.939649 &     -157.743563 &       14.713966 &       55.530511 &     -792.238960 &        7.252549 &\\
\\                                                                                                                                                                               
\ce{^{232}ThO},$^3\Delta_1$     &              HF &       99.771425 &      583.288360 &\\
                                &          BHandH &       82.257609 &      480.305440 &\\
\midrule
\multicolumn{3}{l}{Planned experiments} \\                                                                                 
\midrule
\ce{^{137}BaF},$^2\Sigma_{1/2}$ &              HF &      -13.400990 &      -35.582246 &       -0.753871 &       -2.307369 &       68.017830 &       -0.484274 &       -1.666784\\
                                &          BHandH &      -12.463377 &      -33.040046 &       -0.783702 &       -2.398125 &       69.769000 &       -0.501740 &       -1.539055\\
                                &             LDA &      -10.889079 &      -28.837532 &       -0.770971 &       -2.358616 &       67.329180 &       -0.491920 &       -1.341029\\
\\                                                                                                                                                                               
\ce{^{173}YbOH},$^2\Sigma_{1/2}$&              HF &      -46.577410 &     -168.593642 &       -1.706538 &       -5.988553 &      162.508480 &       -1.130909 &       -4.401125\\
                                &          BHandH &      -46.809408 &     -169.201246 &       -1.939882 &       -6.804555 &      185.670160 &       -1.277310 &       -4.373989\\
\\                                                                                                                                                                               
\ce{^{183}WC},$^3\Delta_1$      &              HF &       68.852554 &      294.483424 &       -1.524177 &       -5.518596 &      135.589200 &       -1.000856 &\\
                                &          BHandH &       45.806440 &      185.007770 &       -3.832282 &      -13.780808 &       92.881996 &       -2.239818 &\\
                                &             LDA &       34.539075 &      136.894010 &       -4.226063 &      -15.198578 &      165.252080 &       -2.477986 &\\
\\                                                                                                                                                                               
\ce{^{223/225/226}RaF},$^2\Sigma_{1/2}$ &      HF &     -111.387900 &     -621.167080 &       -7.562325 &      -29.835445 &      355.584380 &       -4.184940 &       -7.243803\\
                                &          BHandH &     -105.181248 &     -585.993420 &       -7.352359 &      -29.006143 &      337.688900 &       -4.060160 &       -6.688770\\
                                &             LDA &      -93.968522 &     -522.971260 &       -6.684292 &      -26.372397 &      297.229500 &       -3.689860 &       -5.893455\\
\\                                                                                                                                                                               
\ce{^{229}ThF+},$^3\Delta_1$    &              HF &       43.547392 &      247.786470 &      -17.436949 &      -69.078267 &      602.722260 &       -8.368105 &        2.802180\\
                                &          BHandH &       34.696082 &      197.064420 &      -15.223841 &      -60.331129 &      527.092830 &       -7.348072 &        2.508218\\
\\                                                                                                                                                                               
\ce{^{229}PaF^3+},$^2\Phi_{5/2}$&              HF &        2.730413 &       17.310291 &      -27.844047 &     -111.711380 &      937.269580 &      -13.157428 &        0.159535\\
                                &          BHandH &        2.134739 &       13.408145 &      -24.260261 &      -97.358964 &      799.315930 &      -11.514145 &        0.121674\\
\midrule
\multicolumn{3}{l}{Proposed experiments (this work)}\\
\midrule
\ce{^{43}CaOH},$^2\Sigma_{1/2}$ &              HF &       -0.579641 &       -0.623462 &       -0.026911 &       -0.047587 &        6.608102 &       -0.024937 &       -0.223772\\
                                &          BHandH &       -0.568511 &       -0.608963 &       -0.030266 &       -0.053511 &        7.500267 &       -0.028020 &       -0.218846\\
                                &             LDA &       -0.518464 &       -0.553158 &       -0.034164 &       -0.060386 &        8.553227 &       -0.031581 &       -0.199318\\
\\                                                                                                                                                                                
\ce{^{87}SrOH},$^2\Sigma_{1/2}$ &              HF &       -4.215786 &       -7.996744 &       -0.203346 &       -0.500908 &       30.023679 &       -0.144937 &       -0.827103\\
                                &          BHandH &       -4.166802 &       -7.884615 &       -0.220512 &       -0.543164 &       32.447248 &       -0.157067 &       -0.813899\\
                                &             LDA &       -3.913506 &       -7.388063 &       -0.234284 &       -0.576988 &       34.402588 &       -0.166579 &       -0.762061\\
\\                                                                                                                                                                                
\ce{^{89}YO},$^2\Sigma_{1/2}$   &              HF &       -6.376857 &      -12.239485 &       -0.392262 &       -0.979376 &       35.727519 &       -0.269996 & \\
                                &          BHandH &       -5.790409 &      -11.111820 &       -0.369088 &       -0.921849 &       34.641339 &       -0.254918 & \\
                                &             LDA &       -4.987479 &       -9.566226 &       -0.345885 &       -0.864316 &       35.292669 &       -0.240016 & \\
\\                                                                                                                                                                                
\ce{^{111}CdH},$^2\Sigma_{1/2}$ &              HF &      -25.950490 &      -62.418926 &       -0.364571 &       -1.035064 &       81.584417 &       -0.252950 &       -3.639916\\
                                &          BHandH &      -25.302468 &      -60.680260 &       -0.352703 &       -1.002552 &       77.680683 &       -0.247863 &       -3.471097\\
                                &             LDA &      -22.211844 &      -53.113084 &       -0.287262 &       -0.818899 &       64.776662 &       -0.208255 &       -3.006024\\
\bottomrule

\end{tabular}
\end{threeparttable}
\end{adjustbox}
\end{table*}

\section{Accuracy of the ZORA-cGKS-BHandH approach}
Where available we compare our results with \emph{ab initio} results
from literature in Table \ref{tab: litcompare}. Here we chose recent
results we consider as up-to-date reference. Some of the methods used
in literature are less sophisticated than our present approach, others
are more sophisticated. Hence, it is not expected that agreement with
literature gauges the quality of our approach in all cases. For
judging on the quality of our method we take into account comparisons
to literature results that treat electron correlation on a high level,
i.e.\ Coupled Cluster (CC) or Configuration Interaction (CI)
calculations with large basis sets, only.

For all atomic or molecular enhancement factors $\alpha$ or $W$ we compared
our values computed at the level of ZORA-cGKS-BHandH with calculations
at the levels of ZORA-cGKS-LDA and ZORA-cGHF to assess the importance
of electron correlation. Furthermore, for cases where $\alpha$ or $W$ was
computed in literature with a different method we compared our methods
to the literature values. We note, however, that in some cases the
method applied in this work is expected to give more reliable results
than the respective literature approach. In general we see an
agreement of the ZORA-GKS-BHandH method with more sophisticated
coupled cluster or configuration interaction methods within $15~\%$ or
better. 

We will discuss now those cases in the following, which show larger
deviations. Considerably larger deviations are seen for
$\alpha_\mathrm{d}$ for Hg, however we compare here to less
sophisticated DHF-RPA values and we can see that HF gives even the
opposite sign in comparison to DFT. $\alpha_\mathrm{d}$ is a third
order property in Hg and the opposite sign in HF is obtained only
at the level of RPA but not at the level of uncoupled HF. This alteration
of sign seems to be an artifact as a similar change due to the response
effect is not observed on the level of DFT. We therefore consider the
ZORA-GKS results to be more reliable. We also consider our
ZORA-GKS-BHandH results more reliable than DHF results for
$W_\mathrm{T}$ for TlF and the DHF-RPA results for $\alpha_\mathrm{p}$
for Hg (note here the excellent agreement with sophisticated
MRCISD calculations of $\alpha_\mathrm{T}$). 

We see deviations on the order of $\sim20~\%$ for PbO and WC for
$W_\mathrm{d}$. This can have several reasons. We have to note
that in the literature the equilibrium distance was not obtained on a
correlated level of theory in both cases (see
Refs.~\cite{petrov:2005,lee:2013}) and the influence of the bond
length is quite large (see also Table I in Ref.~\cite{petrov:2005} for
PbO).  Concerning this point one has to consider that in the case of
WC, large deviations of HF from DFT have their origin in a considerably
different equilibrium bond length at the level of HF than on the DFT
level of theory (see Table \ref{tab: bond_length}). As we see very good
agreement of ZORA-GKS-BHandH results for molecules that have a similar
electronic structure as WC, i.e.\ ThO, \ce{ThF+} and \ce{HfF+} for
$W_\mathrm{d}$ in comparison to sophisticated CC calculations, we
consider our ZORA-GKS-BHandH data to be more reliable than the
corresponding GRECP-SODCI results \cite{lee:2013}. In case of PbO, the considered meta-stable
electronic state may be more difficult to describe in our DFT
framework and therefore this may contribute to the large
deviations.

A large deviation is found for the Schiff moment enhancement in Ra by
comparison to CI+MBPT values from Ref.~\cite{dzuba:2009a}. In their
calculations the absolute value of the Schiff moment enhancement is
increased by correlation effects, which contradicts the trend of
correlation effects that is seen for Schiff enhancements in other
system and also our ZORA-GKS-BHandH predictions. We, therefore,
consider our results to be more reliable. 

The deviation for $\alpha_\mathrm{d}^\mathrm{m}$ in Xe from RCCSD
calculations of Ref.~\cite{sahoo:2023} could be explained by a
different basis set, as our results on the coupled HF level deviate
also significantly from results of Ref.~\cite{sahoo:2023} on the same
level of theory (RPA) which are given as
$\alpha_\mathrm{d}^\mathrm{m}\approx0.00145/\mu_\mathrm{N}$. Here we
note that our HF values are in excellent agreement with previous calculations
of Ref.~\cite{martensson-pendrill:1987} which also reports the same
sign. We needed to convert the signs of Ref.~\cite{sahoo:2023}, which
were not consistent with the definition in our paper and
Ref.~\cite{martensson-pendrill:1987}. Unfortunately,
Ref.~\cite{sahoo:2023} does not report exact data on the composition
of there basis functions.

For Yb, we find a strangely large deviation to Ref.~\cite{sahoo:2017}
for $\alpha_\mathrm{T}$, which cannot be expected as other parameters
are in much better agreement with the same method. In
Ref.~\cite{sahoo:2017} a huge correlation effect is reported that
explains this deviation, which however, contradicts results of other
works (e.g. Ref.~\cite{dzuba:2009a}). 

A critical and somewhat unexpected large deviation is found for Hg for
$\alpha_\mathrm{s}$ to the MRCI calculations by Fleig
\cite{fleig:2018}. In contradiction an excellent agreement
is found for $\alpha_\mathrm{T}$ by comparison to MRCI calculations.
Furthermore, we note that the value of $\alpha_\mathrm{s}$
computed for Xe in Ref.~\cite{fleig:2021} with the same methods
deviates strongly from our calculations as well and disagrees with results from
Ref.~\cite{sahoo:2023}. In Xe large correlation contributions
are not to be expected, but in Ref. \cite{fleig:2021} it was
surprisingly found that the MRCI results presented therein show a huge
deviation from phenomenological predictions and even the wrong sign in
case of Xe. In contrast our results are in reasonable agreement with
phenomenology. As a more detailed discussion is beyond the scope of our
present paper, we leave a detailed comparison of DFT results and
phenomenology for a future study of $\alpha_\mathrm{d}$ and
$\alpha_\mathrm{s}$ in closed-shell systems.

Finally, there is a surprisingly large deviation between our result and CC
results reported in Ref.~\cite{skripnikov:2015} for $W_\mathcal{M}$
for \ce{ThF+}. As all other properties for \ce{ThF+} as well as
$W_\mathcal{M}$ for other molecules agree much better with corresponding CC results,
we cannot explain the origin of this discrepancy.

\begin{table*}
\begin{adjustbox}{max width=.8\textwidth}
\begin{threeparttable}
\caption{Comparison of electronic structure parameters computed at the
levels of ZORA-GHF or ZORA-GKS to available \emph{ab initio} data from
the literature. Where necessary literature values were converted to
the units and sign conventions which are used in this work.}
\label{tab: litcompare}
\begin{tabular}{
l
l
l
S[table-format = -3.4, round-mode=figures,round-precision=3,
scientific-notation = fixed, fixed-exponent = 0]
S[table-format = -3.1, round-mode=places, round-precision=1]
S[table-format = -3.1, round-mode=places, round-precision=1]
S[table-format = -3.1, round-mode=places, round-precision=1]
}
\toprule
         System &          Method &  & {Value} & {Dev. HF/\%} & {Dev. BHandH/\%} &  {Dev. LDA/\%}                \\
\midrule
\multicolumn{5}{l}{$\alpha_\mathrm{d}$ (open-shell) or $\alpha_\mathrm{d}^\mathrm{m}\mu_\mathrm{N}$ (closed-shell) (Atoms)}\\
\midrule
        \ce{Xe}&                     RCCSD+corr&              \cite{sahoo:2023}&    1.290000e-03&      -16.843931&     -23.162029&     -37.202975\\
        \ce{Cs}&                          RCCSD&            \cite{nataraj:2008}&    2.410600e+02&       18.073715&      -2.507692&     -26.481745\\
        \ce{Hg}&                        DHF-RPA& \cite{martensson-pendrill:1987}&    2.310000e-02&       -7.840677&    1014.026472&     475.250035\\
        \ce{Tl}&                       RCCSD(T)&              \cite{fleig:2020}&   -1.116000e+03&       31.806091&      12.881053&      -9.012148\\
        \ce{Fr}&                          RCCSD&            \cite{shitara:2021}&    1.624380e+03&       23.125697&       3.676687&     -19.304308\\
\midrule
\multicolumn{5}{l}{{$W_\mathrm{d}/(\mathrm{GV}/\mathrm{cm})$} (open-shell) or $W_\mathrm{d}^\mathrm{m}/(\mathrm{GV}/(\mathrm{cm}\,\mu_\mathrm{N}))$ (closed-shell) (Molecules)}\\
\midrule
       \ce{YbF}&                          RCCSD&                \cite{abe:2014}&   -4.620000e+01&        1.009531&       0.437849&\\
      \ce{HfF+}&                         MRCISD&              \cite{fleig:2017}&    2.270000e+01&       13.755461&       3.140023&\\
       \ce{PbO}&           GRECP-RASSCF+CI+CCSD&             \cite{petrov:2005}&   -2.500000e+01&       18.407795&      24.103623&\\
       \ce{ThO}&                         MRCISD&              \cite{denis:2016}&    7.520000e+01&       24.627718&       8.579886&\\
       \ce{BaF}&                       RCCSD(T)&              \cite{haase:2021}&   -1.290000e+01&        3.738457&      -3.503248&     -18.467322\\
      \ce{YbOH}&                           FSCC&              \cite{denis:2019}&   -4.674000e+01&       -0.349075&       0.148278&\\
        \ce{WC}&                    GRECP-SODCI&                \cite{lee:2013}&    3.600000e+01&       47.714358&      21.408431&      -4.229775\\
       \ce{RaF}&              GRECP-FSCCSD+corr&           \cite{kudashov:2014}&   -1.058000e+02&        5.016613&      -0.588272&     -12.590895\\
      \ce{ThF+}&                         MRCISD&              \cite{denis:2015}&    3.520000e+01&       19.168523&      -1.452377&\\
       \ce{CdH}&                          RCCSD&           \cite{talukdar:2020}&   -2.382000e+01&        8.209826&       5.858986&      -7.240083\\
\midrule
\multicolumn{5}{l}{{$\alpha_\mathrm{s}\times10^{20}/(e\,\mathrm{cm})$} (open-shell) or $\alpha_\mathrm{s}^\mathrm{m}\times10^{20}\,\mu_\mathrm{N}/(e\,\mathrm{cm})$ (closed-shell) (Atoms)}\\
\midrule
        \ce{Xe}&                     RCCSD+corr&              \cite{sahoo:2023}&    5.000000e-03&       -1.721239&      -6.105668&     -14.279821\\
        \ce{Cs}&                          RCCSD&              \cite{sahoo:2008}&    6.620000e+01&       14.388449&      -7.109602&     -32.101735\\
        \ce{Hg}&                   MRCISDT+corr&              \cite{fleig:2018}&   -2.224560e-02&       60.288796&      58.392131&      48.538159\\
        \ce{Tl}&                       RCCSD(T)&              \cite{fleig:2020}&   -5.340000e+02&       32.434291&      13.139015&      -9.406258\\
        \ce{Fr}&                       RCCSD(T)&         \cite{skripnikov:2017}&    9.020000e+02&       21.913706&       2.153328&     -21.183727\\
\midrule
\multicolumn{5}{l}{{$W_\mathrm{s}/\mathrm{peV}$} (open-shell) or $W_\mathrm{s}^\mathrm{m}/(\mathrm{peV}/\mu_\mathrm{N})$ (closed-shell) (Molecules)}\\
\midrule
       \ce{YbF}&                   GRECP-RASSCF&              \cite{titov:1996}&   -1.360000e+02&       19.460830&      18.896191&\\
      \ce{HfF+}&                    MRCISD+corr&              \cite{fleig:2017}&    8.270000e+01&       14.394906&       3.508037&\\
       \ce{ThO}&                    MRCISD+corr&              \cite{denis:2016}&    4.380000e+02&       24.908496&       8.808028&\\
       \ce{BaF}&                       RCCSD(T)&              \cite{haase:2021}&   -3.428000e+01&        3.659820&      -3.752882&     -18.872863\\
      \ce{YbOH}&                      ZORA-cGHF&              \cite{gaul:2020a}&   -1.700000e+02&       -0.834170&      -0.472073&\\
       \ce{RaF}&              GRECP-FSCCSD+corr&           \cite{kudashov:2014}&   -5.740000e+02&        7.593300&       2.046682&      -9.757465\\
      \ce{ThF+}&                         MRCISD&              \cite{denis:2015}&    2.000000e+02&       19.285343&      -1.489655&\\
       \ce{CdH}&                          RCCSD&           \cite{talukdar:2020}&   -5.496000e+01&       11.949783&       9.426888&      -3.477328\\
\midrule
\multicolumn{2}{l}{{$\alpha_\mathrm{T}\times10^{20}/(e\,\mathrm{cm})$} (Atoms)}\\
\midrule
        \ce{Xe}&                    MRCISD+corr&              \cite{fleig:2021}&    5.490000e-01&        3.415582&      -0.882642&      -8.024711\\
        \ce{Yb}&                          RCCSD&              \cite{sahoo:2017}&   -2.040000e+00&       40.154593&      34.237607&      20.321392\\
        \ce{Hg}&                    MRCISD+corr& \cite{fleig:2019,fleig:privatecomm}&   -4.970000e+00&       17.163598&      -5.474084&     -38.082973\\
        \ce{Ra}&                    MRCISD+corr& \cite{fleig:2019,fleig:privatecomm}&   -1.620000e+01&        4.827228&     -11.420030&     -31.697970\\
\midrule
\multicolumn{2}{l}{{$W_\mathrm{T}/(\mathrm{peV})$} (Molecules)}\\
\midrule
       \ce{TlF}&                            DHF&             \cite{quiney:1998}&    1.919000e+01&        1.139136&     -17.441882&     -49.949885\\
\midrule
\multicolumn{2}{l}{{$\alpha_\mathrm{p}\times10^{23}/(e\,\mathrm{cm})$} (Atoms)}\\
\midrule
        \ce{Xe}&                     RCCSD+corr&              \cite{sahoo:2023}&    1.442000e+00&       15.900542&      12.247392&       6.110036\\
        \ce{Yb}&                        CI+MBPT&             \cite{dzuba:2009a}&   -1.150000e+01&        3.047822&      -6.584335&     -29.174156\\
        \ce{Hg}&                        CI+MBPT&             \cite{dzuba:2009a}&   -1.840000e+01&       17.784500&      -4.697009&     -37.056666\\
        \ce{Ra}&                        CI+MBPT&             \cite{dzuba:2009a}&   -6.420000e+01&        3.620444&     -12.872790&     -33.465224\\
\midrule
\multicolumn{2}{l}{{$W_\mathrm{m}/(\mathrm{kV}/(\mathrm{cm}\,\mu_\mathrm{N}))$} (Molecules)}\\
\midrule
       \ce{TlF}&                     GRECP-CCSD&             \cite{petrov:2002}&   -6.647000e+02&       14.402581&     -10.355388&     -54.383171\\
\midrule
\multicolumn{2}{l}{{$\alpha_\mathcal{S}\times10^{17}/(\mathrm{cm}/\mathrm{fm}^3)$} (Atoms)}\\
\midrule
        \ce{Xe}&                    MRCISD+corr&             \cite{hubert:2022}&    3.640000e-01&        3.262833&      -0.338622&      -5.825254\\
        \ce{Yb}&                          RCCSD&              \cite{sahoo:2017}&   -1.510000e+00&       19.869646&      11.203385&      -8.122393\\
        \ce{Hg}&                    MRCISD+corr&             \cite{hubert:2022}&   -2.400000e+00&       17.345233&      -4.973083&     -37.368862\\
        \ce{Ra}&                        CI+MBPT&             \cite{dzuba:2009a}&   -8.840000e+00&      -10.438828&     -30.463207&     -55.709307\\
\midrule
\multicolumn{2}{l}{{$W_\mathcal{S}/(\mathrm{nV}/\mathrm{fm}^3)$} (Molecules)}\\
\midrule
       \ce{TlF}&                  GRECP-CCSD(T)&         \cite{skripnikov:2020}&    6.830000e+00&       18.885183&       1.570588&     -28.170063\\
       \ce{RaF}&              GRECP-FSCCSD+corr&           \cite{kudashov:2014}&   -4.046000e+00&        3.320002&       0.348764&      -9.651843\\
\midrule
\multicolumn{2}{l}{{$W_\mathcal{M}/(\mathrm{EV}/(c\,\mathrm{cm}^2))$} (Molecules)}\\
\midrule
       \ce{BaF}&                        CCSD(T)&              \cite{denis:2020}&   -1.590000e+00&        4.606687&      -3.310132&     -18.565711\\
      \ce{YbOH}&                         FSCCSD&             \cite{maison:2019}&   -4.440000e+00&       -0.883306&      -1.509176&\\
      \ce{ThF+}&          GRECP-2c-CCSD(T)+corr&         \cite{skripnikov:2015}&    3.640000e+00&      -29.898860&     -45.122963&\\
       \ce{CdH}&                          RCCSD&           \cite{talukdar:2020}&   -3.350000e+00&        7.964904&       3.488713&     -11.442875\\
\bottomrule
\end{tabular}
\end{threeparttable}
\end{adjustbox}
\end{table*}          

\pagebreak

\section{Numerical results of the global minimization}
The full ranking of choices of systems from various classes is shown
in \prettyref{tab: results_global_min_100} for the range $20\le Z_i\le
\mathbf{100}$ and in \prettyref{tab: results_global_min_90} for the range $20\le Z_i\le
\mathbf{90}$ alongside the minimal and maximal normalized components of the Hessian and
the normalized length of the gradient as measures of the quality of
the achieved minimum.  
\begin{table*}[h!]
\begin{threeparttable}
\caption{
\scriptsize
Results of a global minimization of the volume in the
$\mathcal{P,T}$-odd parameter space with respect to nuclear charges $Z_i$ for
different classes of molecules (I to IV) in the range of $20\le Z_i\le
\mathbf{100}$. For each combination of molecules of each class the minimal
volume is presented alongside the relative gradient and the minimal
($h_\mathrm{min}$) and maximal ($h_\mathrm{max}$) eigenvalues of the
Hessian relative to the absolute minimal volume. Results are given in
order of increasing volume. The number of shown figures is given for
comparability and does not represent the accuracy of the optimization.}
\label{tab: results_global_min_100}
\scriptsize
\begin{tabular}{
SSSSS
S[table-format=+1.1e-1,round-mode=places, scientific-notation=true, round-precision =1]
S[table-format=+1.1e-1,round-mode=places, scientific-notation=true, round-precision =1]
S[table-format=+1.1e-1,round-mode=places, scientific-notation=true, round-precision =1]
S[table-format=+1.1e-1,round-mode=places, scientific-notation=true, round-precision =1]
m{1.2cm}
}
\toprule
{Place} &
{$N_\mathrm{I}$} &
{$N_\mathrm{II}$} &
{$N_\mathrm{III}$} &
{$N_\mathrm{IV}$} &
{$V_\mathrm{min}/\mathrm{arb. u.}$} &
{$\frac{\left|\nabla_Z V_\mathrm{min}\right|}{V_\mathrm{min}}$} &
{$\frac{h_\mathrm{min}}{V_\mathrm{min}}$} &
{$\frac{h_\mathrm{max}}{V_\mathrm{min}}$} &
{Algorithm Index\tnote{a}}
\\
\midrule
1   &   2   &   0   &   1   &   3   &   0.000013783032882965926   &   0.2820327495596826   &   -0.001047803417349651   &   0.08658716881468373   &   5,$\mathbb{R}$   \\
2   &   1   &   1   &   1   &   3   &   0.000014294149157909987   &   0.2747586625811039   &   -0.000519019656638849   &   0.0831652696627106   &   5,$\mathbb{R}$   \\
3   &   2   &   1   &   0   &   3   &   0.000014294574917925613   &   0.27781565682122156   &   -0.0008382034663234195   &   0.08483839352991873   &   2,$\mathbb{R}$   \\
4   &   2   &   0   &   2   &   2   &   0.000021610504777603114   &   0.2858020427556108   &   -0.0010114083893637018   &   0.08920278033484891   &   1,$\mathbb{R}$   \\
5   &   2   &   1   &   1   &   2   &   0.000021851003312640663   &   0.2863651390367818   &   -0.001013295817519292   &   0.08963090579311837   &   2,$\mathbb{R}$   \\
6   &   1   &   1   &   2   &   2   &   0.000022445351297245985   &   0.27902093995221106   &   0.0001851278922905851   &   0.0857562984587676   &   5,$\mathbb{R}$   \\
7   &   1   &   2   &   1   &   2   &   0.000022695111573572716   &   0.2795914546624625   &   -0.0011100548678966982   &   0.08619335956855072   &   4,$\mathbb{R}$   \\
8   &   2   &   2   &   0   &   2   &   0.00002270119076659904   &   0.2819189507117887   &   -0.0008188711160499329   &   0.08734492508635468   &   1,$\mathbb{R}$   \\
9   &   1   &   0   &   2   &   3   &   0.000026401068125429346   &   0.2824126805875663   &   -0.0020656414281092425   &   0.08737893837194868   &   7,$\mathbb{R}$   \\
10   &   0   &   2   &   2   &   2   &   0.000027162638618980345   &   0.29529683195664963   &   0.00016690086503709593   &   0.09955010613371883   &   4,$\mathbb{R}$   \\
11   &   0   &   1   &   2   &   3   &   0.00002739535794291108   &   0.28194066701280135   &   0.00026673975221158186   &   0.08876330618126074   &   4,$\mathbb{R}$   \\
12   &   1   &   2   &   0   &   3   &   0.00002774201856945473   &   0.27399573419741247   &   -0.002088190776173778   &   0.08225369239568699   &   2,$\mathbb{R}$   \\
13   &   0   &   2   &   1   &   3   &   0.000027745331544448853   &   0.27570390455135235   &   -0.0021673295599151337   &   0.0835662464928426   &   1,$\mathbb{R}$   \\
14   &   2   &   1   &   2   &   1   &   0.000048892051582720854   &   0.27887107347997037   &   -0.004306670451906377   &   0.08552726843585826   &   2,$\mathbb{R}$   \\
15   &   1   &   2   &   2   &   1   &   0.00004948927057664735   &   0.2737984868812863   &   -0.004250224406237608   &   0.08038599149722955   &   5,$\mathbb{R}$   \\
16   &   2   &   0   &   3   &   1   &   0.000049590651585082864   &   0.28009063135661616   &   -0.004362052949173131   &   0.08651075261715341   &   1,$\mathbb{R}$   \\
17   &   2   &   2   &   1   &   1   &   0.000049670871101296006   &   0.285383401494015   &   -0.0042520889376705786   &   0.09142941692313934   &   5,$\mathbb{R}$   \\
18   &   1   &   3   &   1   &   1   &   0.000051400920707595975   &   0.27012206812513756   &   -0.003993688536866153   &   0.08189910144526458   &   5,$\mathbb{R}$   \\
19   &   0   &   3   &   2   &   1   &   0.000051401131138158687   &   0.27001000100957195   &   -0.003845086865249087   &   0.08183300844109317   &   4,$\mathbb{R}$   \\
20   &   2   &   3   &   0   &   1   &   0.000051401231248461784   &   0.26933938790700984   &   -0.004037370092755092   &   0.08147038802668645   &   1,$\mathbb{R}$   \\
21   &   0   &   3   &   1   &   2   &   0.00005140126753999818   &   0.2718947913983959   &   -0.001627289888438604   &   0.08279189052135642   &   5,$\mathbb{R}$   \\
22   &   1   &   3   &   0   &   2   &   0.000051402202140467376   &   0.2705698611682907   &   -0.0026070473515055906   &   0.08207500876177827   &   7,$\mathbb{R}$   \\
23   &   1   &   1   &   3   &   1   &   0.00005150602526903771   &   0.27270072968824866   &   -0.004454334400884226   &   0.08310518786800472   &   4,$\mathbb{R}$   \\
24   &   0   &   2   &   3   &   1   &   0.00005425729051938649   &   0.26662856541248575   &   -0.004144934735901415   &   0.07765387680640166   &   5,$\mathbb{R}$   \\
25   &   1   &   0   &   3   &   2   &   0.00005630126954828603   &   0.2779610108960406   &   -0.0019503623685970287   &   0.08367747325014348   &   5,$\mathbb{R}$   \\
26   &   0   &   1   &   3   &   2   &   0.00005847849974833174   &   0.2740804970189483   &   0.000014702376387270329   &   0.0819381698168139   &   7,$\mathbb{R}$   \\
27   &   1   &   0   &   1   &   4   &   0.00006959112686504223   &   0.3264459962230112   &   -0.002835357059270785   &   0.12783310407037007   &   7,$\mathbb{R}$   \\
28   &   0   &   1   &   1   &   4   &   0.00007227955455793525   &   0.3214710428499536   &   -0.002602376239975118   &   0.12533426011892185   &   7,$\mathbb{R}$   \\
29   &   1   &   1   &   0   &   4   &   0.00007227977896378045   &   0.32204551899581335   &   -0.0031955273770199045   &   0.12569143756371615   &   7,$\mathbb{R}$   \\
30   &   1   &   0   &   4   &   1   &   0.00026412924335723686   &   0.32447403459325225   &   -0.0034194614905512437   &   0.12676736378583894   &   7,$\mathbb{R}$   \\
31   &   0   &   1   &   4   &   1   &   0.00027433301711520187   &   0.32095936733599606   &   -0.0033866963432060923   &   0.12514788913622388   &   5,$\mathbb{R}$   \\
32   &   2   &   0   &   0   &   4   &   0.00027751626035145643   &   0.345672708364315   &   -0.0006187607229745094   &   0.14715809339318567   &   2,$\mathbb{R}$   \\
33   &   0   &   0   &   2   &   4   &   0.0010744877225479228   &   0.34414547524304384   &   -0.0014275621711812754   &   0.14654170728535468   &   2,$\mathbb{R}$   \\
34   &   0   &   2   &   0   &   4   &   0.0011828518291437556   &   0.3433633665271499   &   0.0011721758592561893   &   0.14622969555082407   &   2,$\mathbb{R}$   \\
35   &   0   &   0   &   3   &   3   &   0.0018016316370185667   &   0.33952953884707976   &   0.0006374374808402853   &   0.14123917962772709   &   1,$\mathbb{R}$   \\
36   &   0   &   3   &   0   &   3   &   0.0018904959763842918   &   0.338620072328635   &   0.0016049789705778123   &   0.14127897638767134   &   2,$\mathbb{R}$   \\
37   &   0   &   0   &   4   &   2   &   0.00402213736528501   &   0.34905032349049137   &   0.001457661683380064   &   0.15057924708051243   &   7,$\mathbb{R}$   \\
38   &   1   &   2   &   3   &   0   &   0.004097445569616115   &   0.3501274860734794   &   0.00016605930401510303   &   0.15143519481392748   &   2,$\mathbb{R}$   \\
39   &   0   &   3   &   3   &   0   &   0.004255737372295548   &   0.34839748173260277   &   0.0016803956917273363   &   0.15093970593084968   &   1,$\mathbb{R}$   \\
40   &   1   &   1   &   4   &   0   &   0.004282105124570236   &   0.35021026997289634   &   -0.0003193015651267073   &   0.15138399041504896   &   1,$\mathbb{R}$   \\
41   &   2   &   3   &   1   &   0   &   0.004322808976481513   &   0.3474125455578672   &   -0.0005310058916433437   &   0.14841590492966852   &   2,$\mathbb{R}$   \\
42   &   1   &   3   &   2   &   0   &   0.004322815052768358   &   0.34116133975390905   &   -0.002888434753044546   &   0.13923789408882525   &   2,$\mathbb{R}$   \\
43   &   2   &   2   &   2   &   0   &   0.004382552607596631   &   0.3456912697855789   &   -0.0005152812294604535   &   0.14677291116980237   &   4,$\mathbb{R}$   \\
44   &   2   &   0   &   4   &   0   &   0.004390563789828242   &   0.3499469816093883   &   -0.0006073200858963715   &   0.15131725319067144   &   1,$\mathbb{R}$   \\
45   &   0   &   2   &   4   &   0   &   0.004447530651732352   &   0.3464310495450607   &   0.0013231357690807871   &   0.1495837331702905   &   7,$\mathbb{R}$   \\
46   &   1   &   4   &   0   &   1   &   0.0044893660205464255   &   0.3362989873107881   &   -0.0003286862444427864   &   0.1426997965999423   &   4,$\mathbb{R}$   \\
47   &   0   &   4   &   1   &   1   &   0.004489519140642252   &   0.33706093934243214   &   -0.0003296353374314848   &   0.14363302369015174   &   4,$\mathbb{R}$   \\
48   &   0   &   4   &   0   &   2   &   0.004489771947160852   &   0.3362974140692512   &   -0.00032894053654028947   &   0.1426985297266642   &   1,$\mathbb{R}$   \\
49   &   1   &   4   &   1   &   0   &   0.004489806927913565   &   0.3362969782222691   &   -0.00032876612771574177   &   0.14269838709999363   &   1,$\mathbb{R}$   \\
50   &   0   &   4   &   2   &   0   &   0.004489813275065254   &   0.33629595283062336   &   -0.00032878293904137993   &   0.14269721173116426   &   1,$\mathbb{R}$   \\
51   &   2   &   4   &   0   &   0   &   0.0044898220117678715   &   0.3362862895726129   &   -0.00032852073716484507   &   0.14268534897991306   &   7,$\mathbb{R}$   \\
52   &   2   &   1   &   3   &   0   &   0.004541871141235704   &   0.34139581612131525   &   -0.0006521890440371445   &   0.13947450310886927   &   5,$\mathbb{R}$   \\
53   &   1   &   0   &   0   &   5   &   0.00851610744289936   &   0.38253821387543   &   -0.0022296440195478493   &   0.19219305853670063   &   2,$\mathbb{R}$   \\
54   &   0   &   0   &   1   &   5   &   0.015719417686018772   &   0.38214708989887713   &   -0.0022476907195255816   &   0.1916907912870743   &   1,$\mathbb{R}$   \\
55   &   0   &   1   &   0   &   5   &   0.01929578205440574   &   0.38002724293309803   &   -0.0004522948488195615   &   0.19121530382887605   &   2,$\mathbb{R}$   \\
56   &   0   &   0   &   5   &   1   &   0.0982279553683136   &   0.378275248780412   &   -0.002242460323986342   &   0.18732150164101624   &   1,$\mathbb{R}$   \\
57   &   1   &   0   &   5   &   0   &   0.1122574031364264   &   0.3785231248809863   &   -0.0022270562505749426   &   0.1877090456595654   &   7,$\mathbb{R}$   \\
58   &   0   &   1   &   5   &   0   &   0.11659411231194118   &   0.3755648336225092   &   -0.0004782583766095274   &   0.18618383355433477   &   1,$\mathbb{R}$   \\
59   &   0   &   0   &   0   &   6   &   115.84522975516072   &   0.3752636783315904   &   0.004677109434802696   &   0.19295236762159032   &   7,$\mathbb{R}$   \\
60   &   0   &   0   &   6   &   0   &   2268.8241381674993   &   0.36396601605168394   &   0.005012821577463368   &   0.18171993100785638   &   7,$\mathbb{R}$   \\
61   &   0   &   5   &   0   &   1   &   2386.3965512410805   &   0.37580060327560794   &   -0.002045799214649983   &   0.1891947607448222   &   7,$\mathbb{R}$   \\
62   &   0   &   5   &   1   &   0   &   2386.404021744886   &   0.37580634048420636   &   -0.00204564331425337   &   0.18920208192386426   &   1,$\mathbb{R}$   \\
63   &   1   &   5   &   0   &   0   &   2386.411481375562   &   0.3758005051229662   &   -0.002045480697078941   &   0.18919515613590995   &   1,$\mathbb{R}$   \\
64   &   0   &   6   &   0   &   0   &   2.86420595379086e13   &   0.21353964978847087   &   -0.0004860554479214159   &   0.04633272594908236   &   5,$\mathbb{R}$ \\
\bottomrule
\end{tabular}
\begin{tablenotes}
\scriptsize
\item[a] Index of the used minimization algorithm given in
\prettyref{tab: global_algorithms}. $\mathbb{N}$ means optimization
with $Z_i$ being limited to the domain of natural numbers and
$\mathbb{R}$ means optimization with $Z_i$ being limited to the domain of real numbers.
\end{tablenotes}
\end{threeparttable}
\end{table*}
\pagebreak
\begin{table*}[h!]
\begin{threeparttable}
\caption{
\scriptsize
Results of a global minimization of the volume in the
$\mathcal{P,T}$-odd parameter space with respect to nuclear charges $Z_i$ for
different classes of molecules (I to IV) in the range of $20\le Z_i\le
\mathbf{90}$. For each combination of molecules of each class the minimal
volume is presented alongside the relative gradient and the minimal
($h_\mathrm{min}$) and maximal ($h_\mathrm{max}$) eigenvalues of the
Hessian relative to the absolute minimal volume. Results are given in
order of increasing volume. The number of shown figures is given for
comparability and does not represent the accuracy of the optimization.}
\label{tab: results_global_min_90}
\scriptsize
\begin{tabular}{
S
S
S
S
S
S[table-format=+1.1e-1,round-mode=places, scientific-notation=true, round-precision =1]
S[table-format=+1.1e-1,round-mode=places, scientific-notation=true, round-precision =1]
S[table-format=+1.1e-1,round-mode=places, scientific-notation=true, round-precision =1]
S[table-format=+1.1e-1,round-mode=places, scientific-notation=true, round-precision =1]
m{1.2cm}
}
\toprule
{Place} &
{$N_\mathrm{I}$} &
{$N_\mathrm{II}$} &
{$N_\mathrm{III}$} &
{$N_\mathrm{IV}$} &
{$V_\mathrm{min}/\mathrm{arb. u.}$} &
{$\frac{\left|\nabla_Z V_\mathrm{min}\right|}{V_\mathrm{min}}$} &
{$\frac{h_\mathrm{min}}{V_\mathrm{min}}$} &
{$\frac{h_\mathrm{max}}{V_\mathrm{min}}$} &
{Algorithm Index\tnote{a}}
\\
\midrule
1   &   2   &   1   &   0   &   3   &   0.0010083335307280905   &   0.2529585001951594   &   -0.00016886287196576916   &   0.07124324895375446   &   2,$\mathbb{R}$   \\
2   &   1   &   1   &   1   &   3   &   0.0010083348300535238   &   0.24966259202924743   &   -0.0004981638150071303   &   0.06937993169182327   &   1,$\mathbb{R}$   \\
3   &   2   &   0   &   1   &   3   &   0.0010236149957345356   &   0.2532424481921574   &   -0.00019588851914844517   &   0.07133169495308936   &   7,$\mathbb{R}$   \\
4   &   2   &   1   &   1   &   2   &   0.0016609397225367923   &   0.2578531460122786   &   -0.00014718115428243971   &   0.0741152151633627   &   2,$\mathbb{R}$   \\
5   &   1   &   1   &   2   &   2   &   0.0016609418380453848   &   0.25438974235709305   &   0.0006691915399806401   &   0.0721295162341874   &   4,$\mathbb{R}$   \\
6   &   2   &   2   &   0   &   2   &   0.0016728063754689305   &   0.2566983889857181   &   -0.0001613104788793521   &   0.07334893255590394   &   1,$\mathbb{R}$   \\
7   &   1   &   2   &   1   &   2   &   0.0016728085063629412   &   0.2533186411582519   &   -0.0009404538771193569   &   0.07141580596534702   &   5,$\mathbb{R}$   \\
8   &   2   &   0   &   2   &   2   &   0.00168611154280786   &   0.25812605765664337   &   -0.00018212334178762353   &   0.07420351593104885   &   2,$\mathbb{R}$   \\
9   &   0   &   1   &   2   &   3   &   0.0018860447564779737   &   0.2569225460210207   &   0.0005121954151491936   &   0.07469319816679186   &   2,$\mathbb{R}$   \\
10   &   1   &   0   &   2   &   3   &   0.0019146256099881018   &   0.25681362501477767   &   -0.0005826870543056239   &   0.07457476617210342   &   1,$\mathbb{R}$   \\
11   &   1   &   2   &   0   &   3   &   0.0020851471818942933   &   0.2503006685447069   &   -0.0017094078441825824   &   0.06919185536884596   &   1,$\mathbb{R}$   \\
12   &   0   &   2   &   1   &   3   &   0.0020851498718847477   &   0.25116480811788383   &   -0.0016406493657145315   &   0.06959881616734158   &   2,$\mathbb{R}$   \\
13   &   0   &   2   &   2   &   2   &   0.002214127915369174   &   0.27343596557793626   &   0.00010791984674998181   &   0.08708436871399197   &   2,$\mathbb{R}$   \\
14   &   2   &   2   &   1   &   1   &   0.003781642544697258   &   0.25165641619531426   &   -0.0027139012022037922   &   0.07185543701175698   &   1,$\mathbb{R}$   \\
15   &   1   &   2   &   2   &   1   &   0.003781647417483256   &   0.2504656184018434   &   -0.002918292225563424   &   0.07101526643642594   &   4,$\mathbb{R}$   \\
16   &   1   &   3   &   0   &   2   &   0.0037853241835688082   &   0.2520624874169892   &   -0.001668588573635737   &   0.07171937548951175   &   1,$\mathbb{R}$   \\
17   &   0   &   3   &   1   &   2   &   0.003785329061131697   &   0.2533829974880768   &   -0.0009803313103111087   &   0.07233601421156104   &   5,$\mathbb{R}$   \\
18   &   2   &   3   &   0   &   1   &   0.0037853383316804575   &   0.25022072963273867   &   -0.0027201539719198097   &   0.07089237169020964   &   7,$\mathbb{R}$   \\
19   &   1   &   3   &   1   &   1   &   0.0037853394770132197   &   0.25008423484110537   &   -0.0025631073632840676   &   0.07082220270518501   &   1,$\mathbb{R}$   \\
20   &   0   &   3   &   2   &   1   &   0.0037853443543982247   &   0.2507951105689987   &   -0.0024831499926368707   &   0.07115100865383335   &   5,$\mathbb{R}$   \\
21   &   2   &   1   &   2   &   1   &   0.003812871465626233   &   0.2503311523848909   &   -0.0027177865427154006   &   0.07088303626949347   &   7,$\mathbb{R}$   \\
22   &   1   &   1   &   3   &   1   &   0.0038128763786201524   &   0.24929655583952007   &   -0.0029414592446519823   &   0.07009041944635787   &   5,$\mathbb{R}$   \\
23   &   0   &   1   &   3   &   2   &   0.003852437685655084   &   0.24931984693956263   &   -6.247701020276244e-6   &   0.06856172604870674   &   7,$\mathbb{R}$   \\
24   &   2   &   0   &   3   &   1   &   0.0038706561769313126   &   0.25104689164351035   &   -0.002787813715825827   &   0.0709241650645871   &   2,$\mathbb{R}$   \\
25   &   1   &   0   &   3   &   2   &   0.003910824112804764   &   0.248812051581604   &   -0.0006869217077146812   &   0.06820710524530917   &   7,$\mathbb{R}$   \\
26   &   0   &   2   &   3   &   1   &   0.003954435830006858   &   0.2460682947919947   &   -0.002448186445644574   &   0.06701494377441498   &   5,$\mathbb{R}$   \\
27   &   1   &   1   &   0   &   4   &   0.0075250501379124455   &   0.31824258937255706   &   -0.0012981294543807315   &   0.1278665863126703   &   1,$\mathbb{R}$   \\
28   &   0   &   1   &   1   &   4   &   0.007525059807077499   &   0.3170509142088163   &   -0.0008730914287118502   &   0.12731035608137883   &   2,$\mathbb{R}$   \\
29   &   1   &   0   &   1   &   4   &   0.0076390935441719   &   0.3188632988947368   &   -0.0008713437254783803   &   0.1281522002877722   &   1,$\mathbb{R}$   \\
30   &   2   &   0   &   0   &   4   &   0.026277085686148473   &   0.357969321877683   &   0.00037991665096193073   &   0.1635469903784268   &   2,$\mathbb{R}$   \\
31   &   0   &   1   &   4   &   1   &   0.028682653233181577   &   0.31572891796338537   &   -0.0019361055421091588   &   0.126544323009943   &   4,$\mathbb{R}$   \\
32   &   1   &   0   &   4   &   1   &   0.02911730650066996   &   0.3155692597870433   &   -0.001946180064973821   &   0.1264179518082292   &   4,$\mathbb{R}$   \\
33   &   0   &   0   &   2   &   4   &   0.09781818296010907   &   0.3558034500277373   &   -0.000474319528553159   &   0.1625654206423562   &   2,$\mathbb{R}$   \\
34   &   0   &   2   &   0   &   4   &   0.11255526117124182   &   0.36139490075960334   &   0.0015660652489601875   &   0.16577969109055762   &   7,$\mathbb{R}$   \\
35   &   0   &   0   &   3   &   3   &   0.1511370337742557   &   0.34312767118304044   &   0.002241487740582813   &   0.1486076132946392   &   2,$\mathbb{R}$   \\
36   &   0   &   3   &   0   &   3   &   0.16188094915483578   &   0.34619050061492784   &   0.0025052369449665105   &   0.1501597801783759   &   2,$\mathbb{R}$   \\
37   &   0   &   0   &   4   &   2   &   0.37660543785457967   &   0.36413340602590216   &   0.0023753286270945346   &   0.17010453187679916   &   2,$\mathbb{R}$   \\
38   &   0   &   3   &   3   &   0   &   0.4082561320563869   &   0.3686560209389094   &   0.002716659107341077   &   0.17327148927952452   &   2,$\mathbb{R}$   \\
39   &   1   &   3   &   2   &   0   &   0.4097892392699262   &   0.3465307584051999   &   -0.00007211611499598231   &   0.14588777572536854   &   8,$\mathbb{N}$   \\
40   &   1   &   2   &   3   &   0   &   0.41444280774442094   &   0.36537994291477566   &   0.001651235956659855   &   0.17093118707073696   &   4,$\mathbb{R}$   \\
41   &   1   &   4   &   0   &   1   &   0.4273694759007848   &   0.36565688156118603   &   0.0004462085443757983   &   0.17229539552290526   &   4,$\mathbb{R}$   \\
42   &   0   &   4   &   1   &   1   &   0.4273700416640414   &   0.36567926971852005   &   0.0004462722571341363   &   0.17233558160243617   &   5,$\mathbb{R}$   \\
43   &   0   &   4   &   0   &   2   &   0.42737977727110804   &   0.3656684509224239   &   0.0004460507766406229   &   0.1723203982780798   &   2,$\mathbb{R}$   \\
44   &   2   &   4   &   0   &   0   &   0.4273808518459817   &   0.36565251475261046   &   0.0004461784337879567   &   0.17229359350766651   &   7,$\mathbb{R}$   \\
45   &   1   &   4   &   1   &   0   &   0.4273813882049544   &   0.36566368320728465   &   0.00044615205731244863   &   0.1723153526100266   &   7,$\mathbb{R}$   \\
46   &   0   &   4   &   2   &   0   &   0.4273814489275683   &   0.36564670418540873   &   0.00044613096921525464   &   0.17228965697367812   &   1,$\mathbb{R}$   \\
47   &   2   &   2   &   2   &   0   &   0.4280661810437786   &   0.3650269761710979   &   0.0004480462212199852   &   0.1716333956187089   &   7,$\mathbb{R}$   \\
48   &   0   &   2   &   4   &   0   &   0.42958837677993966   &   0.3663511575172596   &   0.0016832500961177078   &   0.17161845803842413   &   2,$\mathbb{R}$   \\
49   &   2   &   3   &   1   &   0   &   0.43099819312998083   &   0.3657152490287079   &   0.0004416434146713841   &   0.17238404581479838   &   7,$\mathbb{R}$   \\
50   &   2   &   1   &   3   &   0   &   0.4345536012742426   &   0.36590928262264305   &   0.00041382690522371056   &   0.1718170776635255   &   7,$\mathbb{R}$   \\
51   &   1   &   1   &   4   &   0   &   0.43609830391123366   &   0.36568273532928414   &   0.0007286461132864575   &   0.17110637269216186   &   7,$\mathbb{R}$   \\
52   &   2   &   0   &   4   &   0   &   0.44072198002038265   &   0.36507463307222304   &   0.00040468299903048444   &   0.17078717472782057   &   7,$\mathbb{R}$   \\
53   &   1   &   0   &   0   &   5   &   0.7871005992371439   &   0.36620036861198746   &   -0.0008042565436199927   &   0.18195846979214053   &   1,$\mathbb{R}$   \\
54   &   0   &   0   &   1   &   5   &   1.4258478851932892   &   0.36307107152770884   &   -0.0008127922402379291   &   0.17867596187390472   &   2,$\mathbb{R}$   \\
55   &   0   &   1   &   0   &   5   &   1.7283213480446544   &   0.3697557946690117   &   -0.0003508510708337847   &   0.1858472234617002   &   2,$\mathbb{R}$   \\
56   &   0   &   0   &   5   &   1   &   8.314907808378758   &   0.3538255883595072   &   -0.0008316899790603488   &   0.1690668040655792   &   2,$\mathbb{R}$   \\
57   &   0   &   1   &   5   &   0   &   9.560026229248964   &   0.35688612059335834   &   -0.0003717334016491034   &   0.1723161769865556   &   7,$\mathbb{R}$   \\
58   &   1   &   0   &   5   &   0   &   9.704897453459889   &   0.3571112253943344   &   -0.0007934505118201712   &   0.17241369264603726   &   1,$\mathbb{R}$   \\
59   &   0   &   0   &   0   &   6   &   5980.1047861229945   &   0.4150965140678629   &   0.0047261499298480265   &   0.2417253784811908   &   7,$\mathbb{R}$   \\
60   &   0   &   0   &   6   &   0   &   109565.24464637873   &   0.4117647156015781   &   0.00489086264360952   &   0.23809640428730305   &   7,$\mathbb{R}$   \\
61   &   1   &   5   &   0   &   0   &   210541.5123796234   &   0.36197009894644655   &   -0.0006202471028151738   &   0.17795400335956285   &   2,$\mathbb{R}$   \\
62   &   0   &   5   &   1   &   0   &   210541.78361775435   &   0.36196848733543163   &   -0.0006202850689399336   &   0.1779520663922757   &   1,$\mathbb{R}$   \\
63   &   0   &   5   &   0   &   1   &   210542.0549797274   &   0.3619699093019406   &   -0.0006203216940528137   &   0.1779538004240763   &   4,$\mathbb{R}$   \\
64   &   0   &   6   &   0   &   0   &   2.2696419475559535e15   &   0.23211914167104156   &   -0.000310239333379622   &   0.05950681590770997   &   4,$\mathbb{R}$ \\ 
\bottomrule
\end{tabular}
\begin{tablenotes}
\scriptsize
\item[a] Index of the used minimization algorithm given in
\prettyref{tab: global_algorithms}. $\mathbb{N}$ means optimization
with $Z_i$ being limited to the domain of natural numbers and
$\mathbb{R}$ means optimization with $Z_i$ being limited to the domain of real numbers.
\end{tablenotes}
\end{threeparttable}
\end{table*}

\end{document}


\title{Supplementary Material to ``Global analysis of $\mathcal{CP}$-violation in atoms, molecules and
role of medium-heavy systems''}
\date{\today}
\author{Konstantin Gaul}
\affiliation{Fachbereich Chemie, Philipps-Universit\"{a}t Marburg, Hans-Meerwein-Stra\ss{}e 4, 35032 Marburg}
\author{Robert Berger}
\affiliation{Fachbereich Chemie, Philipps-Universit\"{a}t Marburg, Hans-Meerwein-Stra\ss{}e 4, 35032 Marburg}
\begin{abstract}
\phantom{a}
\end{abstract}
\maketitle
\section{Nuclear charge number distributions}
\begin{figure*}[h!]
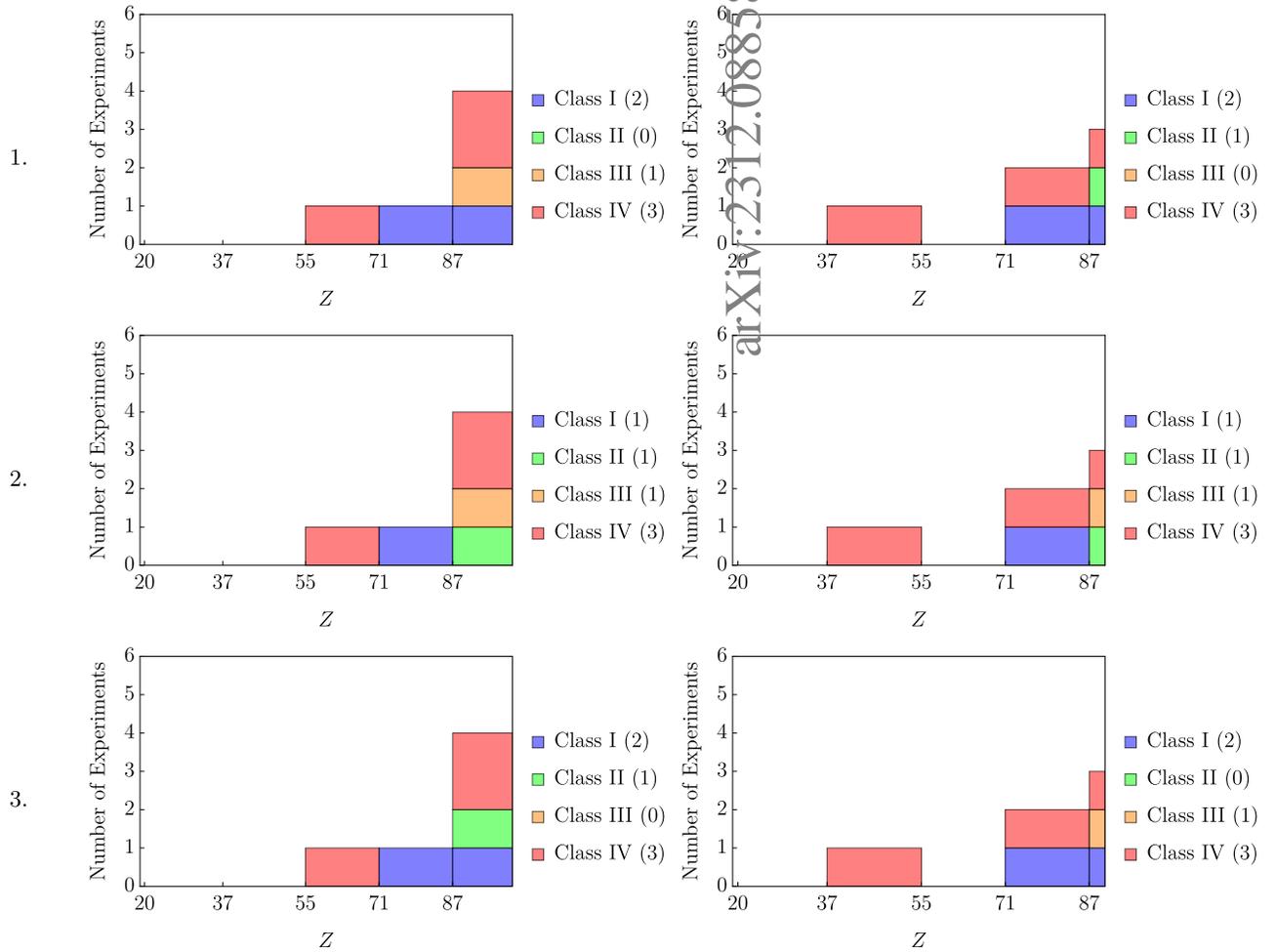

\caption{Optimal nuclear charge number distributions for the global minimum
for different possible choices of atoms and molecules from different
classes (see main text). Results are shown for minimization in the regions of $20\leq Z\leq100$ (left) and
$20\leq Z\leq90$ (right). The binning is chosen such that bins
represent the rows of the periodic table. The first bin corresponds to
the 4th row of the periodic table ($20\leq Z\leq34$), the second bin
corresponds to the 5th row of the periodic table ($35\leq54$), the
third and fourth bin correspond to the first ($55\leq Z\leq70$) and
second half ($71\leq Z\leq86$) of the 6th row of the periodic table and the last bin
includes the elements of the 7th row of the periodic table up to
$Z=90$ or $Z=100$.}
\label{fig: paper_globaledm_fig_suppl}
\foreach \x in {1,...,3}%
{ \begin{minipage}{.1\textwidth}
  \x.
  \end{minipage}%
  \begin{minipage}{.45\textwidth}
  \includegraphics[width=\textwidth]{ptodd_global_charge_distribution100_place\x}
  \end{minipage}%
  \begin{minipage}{.45\textwidth}
  \includegraphics[width=\textwidth]{ptodd_global_charge_distribution90_place\x}
  \end{minipage}%
  \ifnum \x < 3 
    \\%
  \fi
}
\end{figure*}

\begin{figure*}
\ContinuedFloat
\foreach \y in {\the\numexpr4\relax,...,\the\numexpr8\relax}%
{ \begin{minipage}{.1\textwidth}
  \y.
  \end{minipage}%
  \begin{minipage}{.45\textwidth}
  \includegraphics[width=\textwidth]{ptodd_global_charge_distribution100_place\y}
  \end{minipage}%
  \begin{minipage}{.45\textwidth}
  \includegraphics[width=\textwidth]{ptodd_global_charge_distribution90_place\y}
  \end{minipage} %
  \ifnum \y < \the\numexpr8\relax
  \\%
  \fi
}
\end{figure*}

\foreach \x in {3,...,16}%
{ \begin{figure*}
  \ContinuedFloat
  \foreach \y in {\the\numexpr(\x-1)*4+1\relax,...,\the\numexpr\x*4\relax}%
  { \begin{minipage}{.1\textwidth}
    \y.
    \end{minipage}%
    \begin{minipage}{.45\textwidth}
    \includegraphics[width=\textwidth]{ptodd_global_charge_distribution100_place\y}
    \end{minipage}%
    \begin{minipage}{.45\textwidth}
    \includegraphics[width=\textwidth]{ptodd_global_charge_distribution90_place\y}
    \end{minipage} %
    \ifnum \y < \the\numexpr\x*4\relax
    \\%
    \fi
  }
  \end{figure*}
}